\documentclass[aps,prd,10pt,twocolumn,preprintnumbers,superscriptaddress,nofootinbib,letterpaper,longbibliography]{revtex4-1}

\usepackage{hyperref}
\usepackage{graphicx}
\usepackage{amsfonts,amsmath,amssymb,bm}
\usepackage{array}
\usepackage{xcolor}
\usepackage{comment}
\usepackage{soul}
\usepackage[utf8]{inputenc}
\usepackage{tikz}
\usepackage{multirow}
\usepackage{boldline}
\usepackage{svg}
\usepackage{hyperref}
\hypersetup{
  colorlinks  = true,
  urlcolor    = blue,
  linkcolor   = blue,
  citecolor   = red
}

\newcommand{\orcid}[1]
{\begingroup
  \hypersetup{hidelinks}\href{https://orcid.org/#1}{\includegraphics[width=9pt]{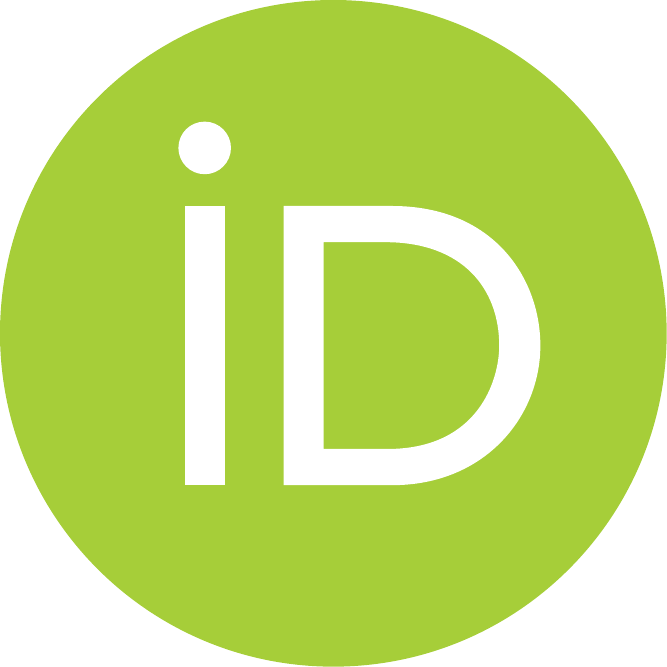}
} \endgroup}

\newcolumntype{P}[1]{>{\centering\arraybackslash}p{#1}}

\graphicspath{{figures/}{figures_appendix/}}

\begin{document}


\title{Old Data, New Forensics: The First Second of SN 1987A Neutrino Emission}

\author{Shirley Weishi Li \orcid{0000-0002-2157-8982}}
\email{shirley.li@uci.edu}
\affiliation{Theoretical Physics Department, Fermi National Accelerator Laboratory, Batavia, IL 60510}
\affiliation{Department of Physics and Astronomy, University of California, Irvine, CA 92697}

\author{John F. Beacom \orcid{0000-0002-0005-2631}}
\email{beacom.7@osu.edu}
\affiliation{Center for Cosmology and AstroParticle Physics (CCAPP), Ohio State University, Columbus, OH 43210}
\affiliation{Department of Physics, Ohio State University, Columbus, OH 43210}
\affiliation{Department of Astronomy, Ohio State University, Columbus, OH 43210}

\author{Luke F. Roberts \orcid{0000-0001-7364-7946}}
\email{lfroberts@lanl.gov}
\affiliation{Computer, Computational, and Statistical Sciences Division, Los Alamos National Laboratory, Los Alamos, NM 87545}

\author{Francesco Capozzi \orcid{0000-0001-6135-1531}}
\email{francesco.capozzi@univaq.it}
\affiliation{Dipartimento di Scienze Fisiche e Chimiche, Universit\`a degli Studi dell'Aquila, 67100 L’Aquila, Italy}
\affiliation{Istituto Nazionale di Fisica Nucleare (INFN), Laboratori Nazionali del Gran Sasso, 67100 Assergi (AQ), Italy}

\date{June 13, 2023}


\begin{abstract}
The next Milky Way supernova will be an epochal event in multi-messenger astronomy, critical to tests of supernovae, neutrinos, and new physics.  Realizing this potential depends on having realistic simulations of core collapse.  We investigate the neutrino predictions of modern models (1-, 2-, and 3-d) over the first $\simeq$1 s, making the first detailed comparisons of these models to each other and to the SN 1987A neutrino data.  Even with different methods and inputs, {\it the models generally agree with each other}.  However, even considering the low neutrino counts, {\it the models generally disagree with data}.  What can cause this?  We show that neither neutrino oscillations nor different progenitor masses appear to be a sufficient solution.  We outline urgently needed work.
\end{abstract}

\preprint{FERMILAB-PUB-23-087-PPD, UCI-HEP-TR-2023-02, LA-UR-23-22079}

\maketitle


\section{Introduction}

As spectacular as SN 1987A was for multi-messenger astronomy --- with detections across the electromagnetic spectrum, plus $\simeq$19 neutrino events~\cite{Hirata:1987hu, Hirata:1988ad, Bionta:1987qt, IMB:1988suc, Arnett:1989tnf, 1993ARA&A..31..175M} --- the next Milky Way core-collapse supernova should be much more so~\cite{Scholberg:2012id, Adams:2013ana, Mirizzi:2015eza, Nakamura:2016kkl}.  We have dramatically better sensitivity to neutrinos, a key observable because they carry the dominant energy release and because they probe the dynamics of the inner core.  And we have dramatically better sensitivity across the electromagnetic spectrum and to gravitational waves.  Because nearby supernovae are rare --- about (2$\pm 1)$/century~\cite{Diehl:2006cf, 2011MNRAS.412.1473L, Rozwadowska:2020nab} --- we likely have just one chance over the next few decades to get this right.

To interpret the supernova data, numerical simulations of core collapse will be essential.  Recently, newly sophisticated approaches --- including 3-d, multi-energy-group radiation-hydrodynamics models of successful explosions --- have become available~\cite{Hanke:2013, Takiwaki:2013cqa, Lentz:2015, Janka:2016fox, OConnor:2018sti, 2013ApJ...767L...6B, Bruenn:2014qea, OConnor:2015rwy, Summa:2015nyk, Kotake:2018ypf, Vartanyan:2018xcd, Ott:2017kxl, OConnor:2018tuw, Glas:2018oyz, Burrows:2019zce}.  These models predict explosion properties (e.g., final energies and remnant masses) as well as neutrino signals up to $\simeq$1~s after core bounce, which is crucial for assessing explodability and which includes a large fraction of the total emission.  However, the readiness of these models for comparison to the next supernova has not been adequately assessed.  

In this paper, we tackle two distinct but related problems.  Our first goal is to compare models to each other, which gives an estimate of the {\it modeling uncertainties}.  Our second goal is to compare models to the SN 1987A data~\cite{Hirata:1987hu, Hirata:1988ad, Bionta:1987qt, IMB:1988suc}, which gives an estimate of the {\it physical uncertainties}.  There is very limited prior work comparing models to each other~\cite{OConnor:2018sti} and models to data~\cite{OConnor:2013, Olsen:2021uvt}.  What makes the present work possible is advances in the breadth and precision of supernova models, which give us new forensic tools to examine the SN 1987A data.

In the following, we first consider a nominal case of a 20$M_\odot$ (initial mass) single-star progenitor with no neutrino oscillations.  This progenitor was initially thought to be appropriate for SN 1987A~\cite{Woosley:1988, Arnett:1989tnf} and thus has the broadest set of supernova models.  While neglecting neutrino oscillations is not realistic, it matches supernova simulation outputs and is well-defined.  We allow other aspects of the simulations, including the dimensionality (1-d, 2-d, and 3-d), to vary freely so that we can include all modern predictions~\cite{OConnor:2018sti, 2013ApJ...767L...6B, Bruenn:2014qea, OConnor:2015rwy, Summa:2015nyk, Kotake:2018ypf, Vartanyan:2018xcd, Ott:2017kxl, OConnor:2018tuw, Glas:2018oyz, Burrows:2019zce}.  Then, to test the impact of changing two key theoretical inputs, we vary the neutrino-oscillation scenario and the progenitor mass.  Last, we conclude and discuss actions needed to prepare for the next Milky Way neutrino burst.  In the appendix, we provide supporting details.


\section{Review of Supernova Models}

In core-collapse supernovae~\cite{Mezzacappa:2005ju, Smartt:2009, Janka:2012wk, Burrows:2020qrp}, the white-dwarf-like iron core of a massive star collapses to form a proto-neutron star (PNS), releasing nearly all of the gravitational binding energy difference, $E_{\rm tot} \simeq (3/5) G M_\textrm{PNS}^2/R_\textrm{PNS} \simeq 3 \times 10^{53} \, \textrm{erg}$, in neutrinos of all flavors with comparable fluences.  Neutrinos diffuse out of the hot, dense, neutron-rich PNS, decoupling at the neutrinospheres, with average energies of $\simeq$10--15~MeV.  About half of the total energy is emitted in the first $\simeq$1~s after core bounce, largely powered by accretion onto the PNS, with the other half released over $\sim$10~s, as the PNS cools and deleptonizes.  In the neutrino mechanism~\citep{Colgate:1966, Bethe:1985}, after decoupling from the PNS, a few percent of the early-time neutrinos interact with the collapsing layers of the star outside the PNS, potentially reversing the infall and driving an explosion.

To understand the detailed physics and astrophysics of core collapse, large-scale multi-dimensional simulations are critical~\cite{Hanke:2013, Takiwaki:2013cqa, Lentz:2015, Janka:2016fox, OConnor:2018sti, 2013ApJ...767L...6B, Bruenn:2014qea, OConnor:2015rwy, Summa:2015nyk, Kotake:2018ypf, Vartanyan:2018xcd, Ott:2017kxl, OConnor:2018tuw, Glas:2018oyz, Burrows:2019zce}.  Starting from a pre-explosion massive-star progenitor model and choices for the equation of state and neutrino opacities of dense matter, modern simulations evolve the equations of non-equilibrium neutrino transport, (magneto-)hydrodynamics, and gravity for as long as is computationally feasible.  In the last decade, the simulation community has made significant progress towards showing the viability of the neutrino mechanism in multi-dimensional simulations and in predicting the observed properties of supernovae.  Nevertheless, these models have shortcomings, including the neglect of neutrino oscillations, significant uncertainties in the progenitor models, often under-resolved hydrodynamic flows, and simulation times of $\lesssim$1~s after bounce, which misses the PNS cooling phase~\cite{Pons:1998mm, Nakazato:2012qf, Nakazato:2019ojk, Li:2020ujl}.


\section{Review of Supernova 1987A}
Multi-messenger observations of SN 1987A confirmed that a Type-II supernova is driven by the collapse of the core of a massive star into a PNS, powering a neutrino burst from the core and a delayed optical burst from the envelope~\cite{Arnett:1989tnf, 1993ARA&A..31..175M}.

The water-Cherenkov experiments Kamiokande-II (Kam-II) and Irvine-Michigan-Brookhaven (IMB) detected $\simeq$19 $\bar{\nu}_e$ events in total via inverse beta decay, $\bar{\nu}_e + p \rightarrow e^+ + n$, over $\simeq$10 s~\cite{Hirata:1987hu, Hirata:1988ad, Bionta:1987qt, IMB:1988suc}.  (We neglect the Baksan experiment because it was $\simeq$8 times smaller than Kam-II and had significant backgrounds~\cite{Alekseev:1988gp, Loredo:2001rx}.)  Though only one flavor was clearly detected, the results were broadly consistent with basic expectations for the total energy, average neutrino energy, and duration of the neutrino pulse.  Theoretical analyses included comparisons to the supernova models of the time~\cite{Burrows:1987zz, Bruenn:1987, Suzuki:1987}, which were far less sophisticated than those available today.

Observations across the electromagnetic spectrum, at the time and since, have also been critical for understanding the explosion~\cite{Arnett:1989tnf, 1993ARA&A..31..175M, Pumo:2023qoy}.  Initially, it was thought that the pre- and post-supernova observations were consistent with those expected for a 20$M_\odot$ single-star progenitor~\cite{Hillenbrandt:1987, Woosley:1987, Saio:1988}.  Later work claimed that a binary-merger scenario is favored~\cite{Posiadlowski:1992}, though there is no consensus on this.  On the one hand, the binary-progenitor models of Ref.~\cite{Menon:2017hva} suggest that the helium core mass may be substantially smaller --- and the envelope mass substantially larger --- than the values found for typical single-star progenitors.  On the other hand, the binary-progenitor models of Refs.~\cite{Urushibata:2017hnl, Nakamura:2022zlc} suggest that the pre-collapse structure of the merger remnant is not so different than that predicted for single-star 20$M_\odot$ progenitors.


\begin{figure}[t]
\centering
\includegraphics[width=\columnwidth]{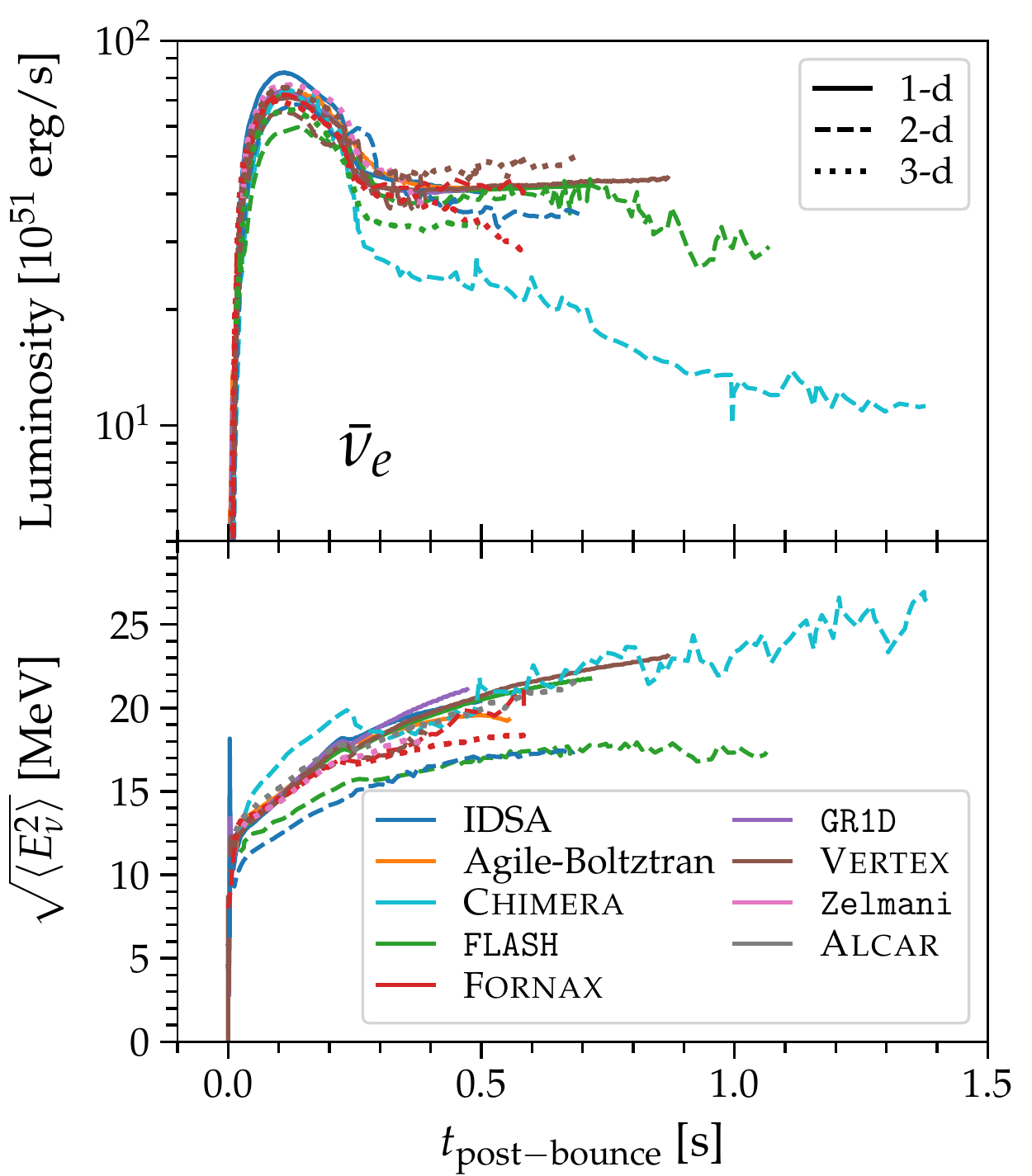}
\caption{Neutrino ($\bar\nu_e$; others shown in S.M.) luminosity and RMS energy profiles from supernova simulations~\cite{OConnor:2018sti, 2013ApJ...767L...6B, Bruenn:2014qea, OConnor:2015rwy, Summa:2015nyk, Kotake:2018ypf, Vartanyan:2018xcd, Ott:2017kxl, OConnor:2018tuw, Glas:2018oyz, Burrows:2019zce}.}
\label{fig:luminosity_comparison}
\end{figure}

\section{Comparing models}
We first consider the nominal case of a 20$M_\odot$ progenitor and no neutrino oscillations.  For all modern 1-, 2-, and 3-d models with 20$M_\odot$ progenitors~\cite{OConnor:2018sti, 2013ApJ...767L...6B, Bruenn:2014qea, OConnor:2015rwy, Summa:2015nyk, Kotake:2018ypf, Vartanyan:2018xcd, Ott:2017kxl, OConnor:2018tuw, Glas:2018oyz, Burrows:2019zce}, we collect information on their neutrino fluxes and spectra.   We seek to assess the full variation between models, though they are not completely distinct, e.g., many share progenitors~\cite{Woosley:2007as}.  These models vary significantly, but we do not attempt to adjudicate between them.  Most multi-d models lead to successful explosions, with explosion times ranging from 0.2--0.8 s.  The model details are given in S.M.

Figure~\ref{fig:luminosity_comparison} shows their time profiles of $\bar\nu_e$ luminosity and root-mean-square (RMS) energy (other flavors in S.M.).  For the spectra, we assume a commonly used form,
$f_\alpha(E_\nu) = 
\mathcal{N}
\left({E_\nu}/{\langle E_\nu\rangle}\right)^{\alpha-2} 
e^{-(\alpha+1)E_\nu/\langle E_\nu\rangle}$,
where $\langle E_\nu\rangle$ is the average energy and $\alpha$ sets the spectrum shape~\cite{Keil:2002in}.  Different groups characterize spectra differently, which we correct for in S.M.

To model the detected spectra, we follow standard calculations (e.g., Refs.~\cite{Jegerlehner:1996kx, Lunardini:2004bj, Costantini:2006xd, Pagliaroli:2008ur, Vissani:2014doa}) and give details in S.M.  The dominant process is $\bar{\nu}_e + p \rightarrow e^+ +n$ with free (hydrogen) protons, for which we take the cross sections and kinematics from Refs.~\cite{Vogel:1999zy, Strumia:2003zx}.  To model the detectors, we take into account their fiducial masses, energy resolutions, and trigger efficiencies~\cite{Hirata:1987hu, Hirata:1988ad, Bionta:1987qt, IMB:1988suc}.  Because of the different detector responses, the detected positron energies are expected to be significantly lower for Kam-II than IMB.  For the distance of SN 1987A, we use 51.4~kpc~\cite{1999IAUS..190..549P}.

We compare the predicted and observed SN 1987A neutrino data using simple, robust observables and statistical tests.  We conduct goodness-of-fit tests (computing p-values) between pairs of models and between each model and the SN 1987A data.  {\it Because we are testing goodness-of-fit, rather than doing parameter estimation, maximum likelihood is not a suitable method; see S.M.}

\begin{figure}[t]
\centering
\includegraphics[width=\columnwidth]{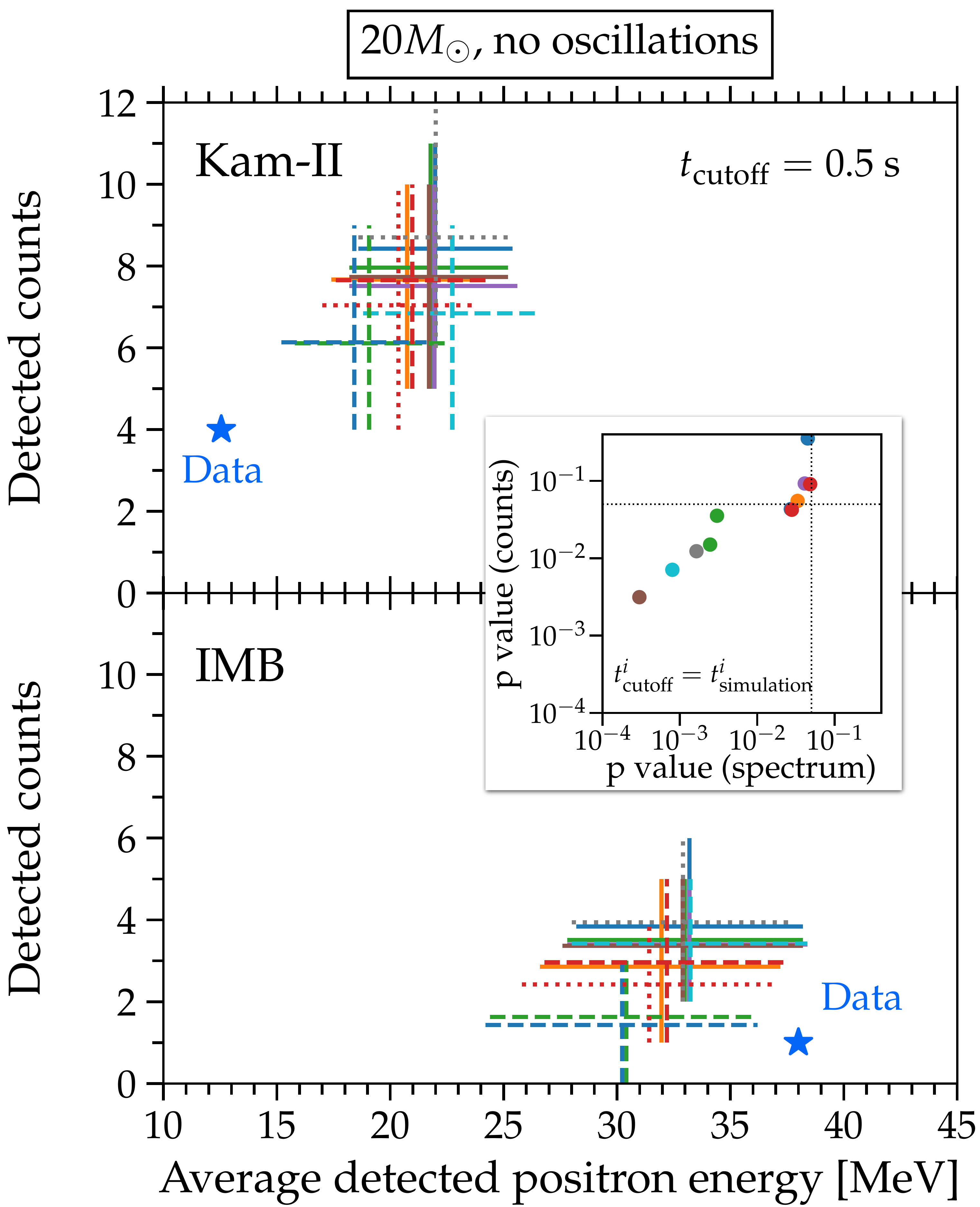}
\caption{Predicted counts and average energies of supernova models (colors as in Fig.~\ref{fig:luminosity_comparison}), compared to each other and to SN 1987A data. {\it Our main calculations are in the inset.}}
\label{fig:comparison}
\end{figure}

The main panels of Figs.~\ref{fig:comparison}--\ref{fig:progenitor} show simple visual comparisons of the counts and average detected energies.  For consistency, here we cut off all models at 0.5~s.  For each model, we forward-model the predicted signals, taking into account properties of the individual detectors and Poisson fluctuations.  Because we are assessing {\it goodness of fit} for many models, we directly calculate $P({\rm data}|{\rm model})$ and show the full uncertainties on the models instead of the data.

{\it The insets of Figs.~\ref{fig:comparison}--\ref{fig:progenitor} show our main statistical calculations.}  Larger p-values indicate agreement; a priori, we defined $p<0.05$ as indicating inconsistency for a given model, {\it though our focus is on what happens for the majority of models.}  Here, we allow each model to go to its full run time (typically 0.5--1.5~s), considering both the counts in the time profile and the shape of the energy spectrum, which we treat separately, given the short timescale and the low statistics.  The spectrum p-value turns out to be the more powerful indicator.  For the counts tests, the p-values are the one-sided cumulative Poisson probabilities.  For the spectrum tests, we use one-dimensional Kolmogorov-Smirnov statistics, following Monte Carlo modeling of the predicted data.  We allow free time offsets between the predictions and the data (core bounce to first event), finding that these values are $\simeq$0.1~s for Kam-II and $\simeq$0.2~s for IMB, both small, so this freedom does not affect our results.


\section{Results for the Nominal Case}
Figure~\ref{fig:comparison} shows the model-to-model comparisons for a 20$M_\odot$ progenitor with no neutrino oscillations.  The p-values (not shown) obtained by comparing pairs of models range over 0.06--0.52 for the counts (Kam-II and IMB combined) and 0.03--0.99 for the spectra (Kam-II only, as IMB has too few counts).  Thus we find general {\it agreement} among the model predictions.  Considering the range and complexity of the inputs and methods in supernova modeling, this agreement is encouraging, though it remains important to understand the residual differences.  Also, this agreement is partially due to restricting all models to a short runtime of 0.5~s (see discussion near Fig.~\ref{fig:oscillation}).

Figure~\ref{fig:comparison} also shows the model-to-data comparisons.  A general {\it disagreement} in both the counts and spectra is evident.  The predicted counts are too high for Kam-II and mostly too high for IMB.  The predicted average detected energies are too high for Kam-II and slightly too low for IMB.  (Because IMB has just one detected event in this time range, we do not use the predicted spectrum in our statistical tests; hence our results do not merely reflect the well-known spectrum tension with Kam-II.)  Quantitatively, no model-to-data comparisons have both p-values larger than 0.05, and many are much worse.  The two p-values tend to be correlated because the counts depend on $E_{\rm tot} \langle E_\nu \rangle$; we show both p-values for illustration.  The results suggest that an overall cooler neutrino spectrum could explain all of the data except the detected average energy for IMB, which might be accommodated by an enhanced spectrum tail~\cite{Yuksel:2007mn, Nagakura:2020gls}.


\section{Possible Solutions: Neutrino Oscillations}
The data are affected by neutrino oscillations~\cite{Duan:2010bg, Mirizzi:2015eza, Horiuchi:2018ofe, Tamborra:2020cul, Capozzi:2022slf, Richers:2022zug}, with the effects depending upon differences in the initial neutrino luminosities and spectra, set by differences in their production processes and opacities.  The details depend on the high densities of matter and other neutrinos, which are uncertain~\cite{Mirizzi:2015eza}. Generally, large flavor conversions due to coherent forward scattering on electrons are expected in the stellar envelope~\cite{Dighe:1999bi}, while conversions induced by forward scattering of neutrinos among themselves and those induced by inelastic collisions can occur starting from the neutrinospheres~\cite{Tamborra:2020cul, Capozzi:2022slf, Richers:2022zug}.

We study the effects of oscillations with a few representative cases.  Considering only matter-induced effects, if the neutrino mass ordering follows the inverted hierarchy (IH), there can be a nearly complete exchange of the $\bar{\nu}_e$ and $\bar{\nu}_x$ ($\bar{\nu}_\mu$ and $\bar{\nu}_\tau$) flavors, with almost no change for the $\nu_e$ and $\nu_x$ flavors~\cite{Dighe:1999bi}.  In the normal hierarchy (NH), the opposite occurs.  With neutrino-induced effects, it is possible to have nearly complete equilibration of all six flavors soon after decoupling, because of rapid flavor conversions induced by interactions of neutrinos among themselves~\cite{Tamborra:2020cul, Capozzi:2022slf, Richers:2022zug}.  Further details in S.M.

\begin{figure}[t]
\centering
\includegraphics[width=\columnwidth]{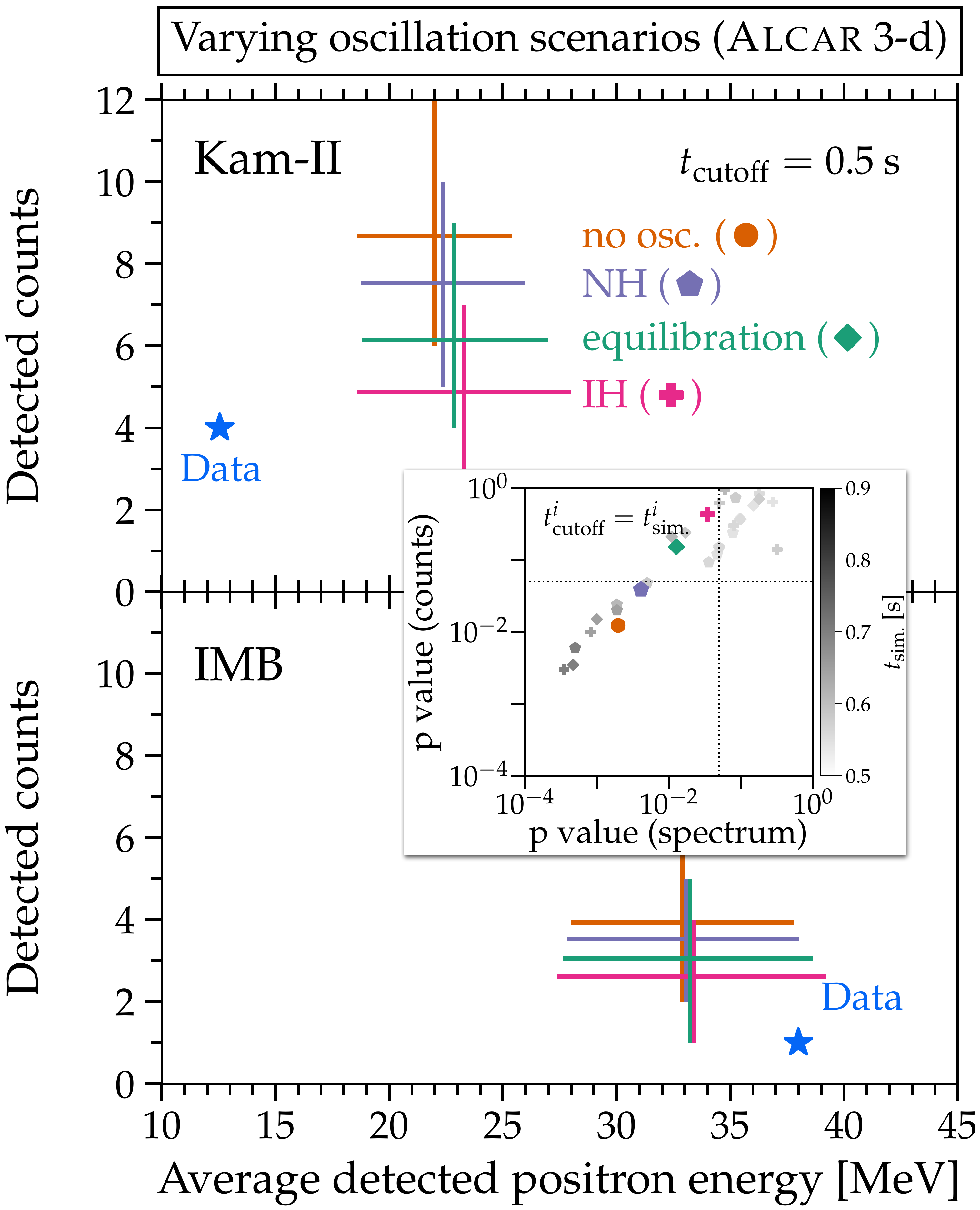}
\caption{Similar to Fig.~\ref{fig:comparison}, with the theory predictions from the \textsc{Alcar} 3-d model for 20 $M_\odot$~\cite{Glas:2018oyz} and different oscillation scenarios.  The simulation cutoff time is 0.68~s.  The grey symbols show the p-values from other models (see S.M.).}
\label{fig:oscillation}
\end{figure}

Figure~\ref{fig:oscillation} shows the range of effects for neutrino-oscillation scenarios (for the \textsc{Alcar} 3-d~\cite{Glas:2018oyz} model and a 20$M_\odot$ progenitor, chosen because it is 3-d and has a long runtime).  In general, oscillations decrease the predicted counts and increase the average energies, as $\bar\nu_x$ has lower fluxes but higher energies than $\bar\nu_e$. For most models (including this one), the reduction in flux is more significant.  The poor agreement may indicate that neutrino oscillations need to be implemented in supernova simulations~\cite{Ehring:2023lcd, Nagakura:2023mhr, Ehring:2023abs}.  Note (see greyscale in inset) that the models with reasonable p-values all have short runtimes.


\begin{figure}[t]
\centering
\includegraphics[width=\columnwidth]{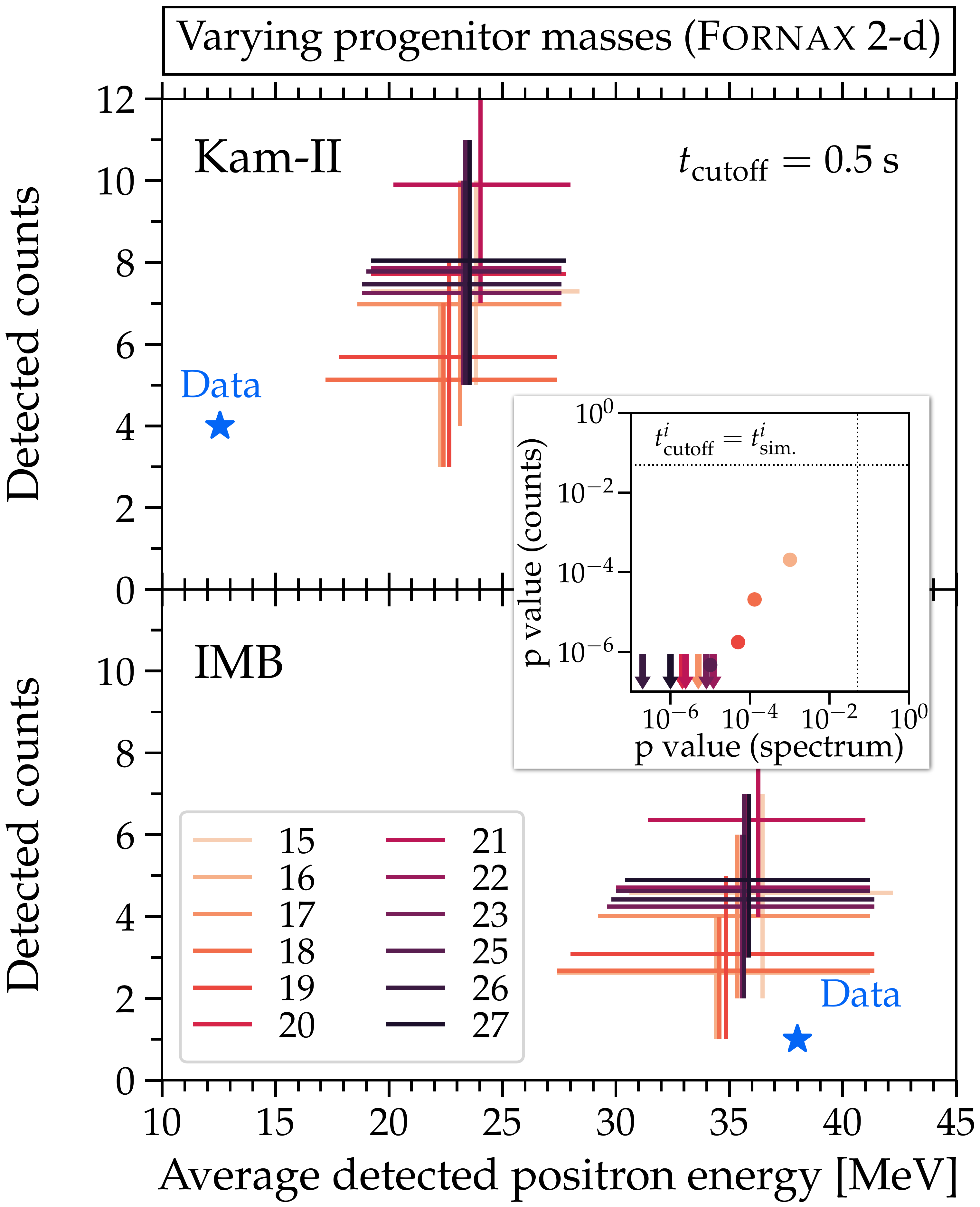}
\caption{Similar to Fig.~\ref{fig:comparison}, with the theory predictions corresponding to different progenitors from F{\sc{ornax}} 2-d models~\cite{Burrows:2020qrp}, with no oscillations included.}
\label{fig:progenitor}
\end{figure}

\section{Possible Solutions: Supernova Progenitors}
The data are also affected by the choice of progenitor~\cite{Woosley:1986, OConnor:2013}. The structure of the star at collapse determines the accretion rate onto the PNS, which strongly influences the neutrino emission.  A key question is if models with progenitors other than the 20$M_\odot$ single-star cases would better fit the SN 1987A data.

Figure~\ref{fig:progenitor} shows the effects of different choices of progenitor mass, using the suite of F{\sc{ornax}} 2-d models~\cite{Burrows:2020qrp}, chosen because of their wide range of progenitor masses (we use 16--30$M_\odot$) and long runtimes.  None of the progenitors provides a good fit to the data.  We have also carried out this analysis (see selected results in S.M.) for the suites of models from Refs.~\cite{Summa:2015nyk, Vartanyan:2018iah, Burrows:2019rtd, Vartanyan:2019ssu, Nagakura:2019gmh, Burrows:2019zce, Warren:2019lgb, SNEWS:2021ewj}, again finding poor agreement with data.  We also considered non-isotropic emission~\cite{Tamborra:2014, Takiwaki:2018, Lin:2019wwm, Walk:2019ier}; however, the predicted changes to the number flux are less than $\pm 20$\% and those on the average energy are even less, too small to consistently explain the observed discrepancies.

The lack of agreement between models and SN 1987A data may indicate missing physics.  Softening the predicted neutrino spectra would require changing the thermodynamic structure of the outer layers of the PNS or changing the neutrino opacity in those regions.  However, it may just be that there are not enough models.  There are not many studies of the same progenitors with different simulation codes, and there are also not many different progenitors in use.  Additionally, binary-merger models for the progenitor of SN 1987A (which do not clearly map onto single-star models) are needed, but only Ref.~\cite{Nakamura:2022zlc} provides neutrino predictions.  A broad range of new simulation work on progenitors and core collapse is needed, especially combining both high sophistication and long runtime.  A big step in that direction is made in Ref.~\cite{Bollig:2020phc} (for $19M_\odot$), but which we find gives a comparably poor match to the SN 1987A data.


\section{Conclusions and Ways Forward}
The SN 1987A neutrino and electromagnetic data, which reasonably agreed with supernova models of the time, have been of critical importance to our understanding of core-collapse supernovae.  It is commonly assumed that modern supernova models --- with 36 years of improvements --- also match the data.  We revisit this assumption.

We show that modern supernova models (for a $20M_\odot$ progenitor and no neutrino oscillations) disagree with 1987A neutrino data in the first $\simeq$0.5--1.5~s, where the highest-precision models end.  When we include the effects of neutrino oscillations or vary the progenitor mass, the tension with data changes somewhat but remains.  We also show that modern supernova models are in good agreement with each other, which suggests that there may be a common solution to the disagreement with SN 1987A data, perhaps even one that also improves explosion energies in simulations.  There is a range of possibilities, including that our implementation or even understanding of the physics in the simulations is incomplete, that not enough progenitor models have been considered, that the initial neutrino spectra are significantly nonthermal, or that neutrino oscillations need to be directly implemented in supernova simulations.  Separately, it would also be interesting to reanalyze the raw data from both detectors, using present detector-modeling and event-reconstruction techniques, both vastly improved over those from 36 years ago.

To realize the full potential of the signals from the next Milky Way supernova, the community will ultimately need a set of modern models that agree (for the same progenitor mass) with each other and the neutrino and electromagnetic data.  Confidence in this would greatly increase if the same were achieved for SN 1987A.  Reaching these goals should be pursued with urgency.  It is especially important that multi-d supernova simulations push their run times out to a few seconds, beyond which PNS cooling simulations may be adequate.

\medskip

{\bf Note added:}  A few months after our work appeared on arXiv, Ref.~\cite{Fiorillo:2023frv} appeared, also investigating how modern supernova simulation predictions compare to 1987A neutrino data.  We discuss this paper in Appendix~\ref{sec:model_selection}.


\section*{Acknowledgements}

We are grateful for helpful discussions with Beno\^it Assi, Elias Bernreuther, Adam Burrows, Nikita Blinov, Basudeb Dasgupta, Sebastian Ellis, Ivan Esteban, Damiano Fiorillo, Chris Fryer, Malte Heinlein, Christopher Hirata, Josh Isaacson, Thomas Janka, Daniel Kresse, Gordan Krnjaic, Bryce Littlejohn, John LoSecco, Sam McDermott, Anthony Mezzacappa, Alessandro Mirizzi, Masayuki Nakahata, Evan O'Connor, Ryan Plestid, Georg Raffelt, Prasanth Shyamsundar, Michael Smy, Todd Thompson, Edoardo Vitagliano, and especially Pedro Machado.  We acknowledge the use of the {\tt KS2D} code.  S.W.L. was supported at FNAL by the Department of Energy under Contract No.\ DE-AC02-07CH11359 during the early stage of this work.  J.F.B. was supported by NSF Grant No.\ PHY-2012955. The work of F.C. was supported in part by Ministero dell’Istruzione, Università e della Ricerca (MIUR) under the program PRIN 2017, Grant 2017X7X85K ``The dark universe: synergic multimessenger approach".




\appendix

\ifx \standalonesupplemental\undefined
\setcounter{figure}{0}
\setcounter{table}{0}
\setcounter{equation}{0}
\setcounter{secnumdepth}{2}
\fi

\renewcommand{\thefigure}{A\arabic{figure}}
\renewcommand{\theHfigure}{A\arabic{figure}}
\renewcommand{\thetable}{A\arabic{table}}


\bigskip
\bigskip
\centerline{\bf \large Appendix}
\medskip

Here we provide additional details that may be useful.  Appendix~\ref{sec:models} summarizes key aspects of the supernova models, Appendix~\ref{sec:data} shows the 1987A data we use, Appendix~\ref{sec:lum_spectrum} focuses on how the average energy and spectrum parameter are calculated for each model, Appendix~\ref{sec:smearing_efficiency} explains the calculation of the detected positron energy spectra, Appendix~\ref{sec:cumu_dist_ks} provides details about {the statistical tests}, Appendix~\ref{sec:model_selection} explains our model selection criteria compared to Ref.~\cite{Fiorillo:2023frv},  Appendix~\ref{sec:progenitors} presents results on progenitor variations, and Appendix~\ref{sec:oscillations} shows details of the oscillation calculations.


\section{Supernova simulations}
\label{sec:models}

\renewcommand{\arraystretch}{1.5}
\begin{table*}[h]
\centering
\begin{tabular}{||c||c|P{1.7cm}|c|c|c|P{1.2cm}|P{1.4cm}|c|}
\hline
Code & Dimension& Progenitor \par Mass [$M_\odot$] & Explosion & $t_\mathrm{exp}$ [s] & $t_{\rm sim}$ [s] & p-value \par (counts) & p-value \par (spectra)  & Reference \\ \hline
\multicolumn{1}{||c||}{3DnSNe-IDSA} & \multirow{6}{*}{1-d} & \multirow{6}{*}{20~\cite{Woosley:2007as}} & \multirow{6}{*}{N/A} & \multirow{6}{*}{N/A} & 0.50 & 0.043 & 0.027 & \multirow{6}{*}{O'Connor et al.~\cite{OConnor:2018sti}} \\\cline{1-1}\cline{6-8}
\multicolumn{1}{||c||}{AGILE-BOLTZTRAN} &  &   & &  & 0.55 & 0.055 & 0.034 & \\\cline{1-1}\cline{6-8}
\multicolumn{1}{||c||}{\texttt{FLASH}-M1} &        &   & &  & 0.71 & 0.015 & 0.002 & \\\cline{1-1}\cline{6-8}
\multicolumn{1}{||c||}{F{\sc{ornax}}} &          &   & &  & 1.10 & -- & -- & \\\cline{1-1}\cline{6-8}
\multicolumn{1}{||c||}{\texttt{GR1D}} &            &   & &  & 0.47 & 0.092 & 0.043 & \\\cline{1-1}\cline{6-8}
\multicolumn{1}{||c||}{{\sc Prometheus-Vertex}} & & & & & 0.87 & 0.003 & 3$\times 10^{-4}$ & \\\hline
{\sc Chimera} & 2-d & 20~\cite{Woosley:2007as} & Yes & 0.21 & 1.37 & 0.007 & 7$\times 10^{-4}$ & Bruenn et al.~\cite{2013ApJ...767L...6B,Bruenn:2014qea} \\ \hline
\texttt{FLASH} & 2-d & 20~\cite{Woosley:2007as} & Yes & 0.82 & 1.06 & 0.035 & 0.003 & O'Connor et al.~\cite{OConnor:2015rwy} \\ \hline
{\sc Prometheus-Vertex} & 2-d & 20~\cite{Woosley:2007as} & Yes & 0.36 & 0.38 & -- & -- & Summa et al.~\cite{Summa:2015nyk} \\ \hline
IDSA & 2-d & 20~\cite{Woosley:2007as} & Yes/No & 0.4--0.6 & 0.68 & 0.35 & 0.047 & Kotake et al.~\cite{Kotake:2018ypf} \\ \hline
F{\sc{ornax}} & 2-d & 20~\cite{Woosley:2007as} & No & N/A & 0.58 & 0.042 & 0.029 & Vartanyan et al.~\cite{Vartanyan:2018xcd} \\ \hline
\texttt{Zelmani} & 3-d & 20~\cite{Woosley:2007as} & Yes & 0.38 & 0.38 & -- & -- & Ott et al.~\cite{Ott:2017kxl} \\ \hline
\texttt{FLASH} & 3-d & 20~\cite{2016ApJS..227...22F} & No & N/A & 0.50 & -- & -- & O'Connor et al.~\cite{OConnor:2018tuw} \\ \hline
\textsc{Alcar} & 3-d & 20~\cite{Woosley:2007as} & No & N/A & 0.68 & 0.012 & 0.002 & Glas et al.~\cite{Glas:2018oyz} \\ \hline
F{\sc{ornax}} & 3-d & 20~\cite{Sukhbold:2015wba} & Yes & 0.45 & 0.59 & 0.091 & 0.047 & Burrows et al.~\cite{Burrows:2019zce} \\ \hline
\end{tabular}
\caption{A summary of the supernova simulations considered. The progenitor mass refers to the zero-age main sequence mass.  We take the explosion time to be the moment when the mean shock radius reaches 500~km. The p-values are obtained from the comparison between the model predictions and SN 1987A data.  The p-values in the counts column are from comparing the total number of events in Kam-II and IMB.  Those in the spectrum column are from comparing the cumulative spectrum distribution in Kam-II only (because IMB has too few counts).  See the text for why some models do not have p-values.} 
\label{tab:simulations}
\end{table*}

Table~\ref{tab:simulations} summarizes the list of supernova models employed in this work.  The p-values are computed up to the maximum time in each simulation.  We exclude four models from our statistical tests: the 1-d F{\sc{ornax}} models from Ref.~\cite{OConnor:2018sti} that have a known bug~\cite{Burrows}, a \texttt{FLASH} 3-d simulation paper~\cite{OConnor:2018tuw} that reports neutrino luminosities but not average energies, and the {\sc Prometheus-Vertex} 2-d~\cite{Summa:2015nyk} and \texttt{Zelmani} 3-d~\cite{Ott:2017kxl} simulations, which only run to 0.38~s. 


\section{SN 1987A data}
\label{sec:data}

Tables~\ref{tab:KamII} and~\ref{tab:IMB} show the 1987A events in Kam-II and IMB that we used in this work. Note that there are later detected events that we did not include because they are after the longest simulation time that we consider, $\simeq 4.6$~s.

\renewcommand{\arraystretch}{1.5}
\begin{table*}[h]
\centering
\begin{tabular}{||c|c|c||}
\hline
Event &  Time [s] & Energy [MeV] \\ \hline
1 & 0.000 & 20.0 \\ \hline
2 & 0.107 & 13.5 \\ \hline
3 & 0.303 & 7.5 \\ \hline
4 & 0.324 & 9.2 \\ \hline
5 & 0.507 & 12.8 \\ \hline
6 & 1.541 & 35.4 \\ \hline
7 & 1.728 & 21.0 \\ \hline
8 & 1.915 & 19.8 \\ \hline
\end{tabular}
\caption{Kam-II data taken from Ref.~\cite{Hirata:1988ad}. The event energy is the total energy of the detected positron.} 
\label{tab:KamII}
\end{table*}

\renewcommand{\arraystretch}{1.5}
\begin{table*}[h]
\centering
\begin{tabular}{||c|c|c||}
\hline
Event &  Time [s] & Energy [MeV] \\ \hline
1 & 0.000 & 38 \\ \hline
2 & 0.412 & 37 \\ \hline
3 & 0.650 & 28 \\ \hline
4 & 1.141 & 39 \\ \hline
5 & 1.562 & 36 \\ \hline
6 & 2.684 & 36 \\ \hline
\end{tabular}
\caption{IMB data taken from Ref.~\cite{IMB:1988suc}. The event energy is the total energy of the detected positron.} 
\label{tab:IMB}
\end{table*}


\section{Neutrino energy spectra and luminosities}
\label{sec:lum_spectrum}

\begin{figure*}[t]
\centering
\includegraphics[width=\textwidth]{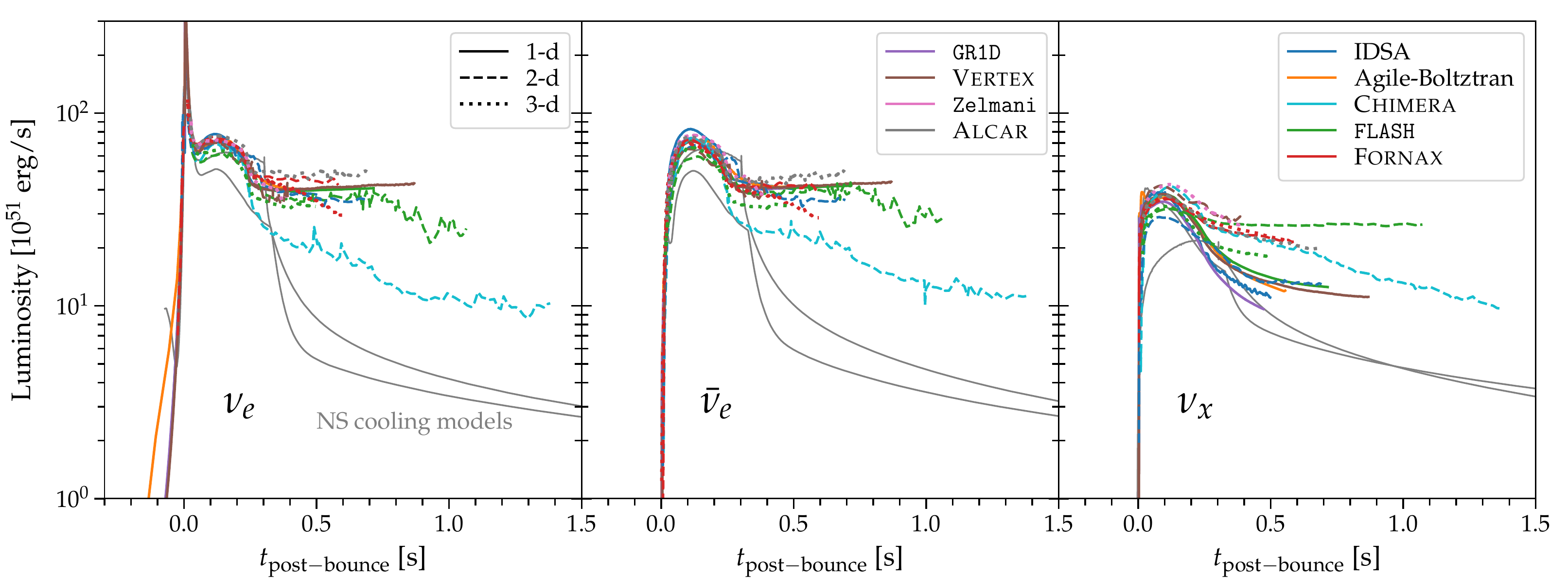}
\includegraphics[width=\textwidth]{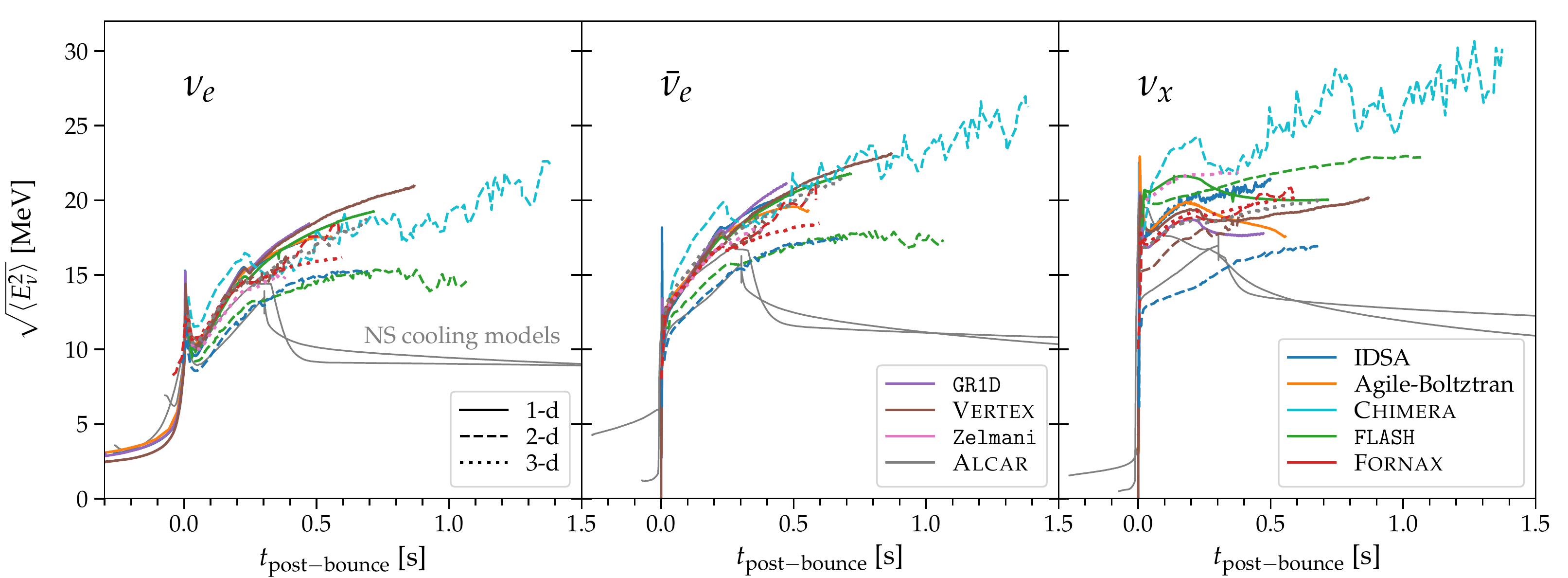}
\caption{Neutrino luminosities (top) and RMS energies (bottom) of $\nu_e$, $\bar\nu_e$, and $\nu_x$ from numerical simulations. The neutronization burst luminosity goes up to $\simeq 5\times 10^{53}$erg/s. The $\nu_x$ luminosity is shown for only one of the four flavors. The two PNS cooling models are from Refs.~\cite{Nakazato:2012qf, Roberts:2016rsf}.}
\label{fig:luminosity_energy}
\end{figure*}

Figure~\ref{fig:luminosity_energy} shows the time evolution of luminosity and root-mean-square (RMS) energy for all flavors and for each model.  It is evident by eye that there is relatively good agreement, with some exceptions.  Most of these quantities are taken directly from publications, but some require a conversion from average energies, $\langle E_\nu\rangle$, to root-mean-square energies, $\sqrt{\langle E^2_\nu\rangle}$. Some require changing from the fluid frame to an infinite observer frame.

Our calculations require the full neutrino energy distribution function, $f(E_\nu)$, which is assumed to be already integrated over neutrino propagation angle.  Sometimes the output $f(E_\nu)$ from a given model is publicly released, usually in the form of large numerical tables. However, this is available only for very few models. Fortunately, in Ref.~\cite{Keil:2002in}, a good analytical approximation for $f(E_\nu)$ has been found (we include the $E_\nu^2$ factor in the phase-space integral, not the distribution function):
\begin{equation}
f(\alpha,E_\nu) = \mathcal{N}\left(\frac{E_\nu}{\langle E_\nu\rangle}\right)^{\alpha-2} e^{-\frac{(\alpha+1)E_\nu}{\langle E_\nu\rangle}},
\label{eq:alpha_fit}
\end{equation}
where $\langle E_\nu\rangle$ is the average neutrino energy, defined as
\begin{equation}
\langle E_\nu\rangle = \frac{\int_0^\infty dE_\nu\, E_\nu^3 f(E_\nu)}{\int_0^\infty dE_\nu\, E_\nu^2 f(E_\nu)}\,,
\end{equation}
and $\alpha$, representing the amount of spectrum pinching, is defined as
\begin{equation}
\frac{\langle E_\nu^2\rangle}{\langle E_\nu\rangle^2} = \frac{2+\alpha}{1+\alpha}\,,
\end{equation}
and $\mathcal{N}$ is a normalization factor that ensures the distribution function integrates to the local neutrino number density.
For a Fermi-Dirac distribution with zero chemical potential, we have $\langle E_\nu \rangle \simeq 3.15T$, where $T$ is the temperature, and $\sqrt{\langle E_\nu^2 \rangle} = 1.14\langle E_\nu \rangle$. 

\begin{figure}[t]
\centering
\includegraphics[width=0.49\textwidth]{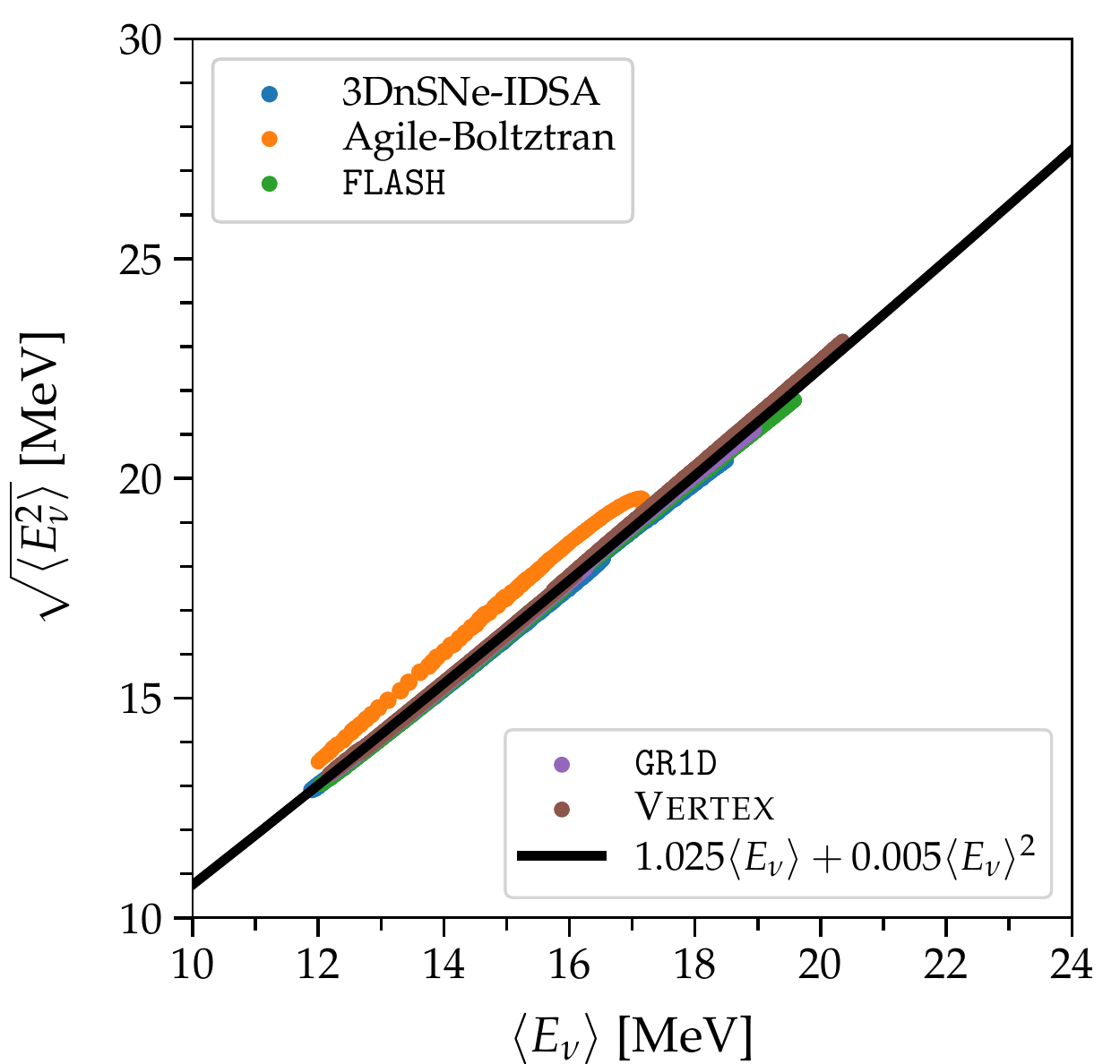}
\caption{$\sqrt{\langle E_{\nu}^2\rangle}$ as a function of $\langle E_\nu\rangle$ obtained from the numerical simulations where both are provided. The black line is the relation we use in this work.} 
\label{fig:ave_vs_rms}
\end{figure}

The pinched spectrum $f(\alpha,E_\nu)$ defined in Eq.~(\ref{eq:alpha_fit}) is what we use to calculate the theoretical predictions for SN 1987A.  We need both $\langle E_{\nu}\rangle$ and $\sqrt{\langle E_\nu^2\rangle}$, but some models only provide one of them, either through numerical tables or figures.  To solve this lack of information, we fit for a simple relation between $\langle E_{\nu}\rangle$ and $\sqrt{\langle E_\nu^2\rangle}$ using the models that provide both. Figure~\ref{fig:ave_vs_rms} shows $\sqrt{\langle E_\nu^2 \rangle}$  as a function of $\langle E_\nu\rangle$ for these models. Most models predict $\sqrt{\langle E_{\nu}^2\rangle}$ to be between 1.08$\langle E_\nu\rangle$ and 1.13$\langle E_\nu\rangle$ in the energy range 12--20~MeV, where the average energies of most models fall. Thus, we adopt the following relation between average and root-mean-square neutrino energy
\begin{equation}
    \sqrt{\langle E_\nu^2 \rangle} = 1.025\langle E_\nu\rangle + 0.005\langle E_\nu\rangle^2 \,,
    \label{eq:average_energy_ratio}
\end{equation}
which fits most of the models well, and is at most 10\% away from the F{\sc{ornax}} results.
Equation~(\ref{eq:average_energy_ratio}) is then used to compute a relation between $\alpha$ and $\langle E_\nu\rangle$. 

Note that some groups define $E_\text{rms} = \sqrt{\langle E_\nu^3\rangle/\langle E_\nu\rangle}$ as opposed to $E_\text{rms} = \sqrt{\langle E_\nu^2\rangle}$. We find that taking this into account leads to negligible changes to the predicted event rates, so we interpret all RMS energies as $\sqrt{\langle E_\nu^2\rangle}$.

Some models provide neutrino predictions in the comoving frame of the stellar fluid. However, we need quantities defined in the laboratory frame. We use the transformations between these frames from Ref.~\cite{Kotake:2018ypf}. For the neutrino luminosity, the transformation reads:
\begin{equation}
    L_\nu = L_\nu^{\mathrm{fluid}} (1+v_r/c)/(1-v_r/c)\,,
\end{equation}
and for the average energy, it is
\begin{equation}
    \sqrt{\langle E_\nu^2\rangle} = \sqrt{\langle E_\nu^2\rangle^{\mathrm{fluid}}} (1+v_r/c) / \sqrt{1-(v_r/c)^2},
\end{equation}
with $v_r$ being the radial velocity and $c$ the speed of light. In Ref.~\cite{Kotake:2018ypf}, it is assumed $v_r = -0.06c$, which is the average infall velocity at 500 km over the entire 250 ms post bounce. Considering that the transformation between frames is at the level of $O(10\%)$, our approximation can be safely applied to all models.


\section{Detected positron energy spectrum}
\label{sec:smearing_efficiency}

\begin{figure*}[t]
\centering
\includegraphics[width=\columnwidth]{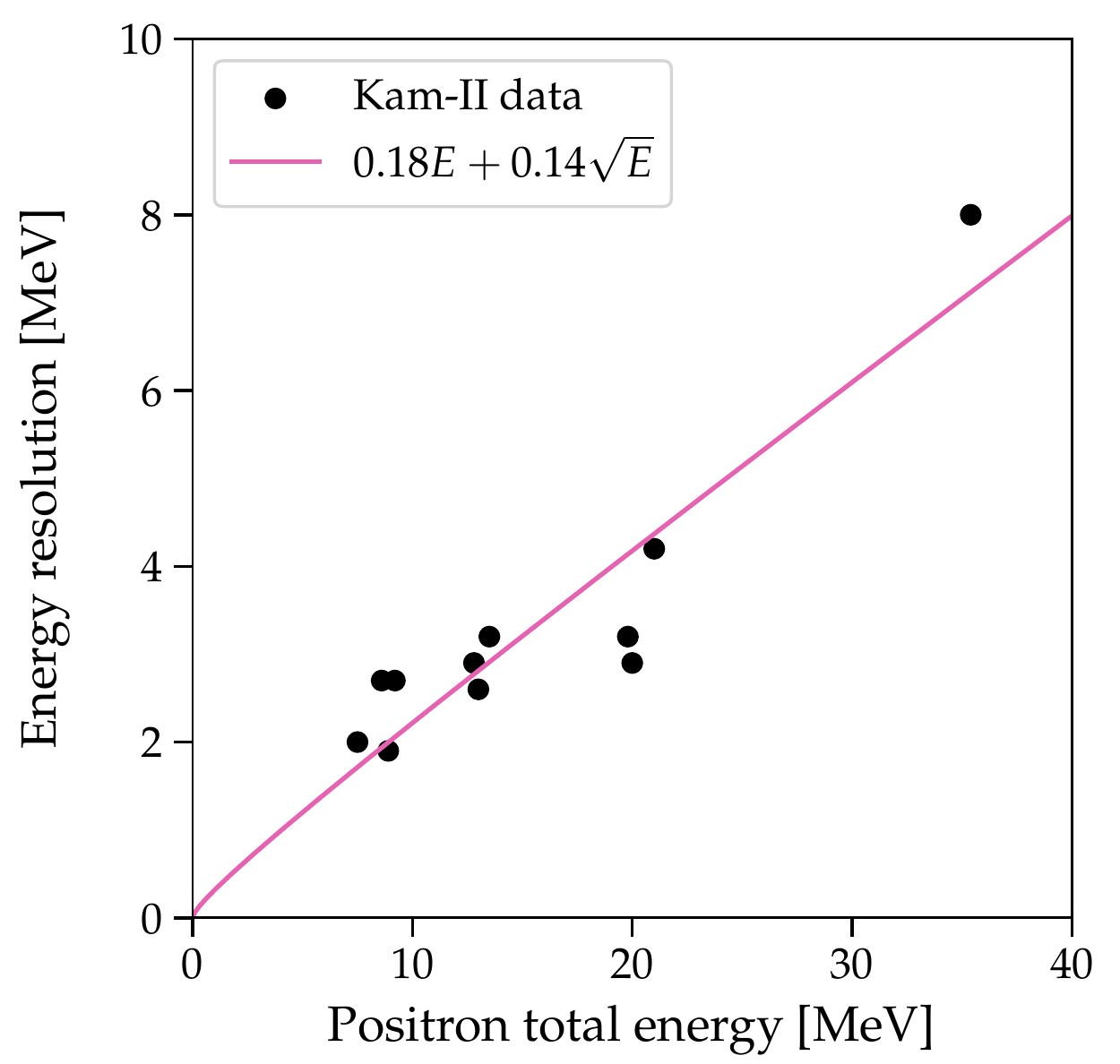}
\includegraphics[width=\columnwidth]{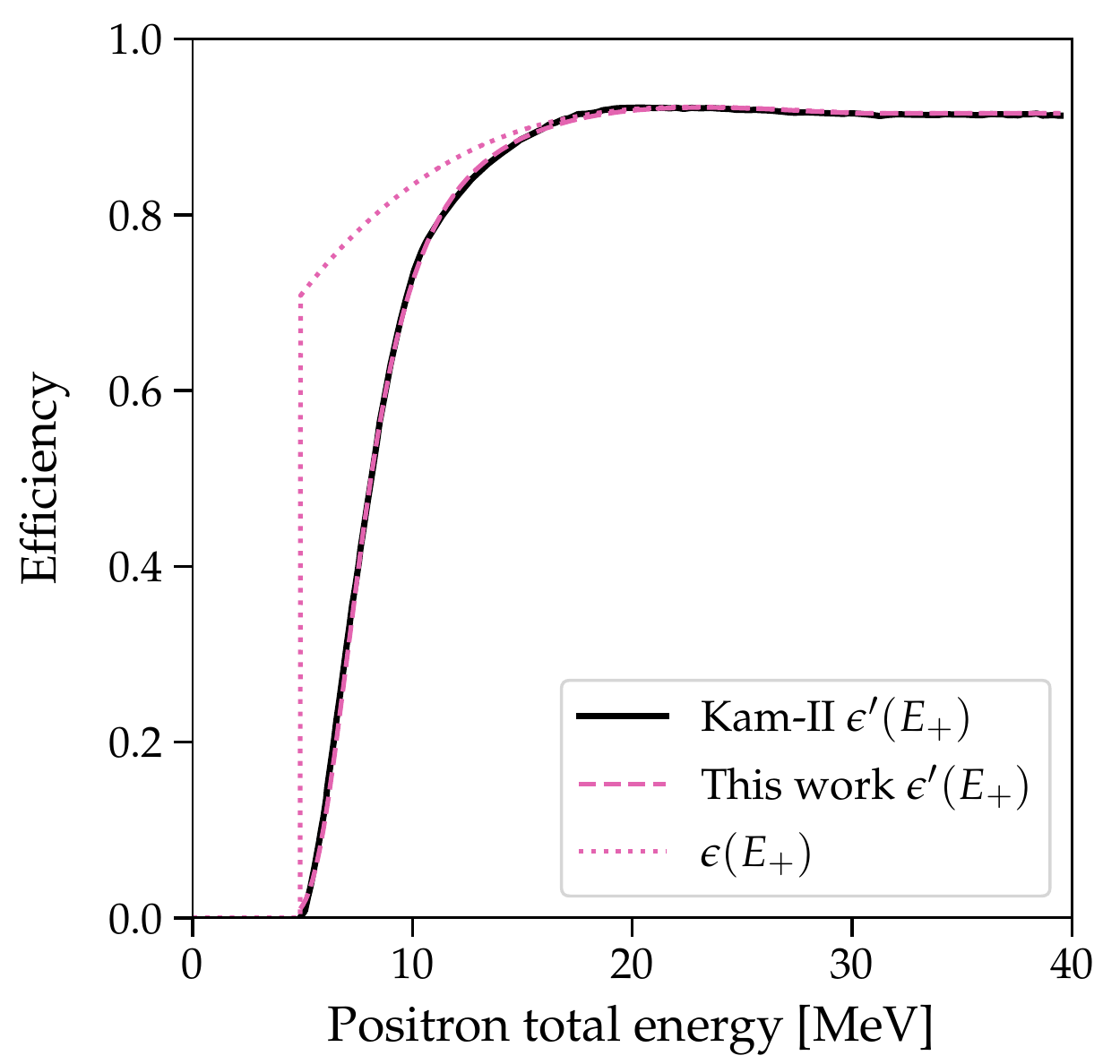}
\caption{(Left) The black dots represent the energy uncertainty for each event detected in Kam-II, whereas the continuous line is a fit to this data; we use the fit to get the width of the Gaussian energy resolution function defined in the text. (Right) The efficiency of Kam-II as a function of the true positron total energy.  Note that our calculation perfectly overlaps the line provided by Kam-II.}
\label{fig:kamII_efficiency}
\end{figure*}

We follow the standard approach to compute the positron spectrum~\cite{Jegerlehner:1996kx, Lunardini:2004bj, Costantini:2006xd, Pagliaroli:2008ur, Vissani:2014doa}.  As a first step to predicting the detected positron energy spectrum, we calculate the spectrum as a function of the true positron total energy $E_+$:
\begin{equation}
    g(E_+) \propto \int dE_\nu E_\nu^2 f(\langle E_\nu\rangle,\alpha) \frac{d\sigma_{\mathrm{IBD}}(E_\nu,E_+)}{dE_+}\,,
    \label{eq:positron_flux}
\end{equation}
where $d\sigma_{\mathrm{IBD}}/dE_+$ is the differential cross section for the inverse beta decay~\cite{Vogel:1999zy, Strumia:2003zx} and $f(\langle E_\nu\rangle,\alpha)$ is defined in Eq.~(\ref{eq:alpha_fit}).

To turn $g(E_+)$ into an observed spectrum, we need to convolve it with the detector energy resolution and efficiency.  We detail the Kam-II case. The trigger has a threshold of $N_\mathrm{hit} \geq 20$ (the number of hit photomultiplier tubes), which is equivalent (on average) to a detected energy, $E_\mathrm{det}$, of 7.5 MeV. The detected spectrum is then
\begin{equation}
    N(E_\mathrm{det}) \propto \int dE_+ g(E_+) \epsilon(E_+) R(E_\mathrm{det}, E_+) \theta(E_\mathrm{det}-7.5),
    \label{eq:smearing}
\end{equation}
where $\epsilon(E_+)$ is the intrinsic detector efficiency (not taking into account the $E_\mathrm{det}$ cut), $R(E_\mathrm{det}, E_+)$ is the energy resolution function and $\theta(E_\mathrm{det}-7.5)$ is the Heaviside step function. We assume $R(E_\mathrm{det},E_+)$ to be a Gaussian function, with the width $\sigma_{R}$ having the form $aE_+ + b\sqrt{E_+}$, where the second term arises from the Poisson fluctuations on $N_{\rm hit}$, while the first term roughly accounts for systematics.  The black dots in Fig.~\ref{fig:kamII_efficiency} (left panel) represent the energy errors of each event detected in Kam-II~\cite{Hirata:1988ad}. We obtain the following expression for the width from a functional fit:
\begin{equation}
\sigma_{R}^{\mathrm{KamII}}(E_+)=0.18E_++0.14\sqrt{E_+}\,.    
\end{equation}
The published trigger efficiency curve from  Fig.~3 of Ref.~\cite{Hirata:1988ad} should be interpreted as $\epsilon'(E_+)$, combining the intrinsic detector efficiency and the trigger cut:
\begin{equation}
    \epsilon'(E_+) = \epsilon(E_+) \int dE_\mathrm{det} R(E_\mathrm{dec}, E_+) \theta(E_\mathrm{det}-7.5)\,.
\end{equation}
Figure~\ref{fig:kamII_efficiency} (right panel) shows the trigger efficiency we have computed (pink dashed line) in comparison with the one published by Kam-II (black solid line), which match perfectly.  In addition, we also show the inferred intrinsic efficiency $\epsilon(E_+)$ with the dotted line.

\begin{figure*}[t]
\centering
\includegraphics[width=\columnwidth]{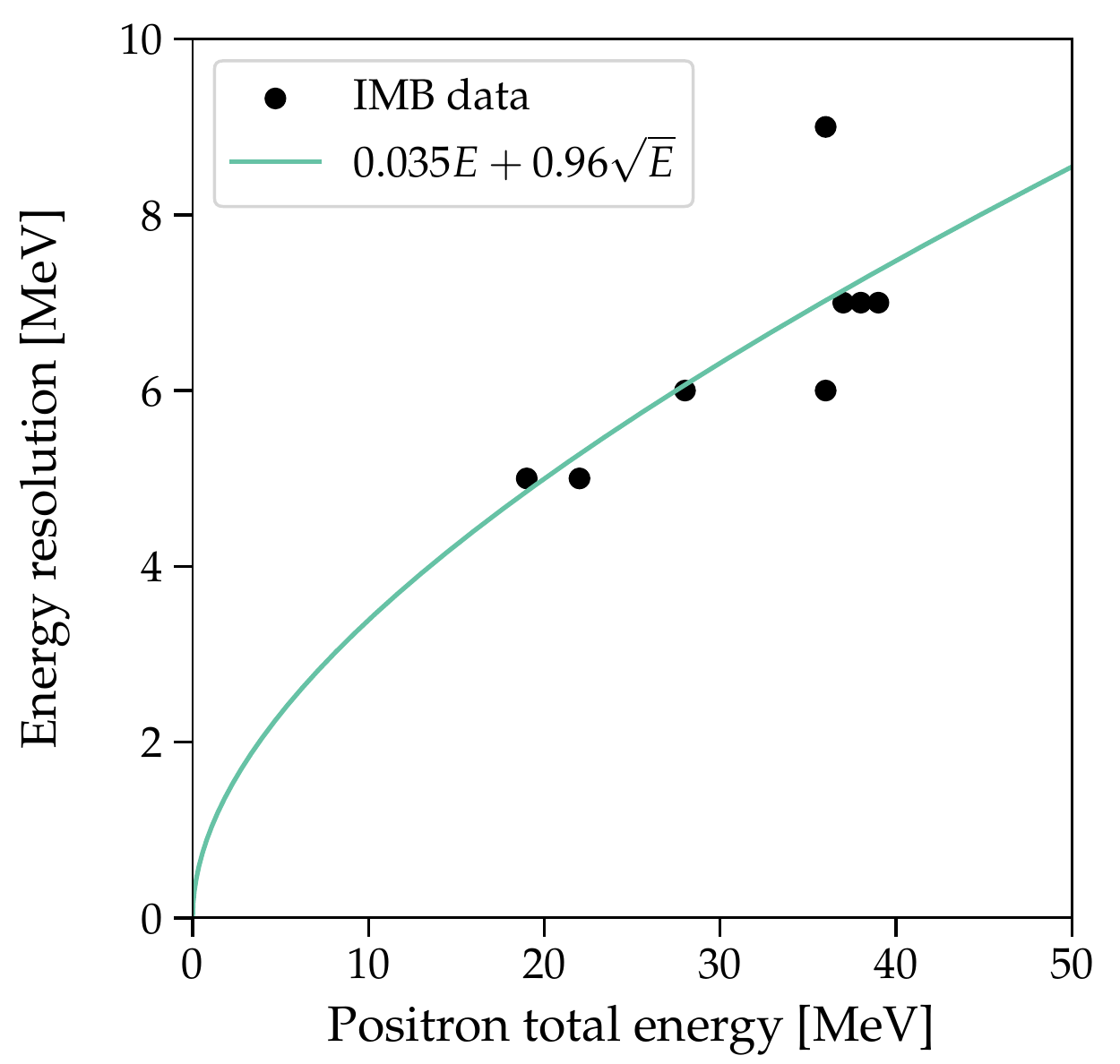}
\includegraphics[width=\columnwidth]{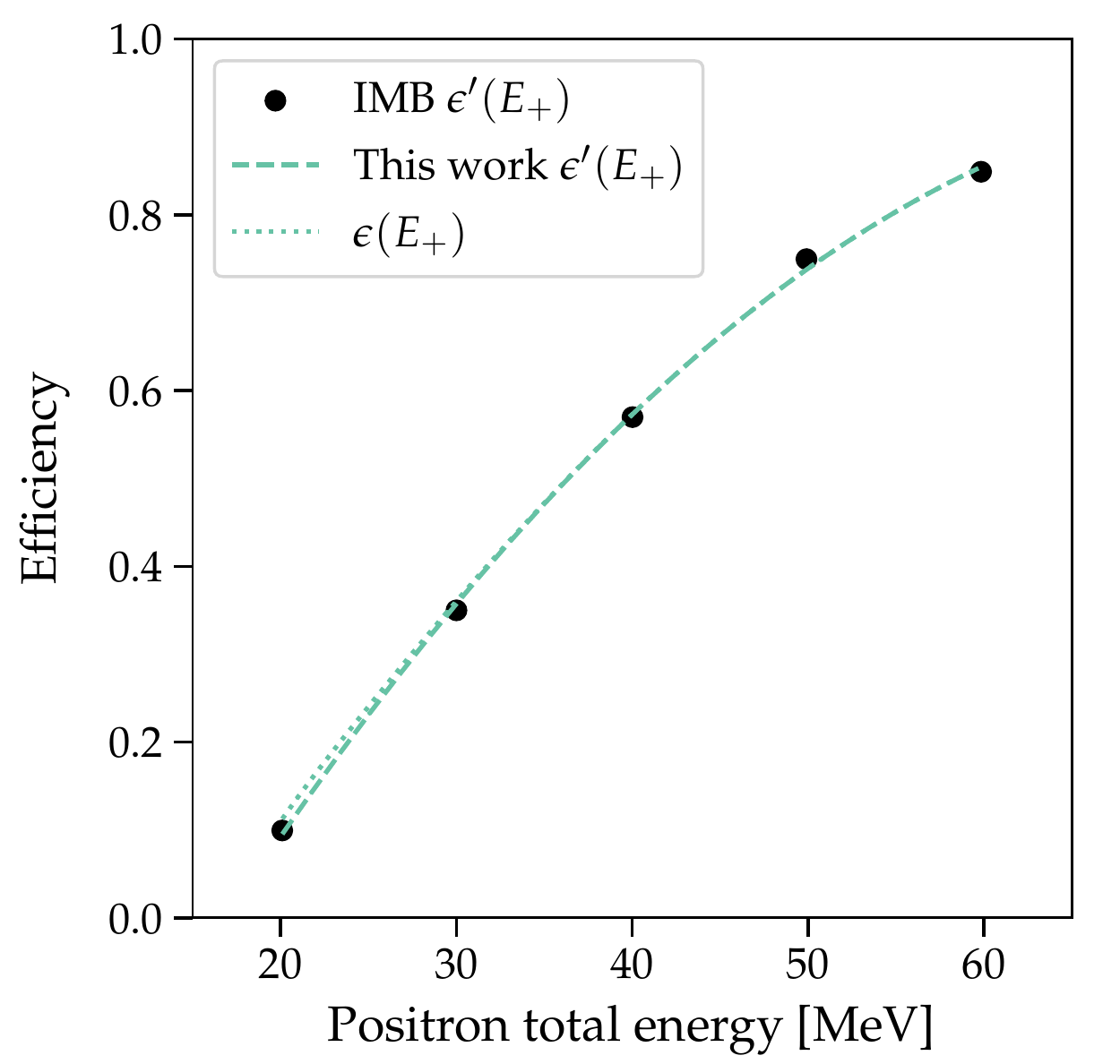}

\caption{Same as Fig.~\ref{fig:kamII_efficiency}, but for IMB, which only provides efficiencies at a few energy points.}
\label{fig:IMB_efficiency}
\end{figure*}

A similar procedure has been applied for IMB. Figure~\ref{fig:IMB_efficiency} displays both the energy resolution (left panel) and the efficiency (right panel) extracted for this experiment.

\section{Statistical methods}
\label{sec:cumu_dist_ks}

\subsection{Calculating the time offset}

As the supernova neutrinos reach Earth, the expected number of events increases as a function of time because the supernova signal is rising. However, due to the limited statistics, the first detected event is generally expected to occur at 0.1--0.2~s after the neutronization burst peak. In addition, because the two detectors, KamII and IMB did not record the absolute time difference of their respective first events, this time offset has to be applied separately to KamII and IMB.

The method we use for calculating the offset time is the following. First, we look only at data as a function of time, and we discard energy information. For example, let us consider one simulation running up to 0.6 s. Let us also assume that KamII data has the following detection times for each event: $(0.0, 0.1, …, t_n)$~s. We test offsets in the range $[0, 0.5]$~s. For every offset ($t_0$), the detection times with respect to the arrival of the neutronization peak, which can be taken as $t=0$ in simulations, become $(0.0+t_0, 0.1+t_0, … , t_n+t_0)$. Next, we compute the p-value as $P_1\times P_2$, where $P_1$ is the Poisson probability of predicting $N$ events and getting $N'$ events, i.e., what we quote as the p-value for rates. $P_2$ is the Kolmogorov-Smirnov test p-value obtained by comparing the detected and predicted temporal shapes (see Fig.~\ref{fig:cumulative_spectra_TS} left panel for the spectrum example). We get an array of p-values, one for each value of $t_0$. We choose the time offset for which the p-value is maximal. In general, in the plausible range of time offset, i.e., $[0.1, 0.2]$~s, the p-value is rather constant because the data is sparse, and we are not including more events to KamII data when shifting the time. Concerning IMB, the number of events can change for some models, which we do use when computing p-values. However, their number would still be so low that their contribution to the combined p-value is only marginal.

\subsection{Cumulative distributions and Kolmogorov-Smirnov tests}

\begin{figure*}[b]
\centering
\includegraphics[width=\columnwidth]{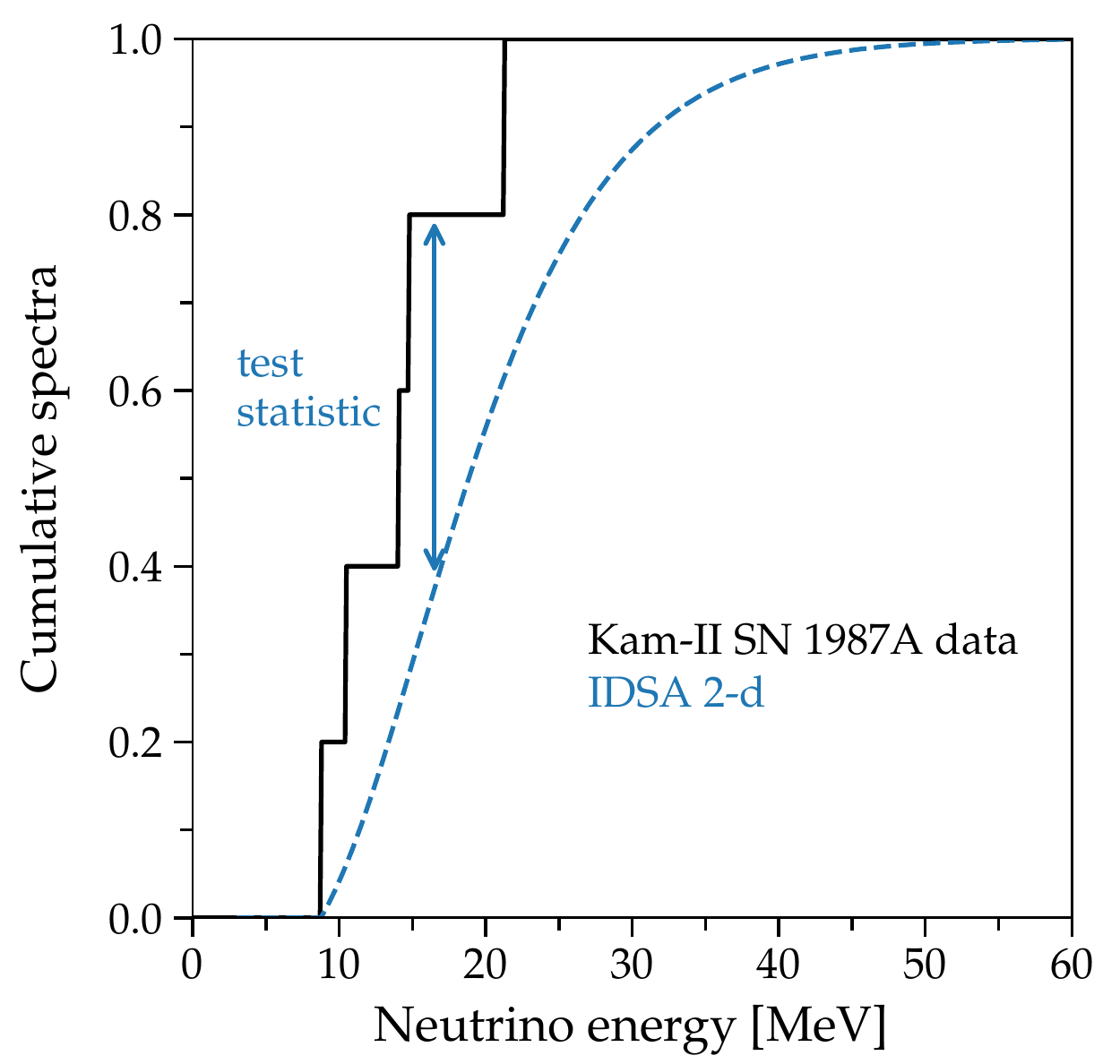}
\includegraphics[width=\columnwidth]{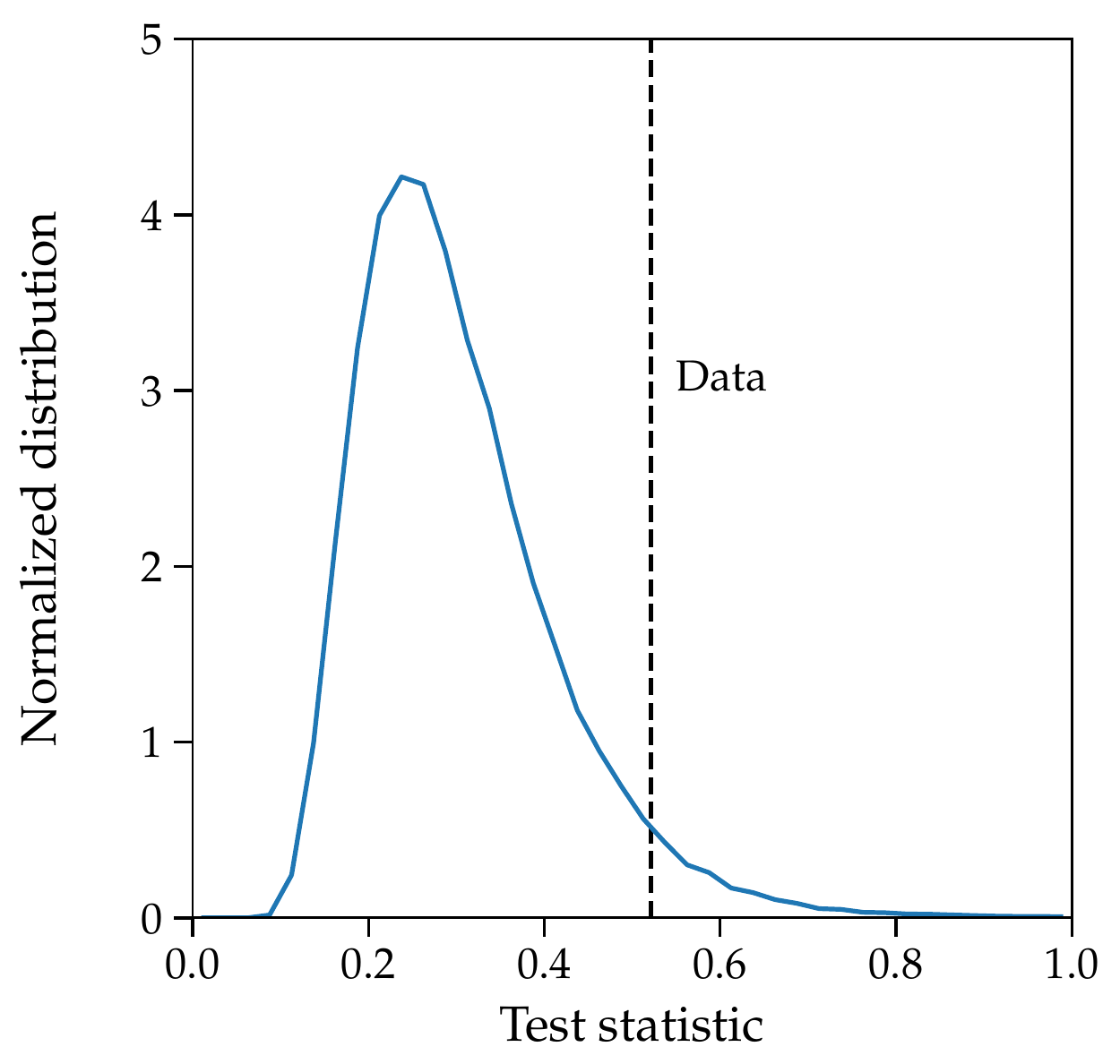}
\caption{A demonstration of p-value calculation. (Left) The cumulative energy distribution of the IDSA 2-d model and 1987A data and the corresponding test statistic. (Right) The test statistic distribution.}
\label{fig:cumulative_spectra_TS}
\end{figure*}

To compute the p-values for the spectrum shape, we use a Kolmogorov-Smirnov test.  Figure~\ref{fig:cumulative_spectra_TS} (left panel) illustrates the test statistic.  We use one specific model, IDSA 2-d~\cite{Kotake:2018ypf}, as a concrete example.  First, we utilize all the available information and set $t_\text{cutoff}$ to be the simulation run time, 0.67~s for this model.  We then integrate the energy spectrum up to $t_\text{cutoff}$ and plot the cumulative distribution (blue dashed line). Next, we cut the 1987A Kam-II data also at $t_\text{cutoff}$, allowing a variable offset time. The black steps show the cumulative spectrum of the data within 0.67~s. The test statistic is the maximum vertical distance between these two curves, TS$_0$.

\begin{figure*}[t]
\centering
\includegraphics[width=0.8\textwidth]{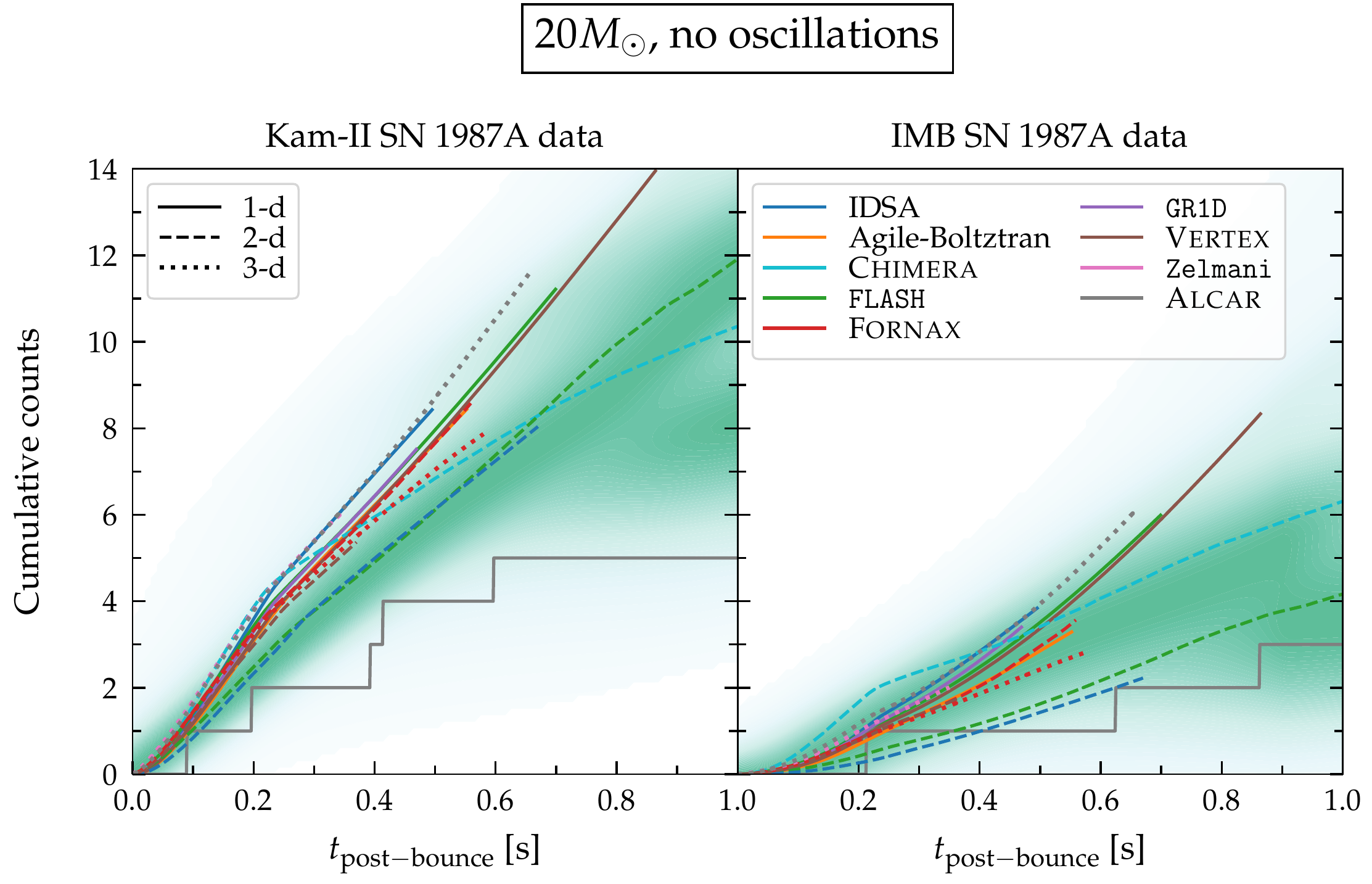}\\
\vspace{2em}
\includegraphics[width=0.8\textwidth]{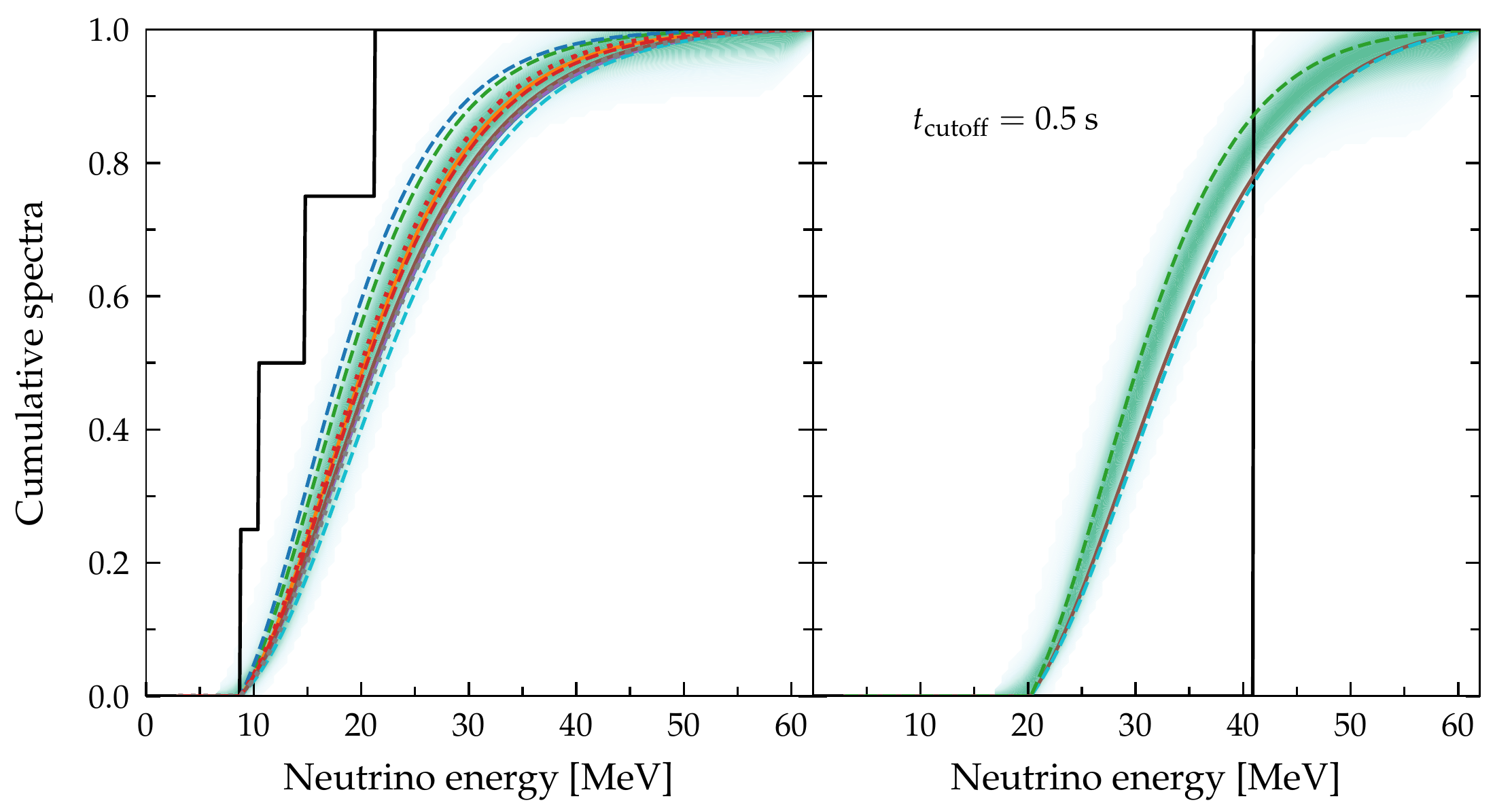}
\caption{Cumulative counts (top panel) and cumulative spectra (bottom panel) of models compared to SN 1987A data.  The green shading shows the density of all models considered in this work, including the alternative scenarios with neutrino oscillation and different progenitor masses.}
\label{fig:cumulative_spectra}
\end{figure*}

To compute p-values, we run Monte Carlo simulations of each detector to build the test statistic distribution, shown in Fig.~\ref{fig:cumulative_spectra_TS} (right panel). For this model, we sample many realizations of the data, letting both the total counts and the energy points vary, and compute the test statistic between the model and each realization. The p-value is determined by the fractional area of the p-value distribution more extreme than TS$_0$.

This Monte Carlo sample is also used to compute the error bars in Figs.~\ref{fig:comparison}--\ref{fig:progenitor}.  The vertical error bars follow Poisson fluctuations, as expected.  The horizontal error bars are the standard deviation of the mean for the detected positron energies in each detector.  As noted in the main text, the results shown in the main panels of those figures are simple visual comparisons, with the insets showing our full statistical calculations.

Figure~\ref{fig:cumulative_spectra} (top panel) shows how the cumulative counts for the models compare to 1987A data. The green shading indicate smearing over all models, including those for different oscillation scenarios and progenitor masses in Ref.~\cite{Warren:2019lgb}.  Here we clearly see the trend of models predicting too high of event counts throughout the entire 1~s.  Figure~\ref{fig:cumulative_spectra} (bottom panel) shows the cumulative energy spectra of all models, and similarly, the green shading indicate the model range.

One may be tempted to use likelihood as a test statistic to compute p-values.  Here we follow the discussions in Ref.~\cite{James:2006zz} (Chapter 11) and explain why this is not a robust statistical procedure. The essential point is that we are doing a goodness-of-fit test.  While maximum likelihood is a good method for parameter estimation, it does not work as a goodness-of-fit test, regardless of whether one works in the frequentist or Bayesian framework.  A simple demonstration of why is to consider a uniform distribution $h(x)=1$ between (0, 1), where we want to test whether some random numbers $\{x_i\}$ are drawn from this distribution. Clearly any data set would generate a likelihood of 1, so this fails as a goodness-of-fit test.  The lesson is generically applicable to other distributions because we can always transform a smooth distribution to a uniform one by changing variables.  Attempts at goodness-of-fit tests with maximum likelihood thus give different results depending on if one uses $x$, $\log x$, or another choice as the dependent variable.  This model dependence renders the results uninformative.


\section{Model selection compared to Fiorillo et al. (2023)}
\label{sec:model_selection}

A few months after our work appeared on arXiv, Ref.~\cite{Fiorillo:2023frv} appeared, also investigating how modern supernova simulation predictions compare to 1987A neutrino data.  We view this paper as largely complementary, as we focus on explosion models (only available at early times) while they focus on PNS cooling models (most suitable for late times) and statistical tests of the complete signal, which seems to be dominated by late times.

Reference~\cite{Fiorillo:2023frv} also comments on the early time signals, claiming that there is no tension between modern simulations and the 1987A neutrino data, in contrast to our results.  Even though there are differences in the treatments regarding experimental data, statistical procedure, etc., we think the crucial difference between the two groups is in the supernova simulations considered.  We explain this point here.

\begin{figure*}[h]
\centering
\includegraphics[width=\columnwidth]{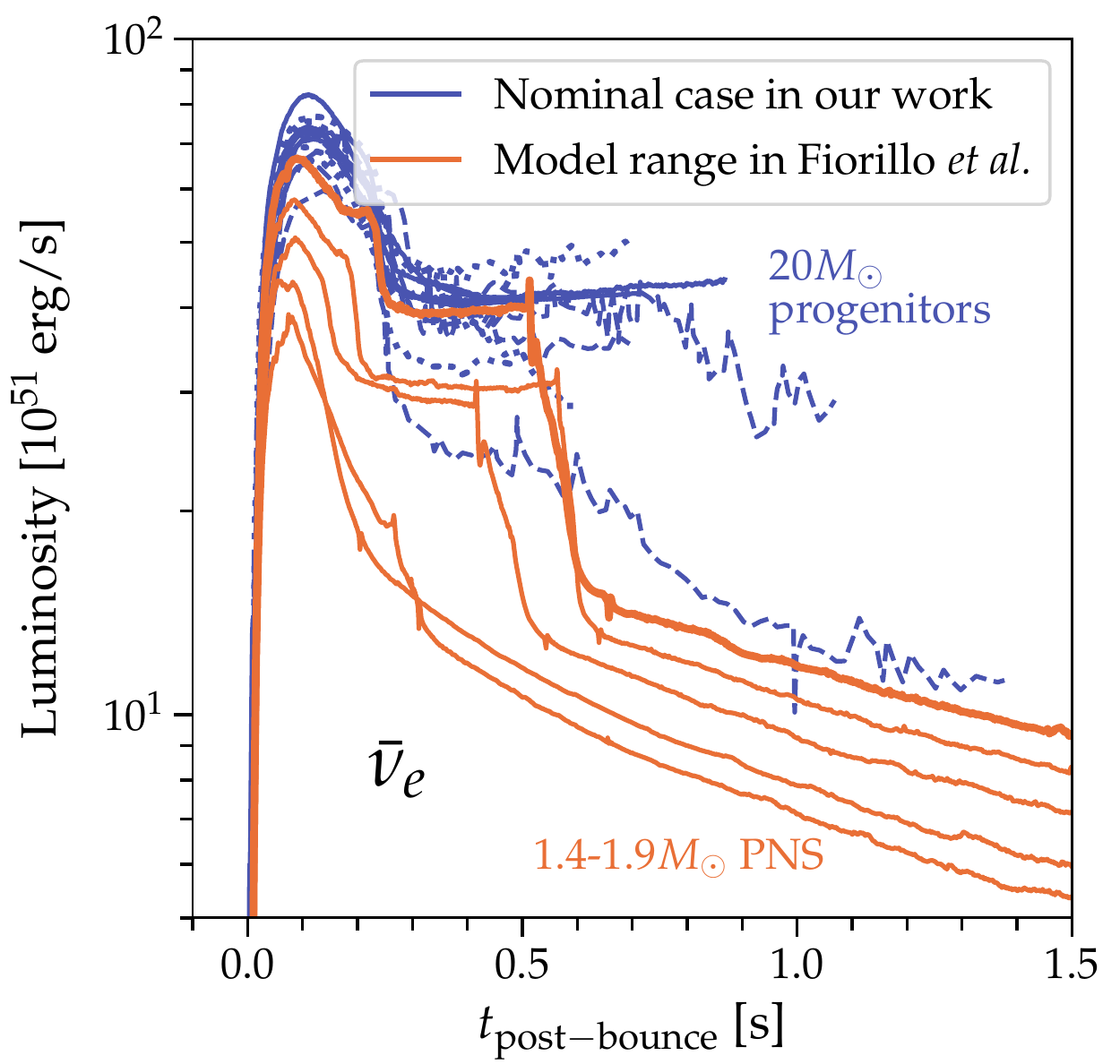}
\includegraphics[width=\columnwidth]{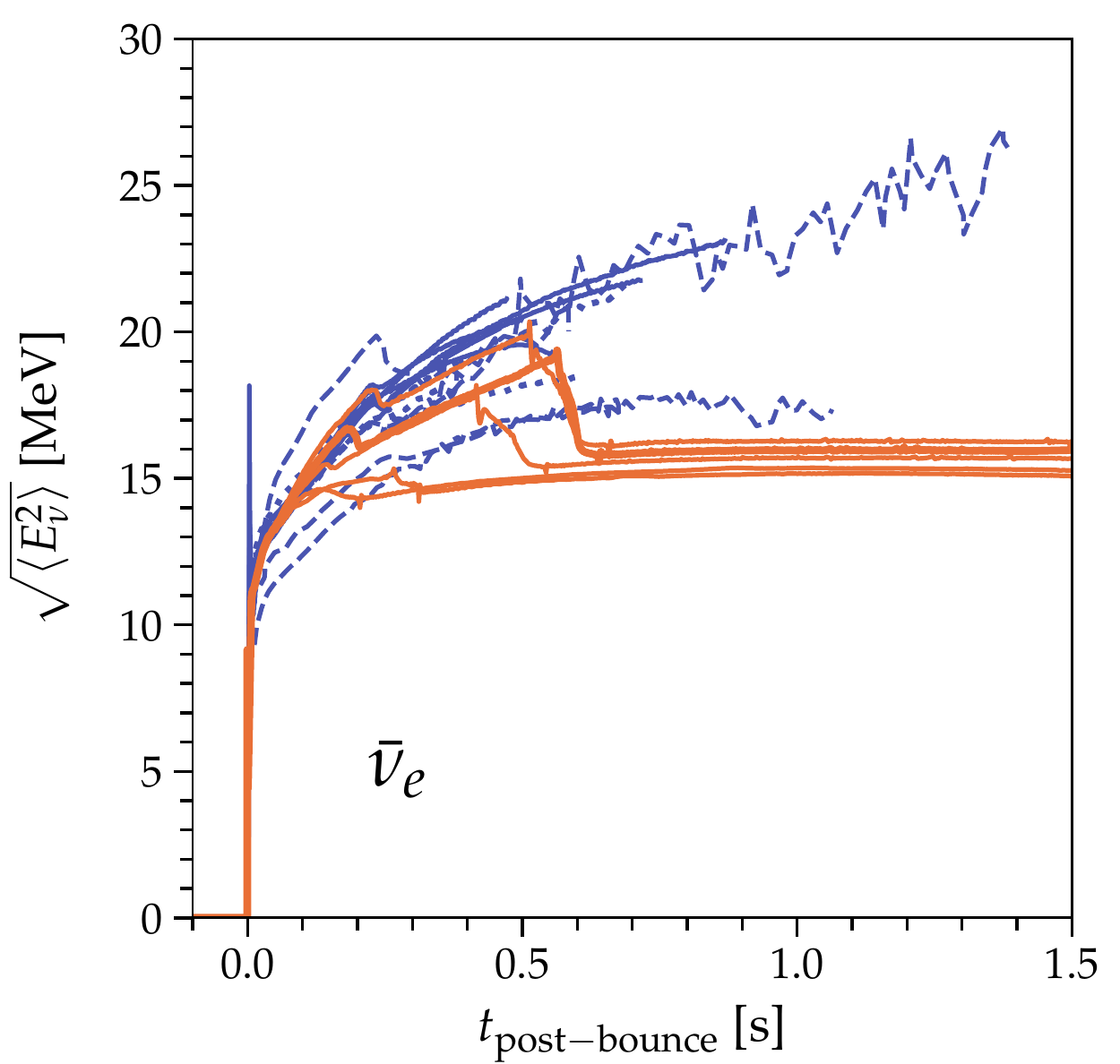}
\caption{Comparison of supernova models: the 1-, 2-, 3-d models used in our work (purple lines; see Fig.~\ref{fig:luminosity_comparison}) and the 1-d models used in Ref.~\cite{Fiorillo:2023frv} (orange lines, corresponding to various PNS masses).  From bottom to top at a time of 1.0~s, these are 1.36, 1.44, 1.62, 1.77, and 1.93$M_\odot$; we show the 1.93$M_\odot$ case with a thick line.  These correspond to progenitor masses of 9.0, 18.8, 18.6, 27, and 20$M_\odot$ (note that the order is non-monotonic); we show the 20$M_\odot$ case with a thick line.  {\it Note that in their fits they emphasize models with substantially lower neutrino luminosities (and somewhat lower average energies) than found in the multi-d models we consider, either because their models employ progenitors that give intrinsically lower luminosities or because they are artificially cut off accretion near 0.6~s by their explosion prescription.}} 
\label{fig:model_selection}
\end{figure*}

Figure~\ref{fig:model_selection} shows that the nominal models in Ref.~\cite{Fiorillo:2023frv} \textit{have much lower luminosities (and significantly lower average energies) than the models we consider}. This immediately brings their theoretical predictions for the detected signals closer to the early time SN 1987A data, which are dominated by the KamII data.

The main reason for this difference in luminosities and average energies is that we consider models focused on the explosion phase and, therefore, consider primarily the results of 2- and 3-d simulations (although we also compare to the results of 1-d simulations).  In contrast, Ref.~\cite{Fiorillo:2023frv} considers only 1-d models, which are most suited for studying the PNS cooling phase. Because these 1-d models do not naturally explode, an explosion must be artificially imposed to prevent collapse to a black hole and have the models give a long-term neutrino signal. In Ref.~\cite{Fiorillo:2023frv}, they trigger explosions by removing most of the infalling material outside of the supernova shock at a prescribed time after bounce. As a result, the mass of the PNS in their models and the time of shock runaway are determined by this tunable parameter. Abruptly shutting off accretion by removing most of the outer layers of the star causes the neutrino luminosity to drop rapidly, which is evident in Fig.~\ref{fig:model_selection}. 

For the five luminosity profiles from Ref.~\cite{Fiorillo:2023frv}, as shown in Fig.~\ref{fig:model_selection}, two are significantly lower than ours at all times, making them closer to the Kam-II data.  For one of these profiles, the progenitor mass is very small (9$M_\odot$).  For the other three profiles, while the initial luminosities are more like ours, they drop precipitously in the first second, the time period we focus on, again making the predicted counts closer to the Kam-II data and rather unlike what is expected from most of the 2- and 3-d models.

\section{Progenitors from different groups}
\label{sec:progenitors}

In Fig.~\ref{fig:progenitor}, we show a comparison between model predictions and data for different values of the progenitor mass, but only for F{\sc{ornax}} 2-d. Here we provide further details.

\begin{figure*}[h]
\centering
\includegraphics[width=0.9\textwidth]{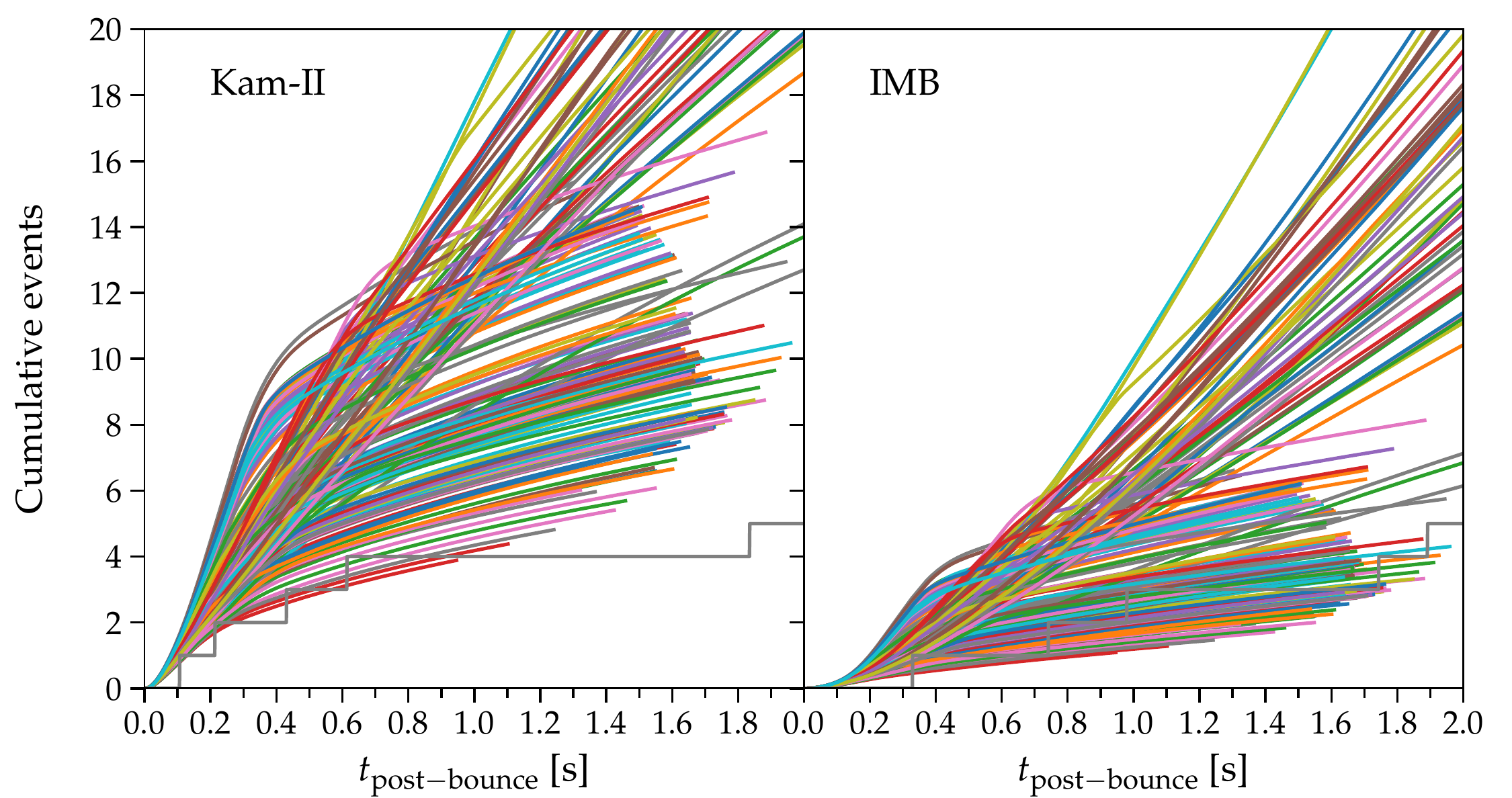}
\caption{Cumulative numbers of events for different progenitors simulated with \texttt{FLASH} 1-d~\cite{Warren:2019lgb}.}
\label{fig:cumulative_rates_FLASH}
\end{figure*}

Figure~\ref{fig:cumulative_rates_FLASH} shows the cumulative distributions for different progenitors by \texttt{FLASH} 1-d~\cite{Warren:2019lgb} compared to 1987A data~\cite{SNEWS:2021ewj}.  This is the study with the largest suites of progenitor models.  Because of the huge number of models, we decide not to label any specific progenitors but rather focus on some general points.  First, there are clearly two families of curves, one with steeply rising event rates and one with flattening event rates beyond $\simeq$ 0.5~s.  The first family corresponds to models where the supernova failed to explode and the second to successful explosions.  Interestingly, models with successful explosions are thus closer to the SN 1987A data.  {\it This means there may be a strong connection between the two questions  ``Does the model explode?'' and ``Does the model match the data?''}  Further exploration of that conjecture is needed.

\begin{figure*}[h]
\centering
\includegraphics[width=\columnwidth]{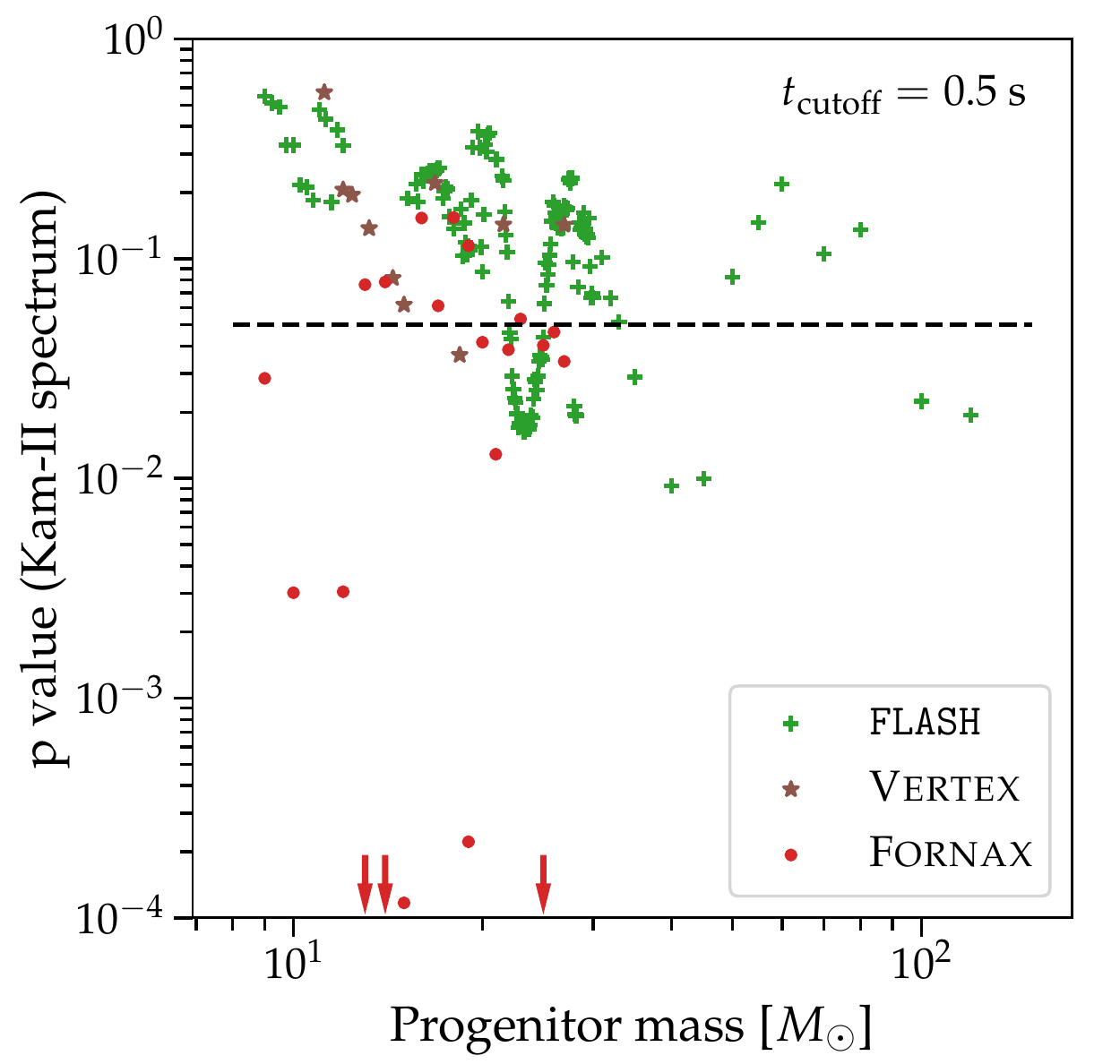}
\includegraphics[width=\columnwidth]{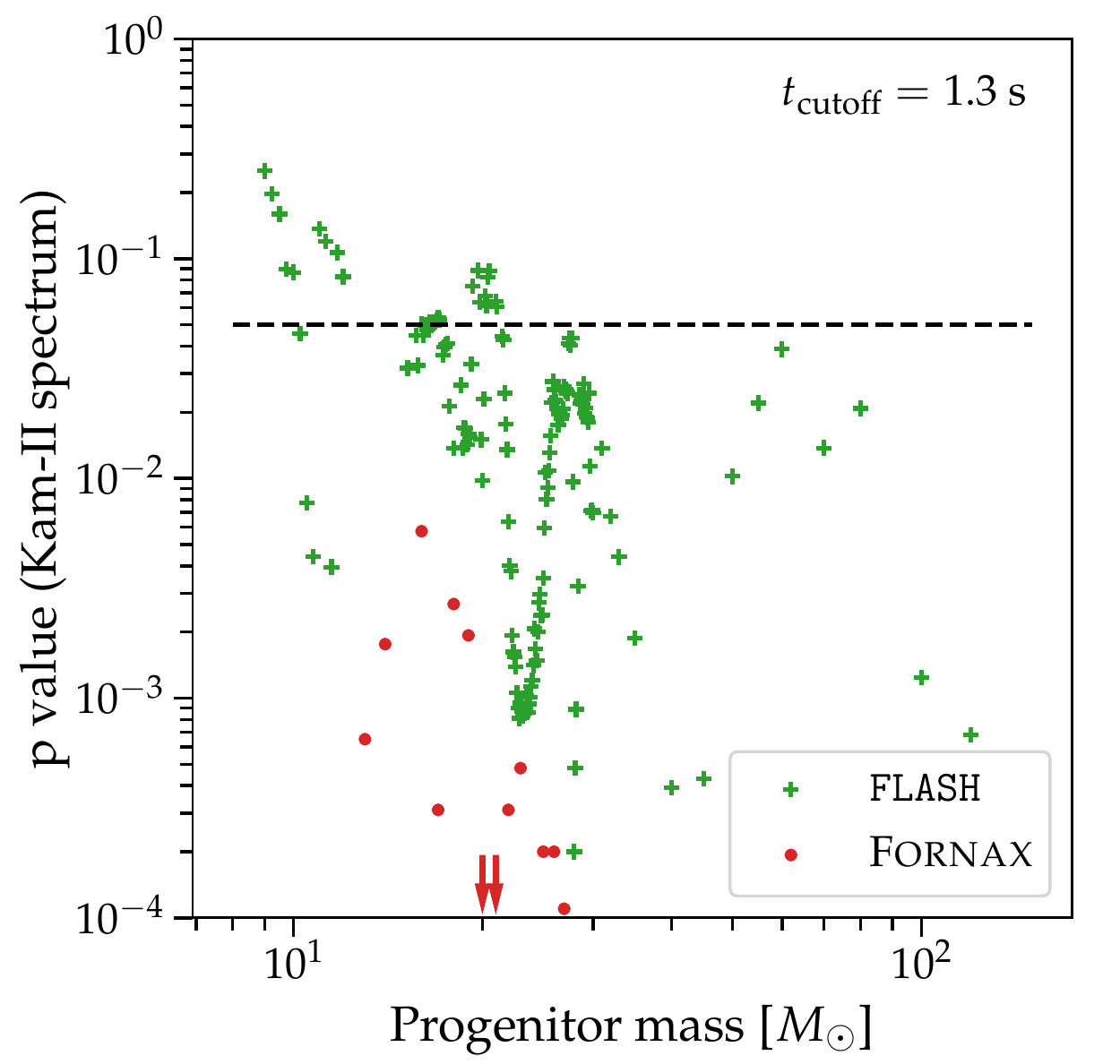}
\caption{The p-values obtained through an energy spectrum comparison with Kam-II data for different progenitors masses, simulated with \texttt{FLASH}~\cite{Warren:2019lgb},  {\sc Vertex}~\cite{Summa:2015nyk} and F{\sc{ornax}}~\cite{Vartanyan:2018iah, Burrows:2019rtd, Vartanyan:2019ssu, Nagakura:2019gmh, Burrows:2019zce, Burrows:2020qrp}, with a cutoff time of 0.5~s (left) and 1.3~s (right).}
\label{fig:progenitor_mass_complete}
\end{figure*}

Combining results from all groups with long-term simulations for a range of progenitor models, Fig.~\ref{fig:progenitor_mass_complete} (left panel) shows how the p-values for the energy spectrum for \texttt{FLASH}~\cite{Warren:2019lgb},  {\sc Vertex}~\cite{Summa:2015nyk} and F{\sc{ornax}}~\cite{Burrows:2020qrp} change with the progenitor mass in 10--80$M_\odot$. This panel assumes a cutoff time of $t_{\rm cutoff}=0.5$ s. In this case, most of the p-values lie above the dashed line that represents our $p = 0.05$ threshold. Then most of these models appear to be in agreement with data, especially in the range 10--20$M_\odot$. On the other hand, the right panel uses $t_{\rm cutoff}=1.3$ s. In this case, a general disagreement with data is clearly visible. We stress that the choice of $t_{\rm cutoff}=0.5$ s is only to make the time range for each simulation equal.  It is unsurprising that restricting the time range lessens the statistical tension.  The p-values for the case where $t_{\rm cutoff}$ is equal to the cutoff time of each simulations are the ones to be taken as references, since ultimately we would like all simulations to agree with data for the entire duration of the neutrino burst.

\section{Supernova neutrino oscillation treatment}
\label{sec:oscillations}

We consider a few illustrative oscillation scenarios in the main text, focusing on exploring the possible impact of oscillations on the model-data comparison.  We use the following equation for calculating the oscillated flux $\tilde{f}_{\bar\nu_e}$ of $\bar{\nu}_e$ as a function of the unoscillated fluxes.
\begin{equation}
\tilde{f}_{\bar\nu_e} = f_{\bar\nu_e}P(\bar{\nu}_e\to\bar{\nu}_e)+f_{\bar\nu_x} [1-P(\bar{\nu}_e\to\bar{\nu}_e)]\,,
\end{equation}
where $P(\bar{\nu}_e\to\bar{\nu}_e)=\cos^2\theta_{12}\simeq 0.7$ in NH, $P(\bar{\nu}_e\to\bar{\nu}_e)=0$ in IH and $P(\bar{\nu}_e\to\bar{\nu}_e)=\frac{1}{3}$ for flavor equilibration.

\begin{figure*}[h]
\centering
\includegraphics[width=0.9\columnwidth]{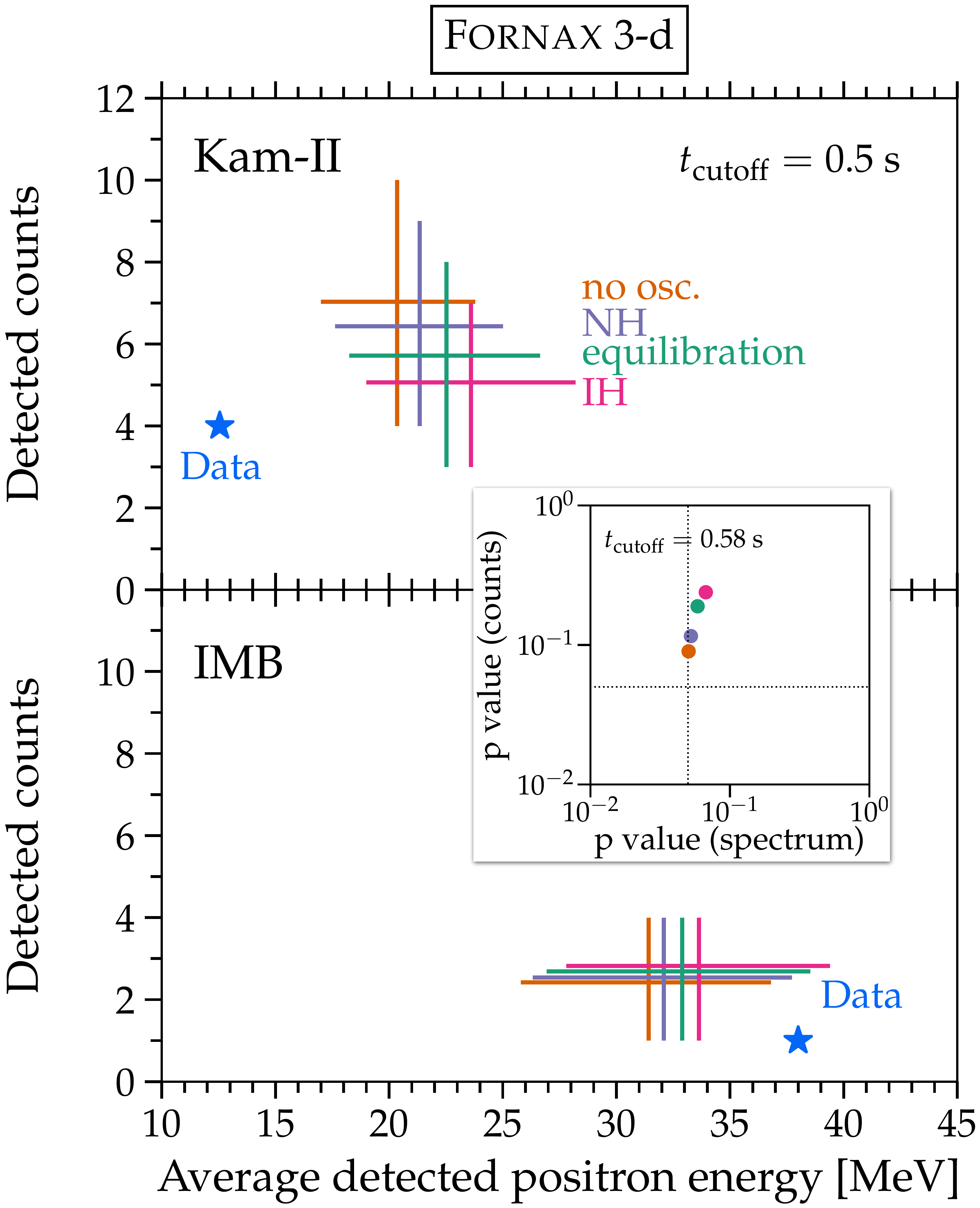}
\includegraphics[width=0.9\columnwidth]{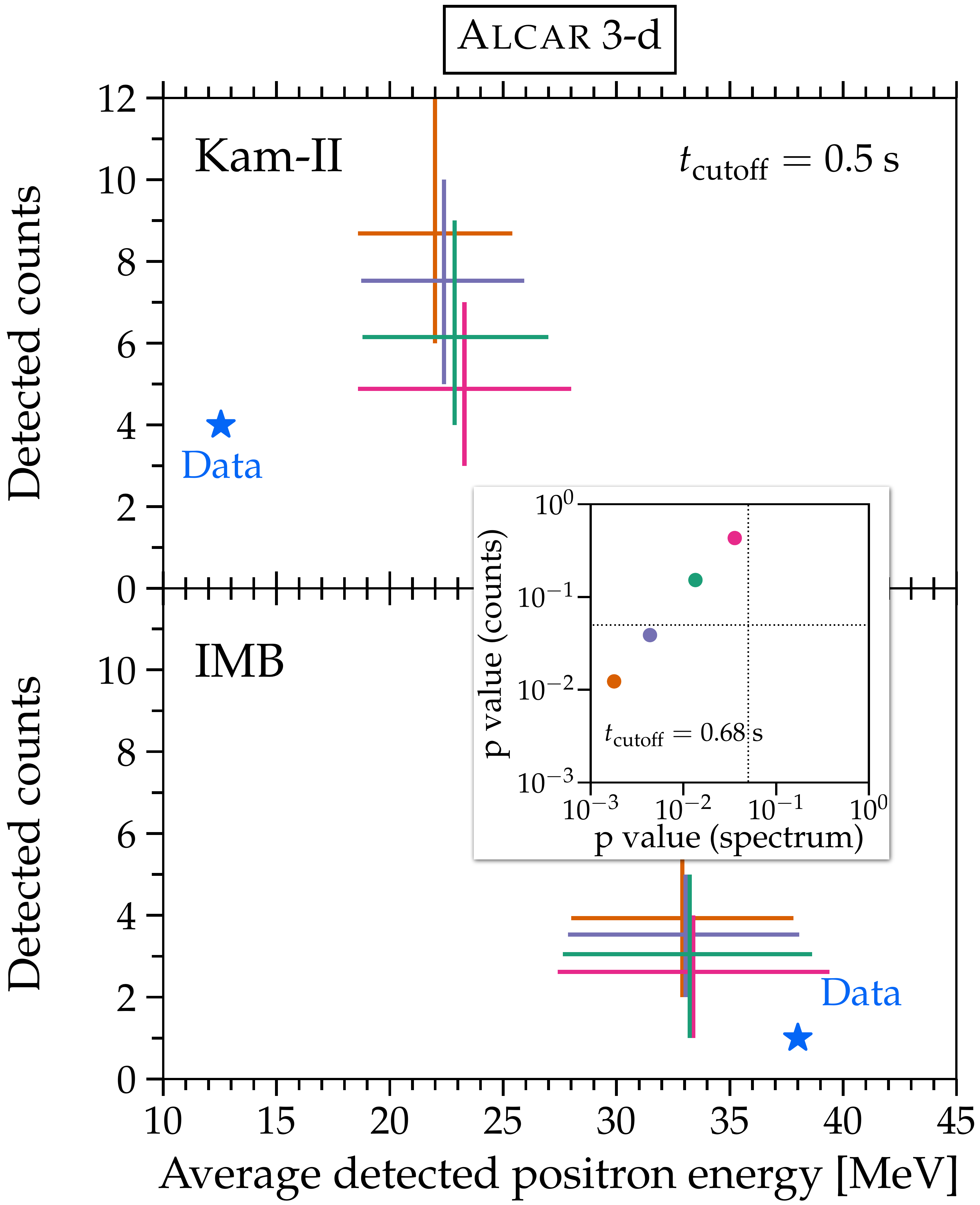}
\caption{Different oscillation scenarios for 3-d simulations.}
\label{fig:oscillation_3d}
\end{figure*}

\begin{figure*}[h]
\centering
\includegraphics[width=0.9\columnwidth]{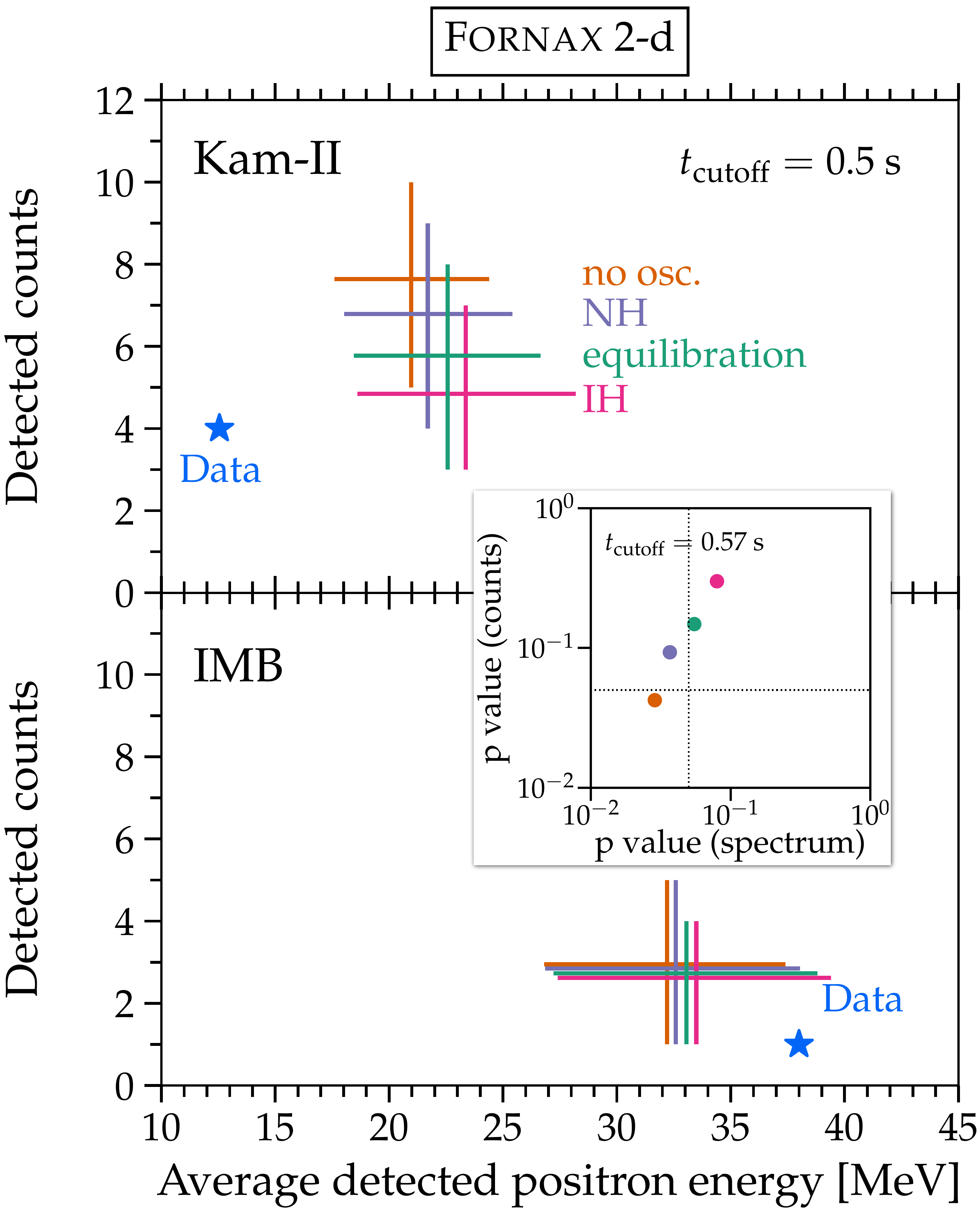}
\includegraphics[width=0.9\columnwidth]{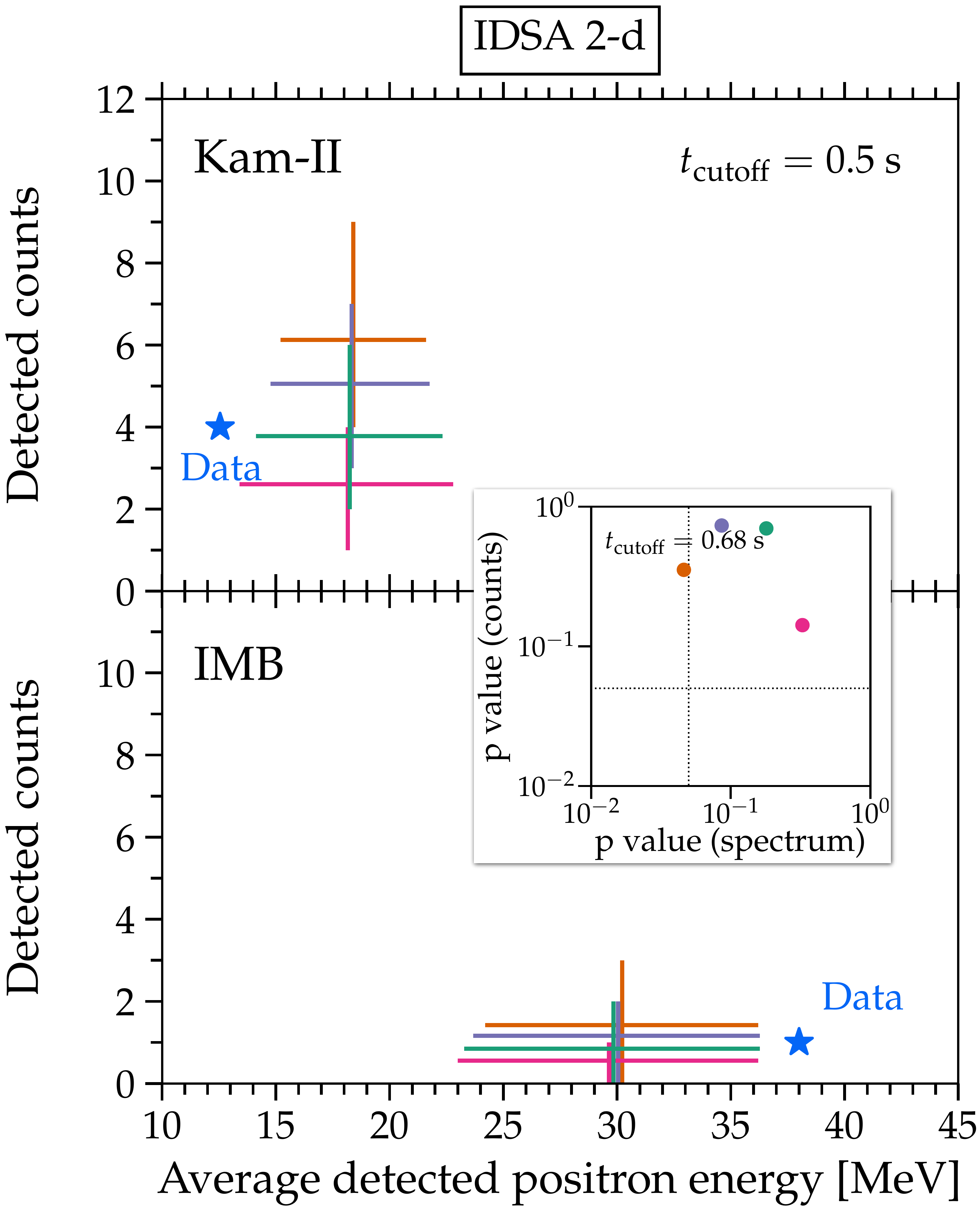} \\
\vspace{2em}
\includegraphics[width=0.9\columnwidth]{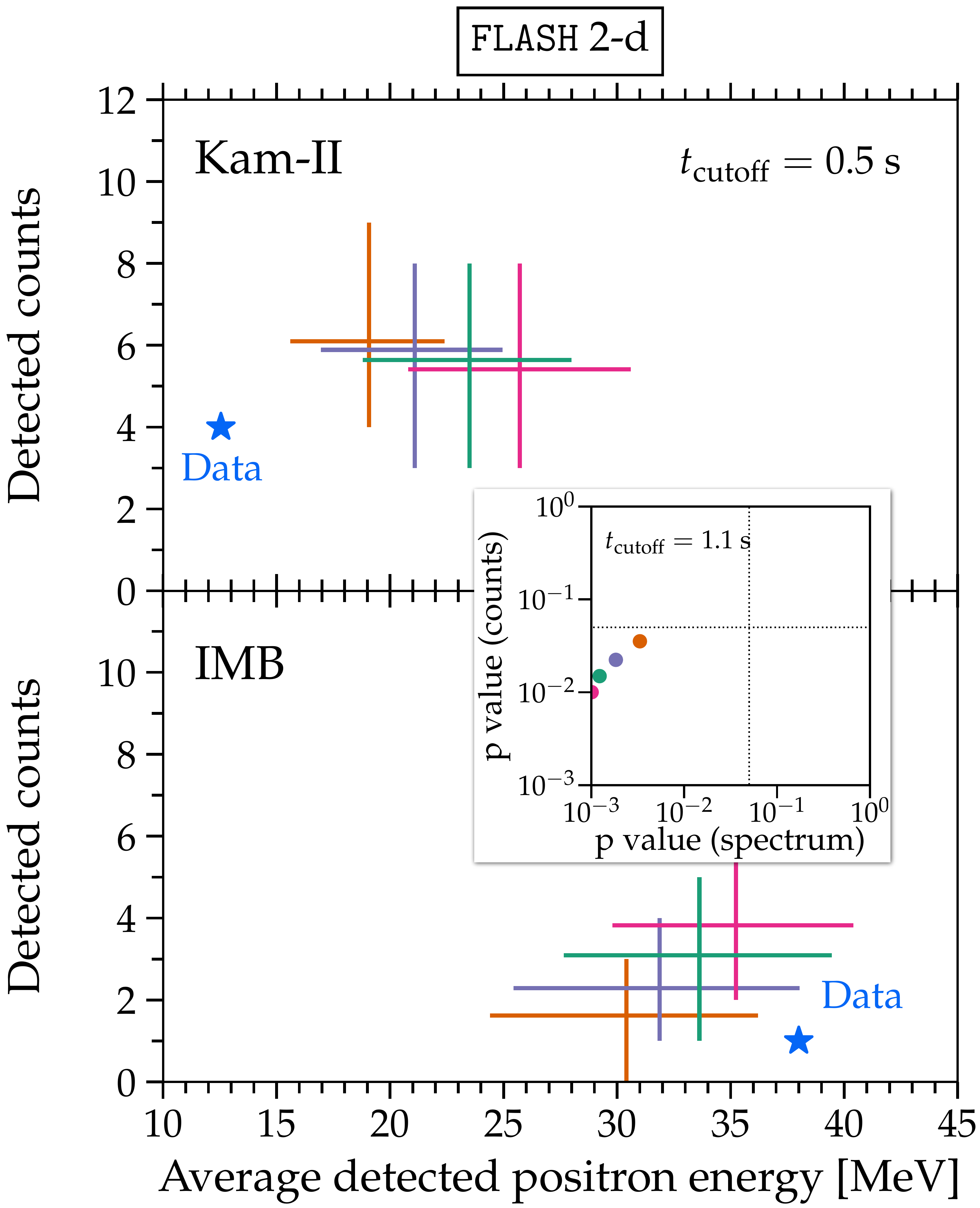}
\includegraphics[width=0.9\columnwidth]{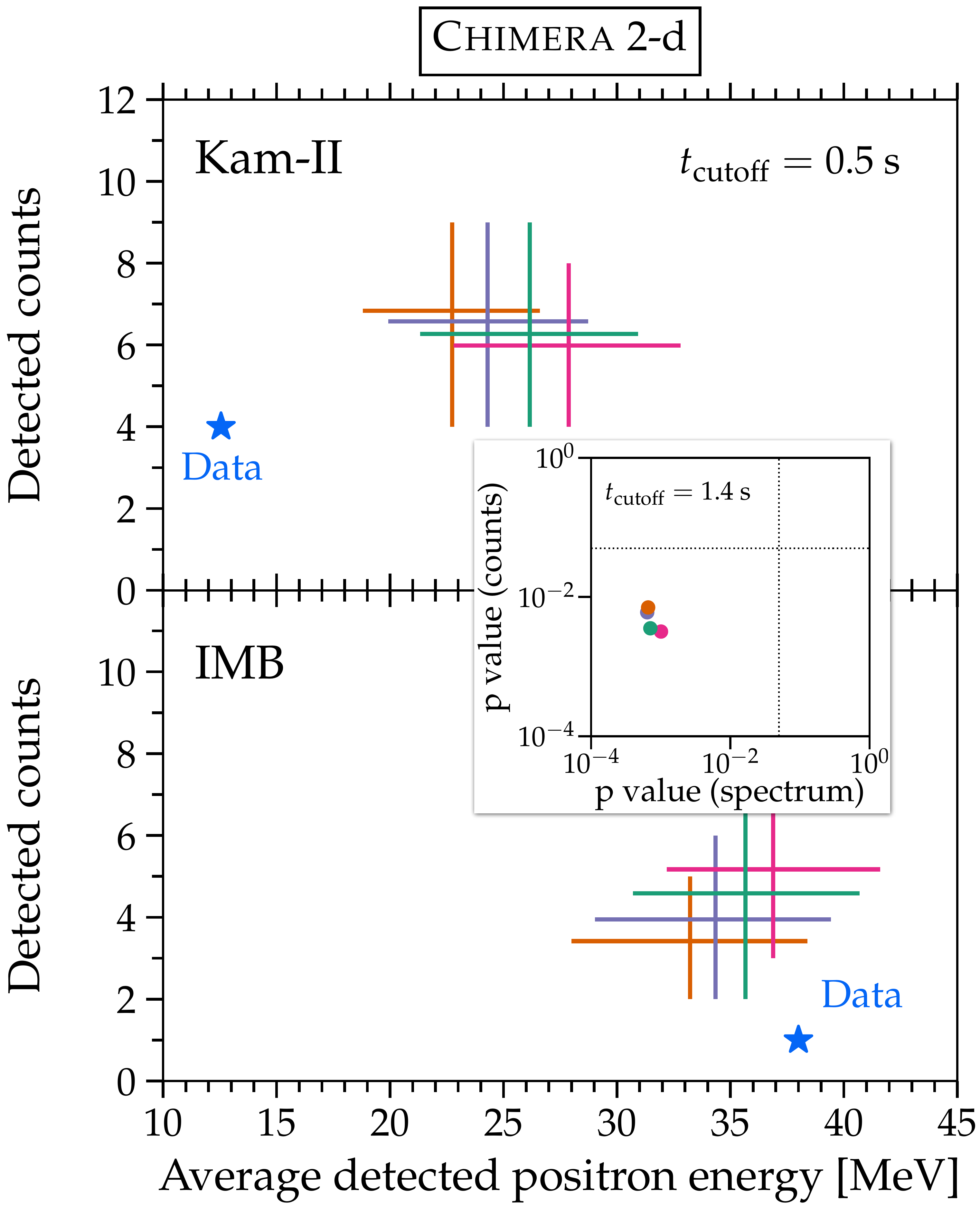}
\caption{Different oscillation scenarios for 2-d simulations.}
\label{fig:oscillation_2d}
\end{figure*}

\begin{figure*}[h]
\centering
\includegraphics[width=0.9\columnwidth]{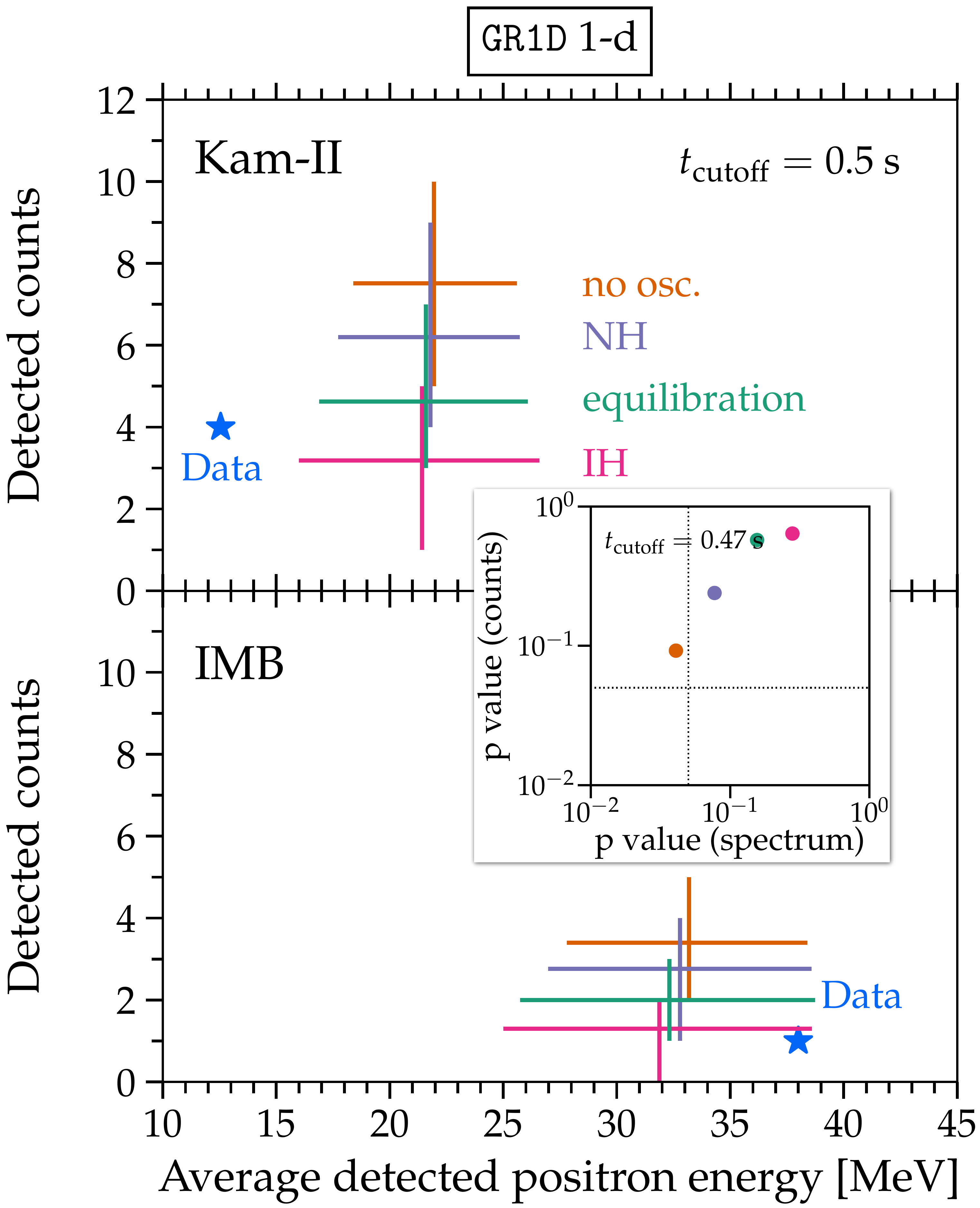}
\includegraphics[width=0.9\columnwidth]{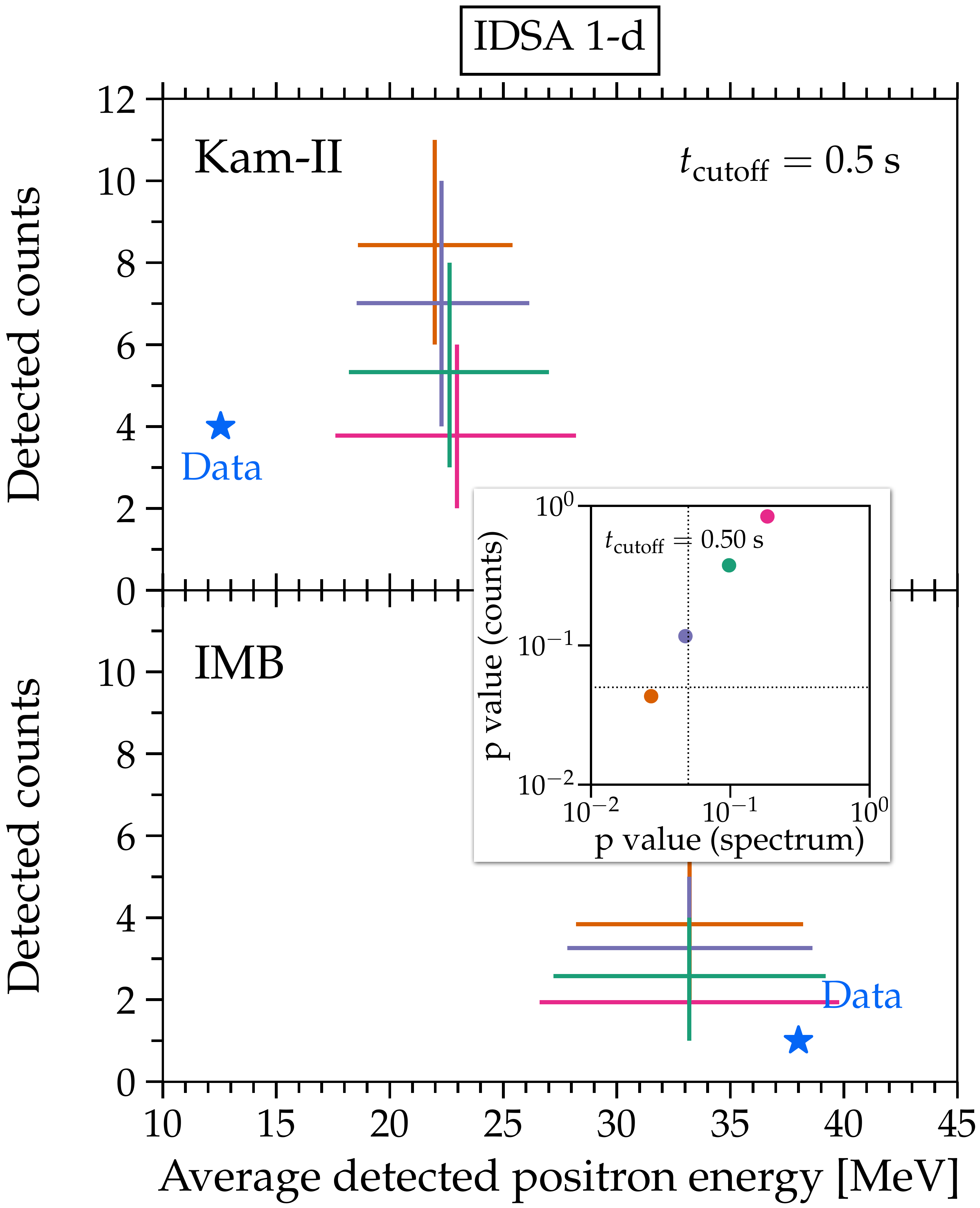} \\
\vspace{2em}
\includegraphics[width=0.9\columnwidth]{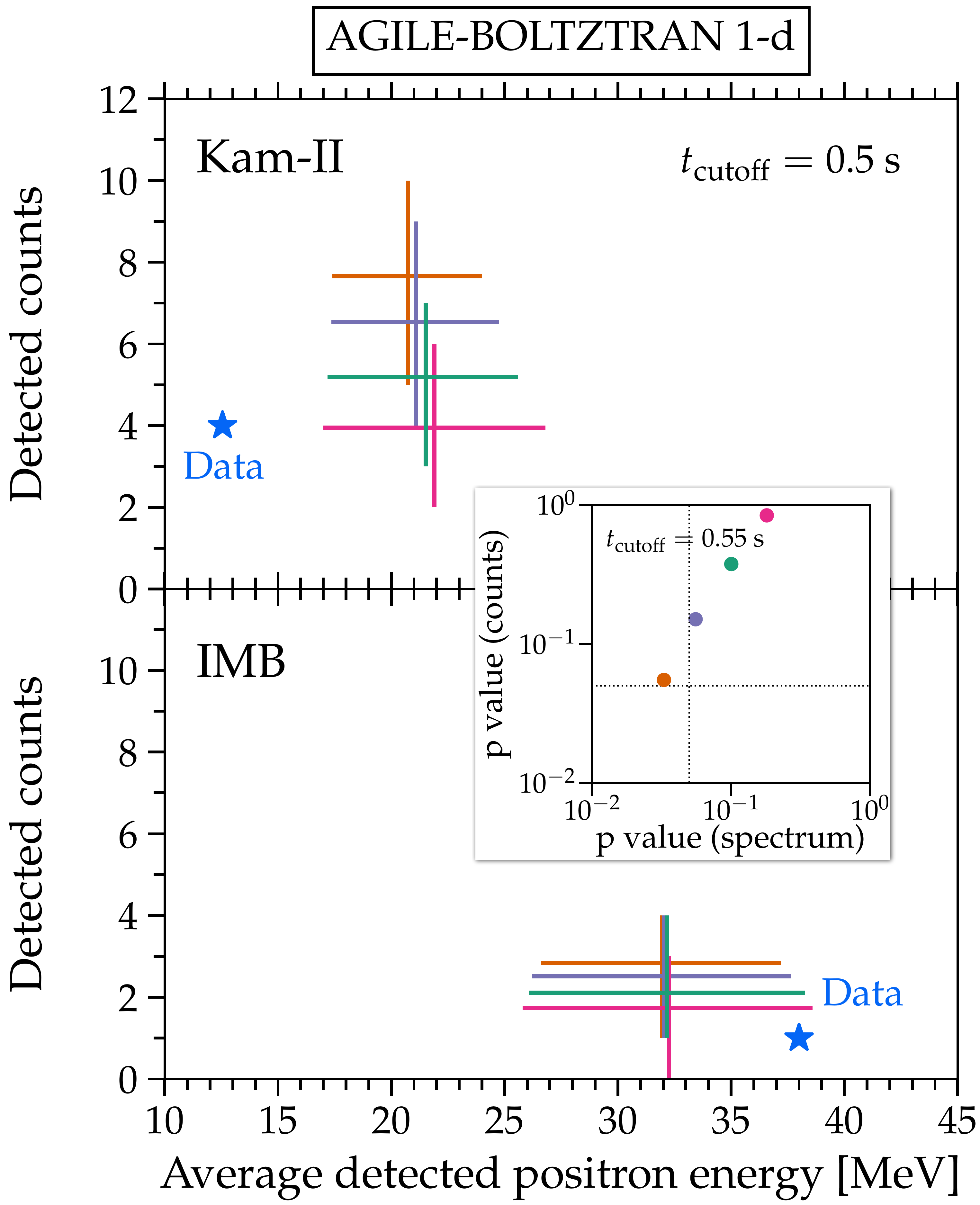}
\includegraphics[width=0.9\columnwidth]{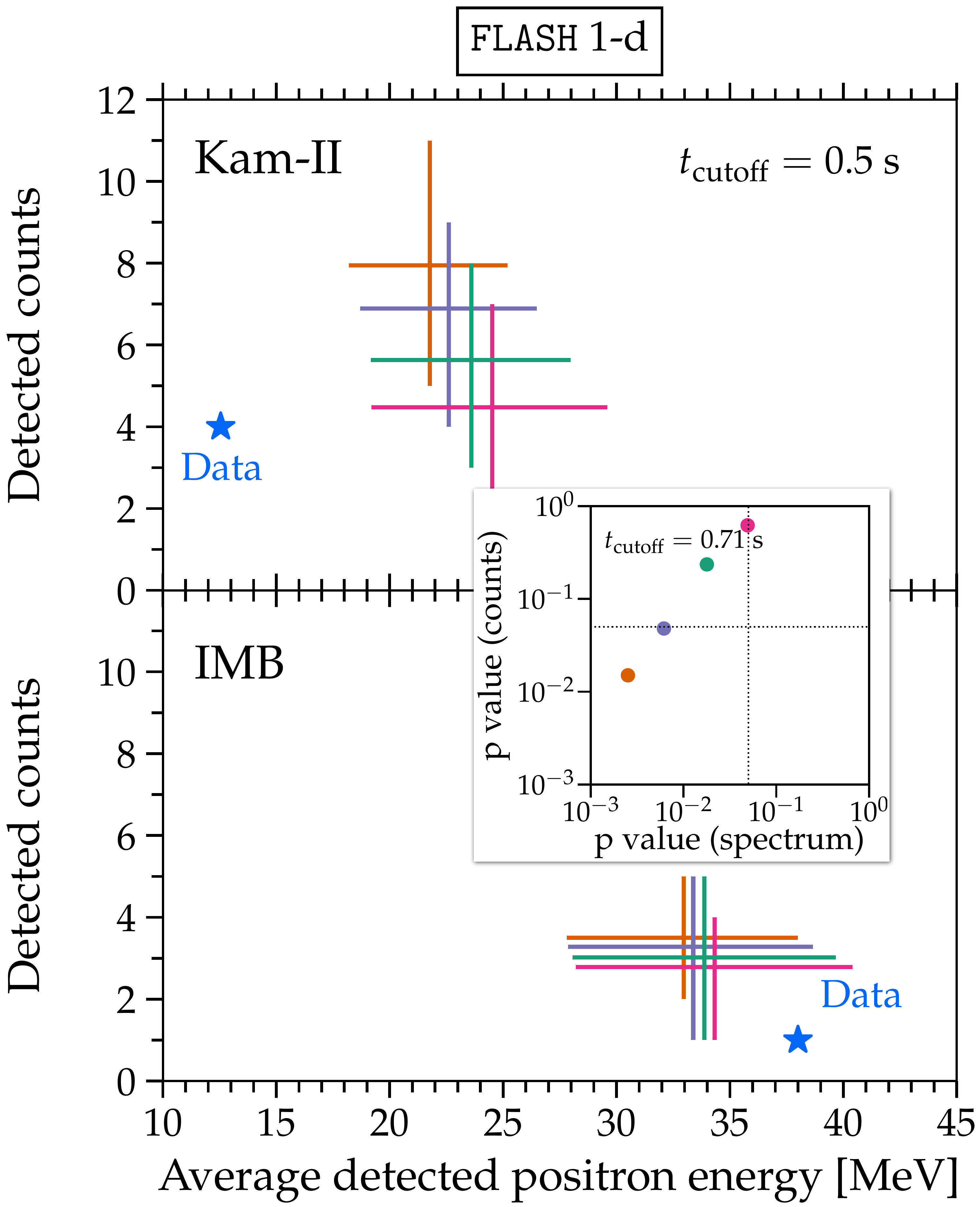}
\caption{Different oscillation scenarios for 1-d simulations.}
\label{fig:oscillation_1d}
\end{figure*}

\begin{figure*}[h]
\centering
\includegraphics[width=0.9\columnwidth]{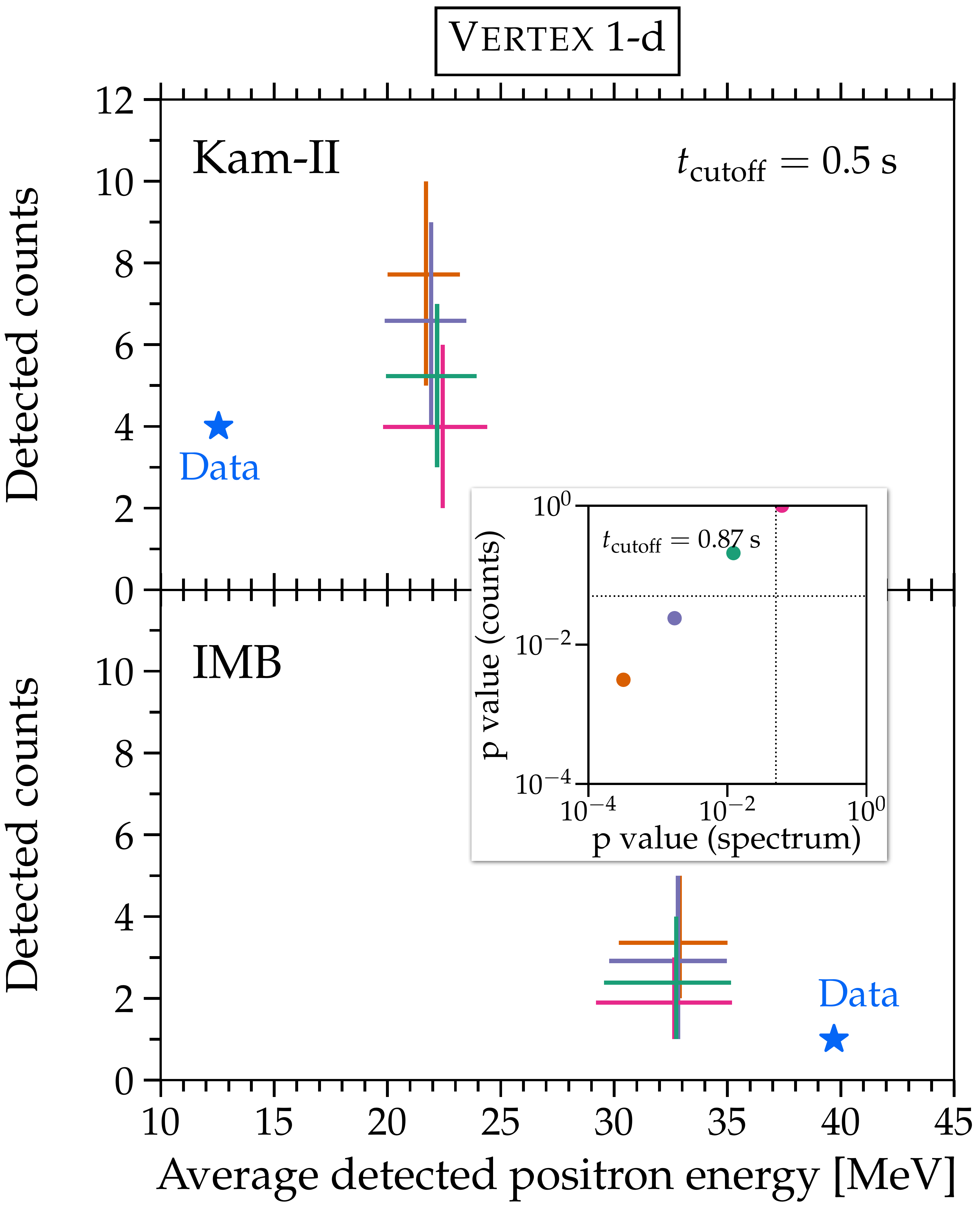}
\caption{Figure~\ref{fig:oscillation_1d} continued.}
\label{fig:oscillation_1d_cont}
\end{figure*}

In Figs.~\ref{fig:oscillation_3d} to~\ref{fig:oscillation_1d_cont}, we show the effects of oscillations on the predicted event rates, spectra, and p-values for models from different groups. For simulations with the same dimensions, we sort the results in ascending order of simulation time.  While some models have reasonable p-values, in general the fit worsen as the simulation time increases; when the runtimes are short, models can appear better than they likely are.  This point seems to be confirmed by the results of Ref.~\cite{Bollig:2020phc} (for $19M_\odot$), a sophisticated model with a long runtime, which also gives a poor match to the SN 1987A data.  It would be desirable to have more models that run longer.  Finally, the goal is to have nearly all models fit the data well.

The p-values for counts and spectra tend to be correlated.  The most basic reason is that the detected average energy depends on the neutrino average energy $\langle E\rangle$ and the counts depends on $E_\text{tot} \langle E\rangle$.  There is also a nontrivial interplay in the p-values.  As an example, let us consider the \textsc{Alcar} 3-d results, as shown in the main text in Fig.~\ref{fig:oscillation}.  Oscillations convert $\bar\nu_x$ and $\bar{\nu}_e$ into each other, which means that the detected $\bar{\nu}_e$ spectrum has a lower number of neutrinos and an increased average energy with respect to the one at production.  Because without oscillations theoretical predictions give a higher flux compared to data, their inclusion makes the counts p-value better.  On the other hand, the unoscillated value of the average energy for $\bar\nu_e$ is higher than that the observed one in the Kam-II data, thus oscillations should, in principle, make the spectrum p-value worse.  But the opposite happens because the spectrum p-value includes information on counts in a subtle way.  To explain why, let us look at Fig.~\ref{fig:cumulative_spectra_TS}.  If we increase the predicted average energy, the predicted spectrum (blue dashed line in the left panel) shifts to the right, increasing the test statistic relative to data.  However, to calculate a p-value for a test statistic, we run Monte Carlo samples based on the blue dashed curve.  With oscillations, because of the lower flux, we would generally sample fewer events for the mock data samples, which would increase the mock test statistic because of the coarse steps.  In combination, it is not clear whether the spectrum p-value would increase or decrease, but what seems to happen in most of the models is that the spectrum p-values end up increasing because they are dominated by the bad fit due to the counts.

\begin{thebibliography}{93}%
\makeatletter
\providecommand \@ifxundefined [1]{%
 \@ifx{#1\undefined}
}%
\providecommand \@ifnum [1]{%
 \ifnum #1\expandafter \@firstoftwo
 \else \expandafter \@secondoftwo
 \fi
}%
\providecommand \@ifx [1]{%
 \ifx #1\expandafter \@firstoftwo
 \else \expandafter \@secondoftwo
 \fi
}%
\providecommand \natexlab [1]{#1}%
\providecommand \enquote  [1]{``#1''}%
\providecommand \bibnamefont  [1]{#1}%
\providecommand \bibfnamefont [1]{#1}%
\providecommand \citenamefont [1]{#1}%
\providecommand \href@noop [0]{\@secondoftwo}%
\providecommand \href [0]{\begingroup \@sanitize@url \@href}%
\providecommand \@href[1]{\@@startlink{#1}\@@href}%
\providecommand \@@href[1]{\endgroup#1\@@endlink}%
\providecommand \@sanitize@url [0]{\catcode `\\12\catcode `\$12\catcode
  `\&12\catcode `\#12\catcode `\^12\catcode `\_12\catcode `\%12\relax}%
\providecommand \@@startlink[1]{}%
\providecommand \@@endlink[0]{}%
\providecommand \url  [0]{\begingroup\@sanitize@url \@url }%
\providecommand \@url [1]{\endgroup\@href {#1}{\urlprefix }}%
\providecommand \urlprefix  [0]{URL }%
\providecommand \Eprint [0]{\href }%
\providecommand \doibase [0]{http://dx.doi.org/}%
\providecommand \selectlanguage [0]{\@gobble}%
\providecommand \bibinfo  [0]{\@secondoftwo}%
\providecommand \bibfield  [0]{\@secondoftwo}%
\providecommand \translation [1]{[#1]}%
\providecommand \BibitemOpen [0]{}%
\providecommand \bibitemStop [0]{}%
\providecommand \bibitemNoStop [0]{.\EOS\space}%
\providecommand \EOS [0]{\spacefactor3000\relax}%
\providecommand \BibitemShut  [1]{\csname bibitem#1\endcsname}%
\let\auto@bib@innerbib\@empty
\bibitem [{\citenamefont {Hirata}\ \emph {et~al.}(1987)\citenamefont {Hirata}
  \emph {et~al.}}]{Hirata:1987hu}%
  \BibitemOpen
  \bibfield  {author} {\bibinfo {author} {\bibfnamefont {K.}~\bibnamefont
  {Hirata}} \emph {et~al.} (\bibinfo {collaboration} {Kamiokande-II}),\
  }\bibfield  {title} {\enquote {\bibinfo {title} {{Observation of a Neutrino
  Burst from the Supernova SN 1987a}},}\ }\href {\doibase
  10.1103/PhysRevLett.58.1490} {\bibfield  {journal} {\bibinfo  {journal}
  {Phys. Rev. Lett.}\ }\textbf {\bibinfo {volume} {58}},\ \bibinfo {pages}
  {1490--1493} (\bibinfo {year} {1987})}\BibitemShut {NoStop}%
\bibitem [{\citenamefont {Hirata}\ \emph {et~al.}(1988)\citenamefont {Hirata}
  \emph {et~al.}}]{Hirata:1988ad}%
  \BibitemOpen
  \bibfield  {author} {\bibinfo {author} {\bibfnamefont {K.~S.}\ \bibnamefont
  {Hirata}} \emph {et~al.},\ }\bibfield  {title} {\enquote {\bibinfo {title}
  {{Observation in the Kamiokande-II Detector of the Neutrino Burst from
  Supernova SN 1987a}},}\ }\href {\doibase 10.1103/PhysRevD.38.448} {\bibfield
  {journal} {\bibinfo  {journal} {Phys. Rev. D}\ }\textbf {\bibinfo {volume}
  {38}},\ \bibinfo {pages} {448--458} (\bibinfo {year} {1988})}\BibitemShut
  {NoStop}%
\bibitem [{\citenamefont {Bionta}\ \emph {et~al.}(1987)\citenamefont {Bionta}
  \emph {et~al.}}]{Bionta:1987qt}%
  \BibitemOpen
  \bibfield  {author} {\bibinfo {author} {\bibfnamefont {R.~M.}\ \bibnamefont
  {Bionta}} \emph {et~al.},\ }\bibfield  {title} {\enquote {\bibinfo {title}
  {{Observation of a Neutrino Burst in Coincidence with Supernova SN 1987a in
  the Large Magellanic Cloud}},}\ }\href {\doibase 10.1103/PhysRevLett.58.1494}
  {\bibfield  {journal} {\bibinfo  {journal} {Phys. Rev. Lett.}\ }\textbf
  {\bibinfo {volume} {58}},\ \bibinfo {pages} {1494} (\bibinfo {year}
  {1987})}\BibitemShut {NoStop}%
\bibitem [{\citenamefont {Bratton}\ \emph {et~al.}(1988)\citenamefont {Bratton}
  \emph {et~al.}}]{IMB:1988suc}%
  \BibitemOpen
  \bibfield  {author} {\bibinfo {author} {\bibfnamefont {C.~B.}\ \bibnamefont
  {Bratton}} \emph {et~al.} (\bibinfo {collaboration} {IMB}),\ }\bibfield
  {title} {\enquote {\bibinfo {title} {{Angular Distribution of Events From
  Sn1987a}},}\ }\href {\doibase 10.1103/PhysRevD.37.3361} {\bibfield  {journal}
  {\bibinfo  {journal} {Phys. Rev. D}\ }\textbf {\bibinfo {volume} {37}},\
  \bibinfo {pages} {3361} (\bibinfo {year} {1988})}\BibitemShut {NoStop}%
\bibitem [{\citenamefont {Arnett}\ \emph {et~al.}(1989)\citenamefont {Arnett},
  \citenamefont {Bahcall}, \citenamefont {Kirshner},\ and\ \citenamefont
  {Woosley}}]{Arnett:1989tnf}%
  \BibitemOpen
  \bibfield  {author} {\bibinfo {author} {\bibfnamefont {W.~D.}\ \bibnamefont
  {Arnett}}, \bibinfo {author} {\bibfnamefont {John~N.}\ \bibnamefont
  {Bahcall}}, \bibinfo {author} {\bibfnamefont {R.~P.}\ \bibnamefont
  {Kirshner}}, \ and\ \bibinfo {author} {\bibfnamefont {S.~E.}\ \bibnamefont
  {Woosley}},\ }\bibfield  {title} {\enquote {\bibinfo {title} {{Supernova
  1987A}},}\ }\href {\doibase 10.1146/annurev.aa.27.090189.003213} {\bibfield
  {journal} {\bibinfo  {journal} {Ann. Rev. Astron. Astrophys.}\ }\textbf
  {\bibinfo {volume} {27}},\ \bibinfo {pages} {629--700} (\bibinfo {year}
  {1989})}\BibitemShut {NoStop}%
\bibitem [{\citenamefont {{McCray}}(1993)}]{1993ARA&A..31..175M}%
  \BibitemOpen
  \bibfield  {author} {\bibinfo {author} {\bibfnamefont {Richard}\ \bibnamefont
  {{McCray}}},\ }\bibfield  {title} {\enquote {\bibinfo {title} {{Supernova
  1987A revisited.}}}\ }\href {\doibase 10.1146/annurev.aa.31.090193.001135}
  {\bibfield  {journal} {\bibinfo  {journal} {Ann. Rev. Astron. Astrophys.}\
  }\textbf {\bibinfo {volume} {31}},\ \bibinfo {pages} {175--216} (\bibinfo
  {year} {1993})}\BibitemShut {NoStop}%
\bibitem [{\citenamefont {Scholberg}(2012)}]{Scholberg:2012id}%
  \BibitemOpen
  \bibfield  {author} {\bibinfo {author} {\bibfnamefont {Kate}\ \bibnamefont
  {Scholberg}},\ }\bibfield  {title} {\enquote {\bibinfo {title} {{Supernova
  Neutrino Detection}},}\ }\href {\doibase 10.1146/annurev-nucl-102711-095006}
  {\bibfield  {journal} {\bibinfo  {journal} {Ann. Rev. Nucl. Part. Sci.}\
  }\textbf {\bibinfo {volume} {62}},\ \bibinfo {pages} {81--103} (\bibinfo
  {year} {2012})},\ \Eprint {http://arxiv.org/abs/1205.6003} {arXiv:1205.6003
  [astro-ph.IM]} \BibitemShut {NoStop}%
\bibitem [{\citenamefont {Adams}\ \emph {et~al.}(2013)\citenamefont {Adams},
  \citenamefont {Kochanek}, \citenamefont {Beacom}, \citenamefont {Vagins},\
  and\ \citenamefont {Stanek}}]{Adams:2013ana}%
  \BibitemOpen
  \bibfield  {author} {\bibinfo {author} {\bibfnamefont {Scott~M.}\
  \bibnamefont {Adams}}, \bibinfo {author} {\bibfnamefont {C.~S.}\ \bibnamefont
  {Kochanek}}, \bibinfo {author} {\bibfnamefont {John~F.}\ \bibnamefont
  {Beacom}}, \bibinfo {author} {\bibfnamefont {Mark~R.}\ \bibnamefont
  {Vagins}}, \ and\ \bibinfo {author} {\bibfnamefont {K.~Z.}\ \bibnamefont
  {Stanek}},\ }\bibfield  {title} {\enquote {\bibinfo {title} {{Observing the
  Next Galactic Supernova}},}\ }\href {\doibase 10.1088/0004-637X/778/2/164}
  {\bibfield  {journal} {\bibinfo  {journal} {Astrophys. J.}\ }\textbf
  {\bibinfo {volume} {778}},\ \bibinfo {pages} {164} (\bibinfo {year}
  {2013})},\ \Eprint {http://arxiv.org/abs/1306.0559} {arXiv:1306.0559
  [astro-ph.HE]} \BibitemShut {NoStop}%
\bibitem [{\citenamefont {Mirizzi}\ \emph {et~al.}(2016)\citenamefont
  {Mirizzi}, \citenamefont {Tamborra}, \citenamefont {Janka}, \citenamefont
  {Saviano}, \citenamefont {Scholberg}, \citenamefont {Bollig}, \citenamefont
  {Hudepohl},\ and\ \citenamefont {Chakraborty}}]{Mirizzi:2015eza}%
  \BibitemOpen
  \bibfield  {author} {\bibinfo {author} {\bibfnamefont {Alessandro}\
  \bibnamefont {Mirizzi}}, \bibinfo {author} {\bibfnamefont {Irene}\
  \bibnamefont {Tamborra}}, \bibinfo {author} {\bibfnamefont {Hans-Thomas}\
  \bibnamefont {Janka}}, \bibinfo {author} {\bibfnamefont {Ninetta}\
  \bibnamefont {Saviano}}, \bibinfo {author} {\bibfnamefont {Kate}\
  \bibnamefont {Scholberg}}, \bibinfo {author} {\bibfnamefont {Robert}\
  \bibnamefont {Bollig}}, \bibinfo {author} {\bibfnamefont {Lorenz}\
  \bibnamefont {Hudepohl}}, \ and\ \bibinfo {author} {\bibfnamefont {Sovan}\
  \bibnamefont {Chakraborty}},\ }\bibfield  {title} {\enquote {\bibinfo {title}
  {{Supernova Neutrinos: Production, Oscillations and Detection}},}\ }\href
  {\doibase 10.1393/ncr/i2016-10120-8} {\bibfield  {journal} {\bibinfo
  {journal} {Riv. Nuovo Cim.}\ }\textbf {\bibinfo {volume} {39}},\ \bibinfo
  {pages} {1--112} (\bibinfo {year} {2016})},\ \Eprint
  {http://arxiv.org/abs/1508.00785} {arXiv:1508.00785 [astro-ph.HE]}
  \BibitemShut {NoStop}%
\bibitem [{\citenamefont {Nakamura}\ \emph {et~al.}(2016)\citenamefont
  {Nakamura}, \citenamefont {Horiuchi}, \citenamefont {Tanaka}, \citenamefont
  {Hayama}, \citenamefont {Takiwaki},\ and\ \citenamefont
  {Kotake}}]{Nakamura:2016kkl}%
  \BibitemOpen
  \bibfield  {author} {\bibinfo {author} {\bibfnamefont {Ko}~\bibnamefont
  {Nakamura}}, \bibinfo {author} {\bibfnamefont {Shunsaku}\ \bibnamefont
  {Horiuchi}}, \bibinfo {author} {\bibfnamefont {Masaomi}\ \bibnamefont
  {Tanaka}}, \bibinfo {author} {\bibfnamefont {Kazuhiro}\ \bibnamefont
  {Hayama}}, \bibinfo {author} {\bibfnamefont {Tomoya}\ \bibnamefont
  {Takiwaki}}, \ and\ \bibinfo {author} {\bibfnamefont {Kei}\ \bibnamefont
  {Kotake}},\ }\bibfield  {title} {\enquote {\bibinfo {title} {{Multimessenger
  signals of long-term core-collapse supernova simulations: synergetic
  observation strategies}},}\ }\href {\doibase 10.1093/mnras/stw1453}
  {\bibfield  {journal} {\bibinfo  {journal} {Mon. Not. Roy. Astron. Soc.}\
  }\textbf {\bibinfo {volume} {461}},\ \bibinfo {pages} {3296--3313} (\bibinfo
  {year} {2016})},\ \Eprint {http://arxiv.org/abs/1602.03028} {arXiv:1602.03028
  [astro-ph.HE]} \BibitemShut {NoStop}%
\bibitem [{\citenamefont {Diehl}\ \emph {et~al.}(2006)\citenamefont {Diehl}
  \emph {et~al.}}]{Diehl:2006cf}%
  \BibitemOpen
  \bibfield  {author} {\bibinfo {author} {\bibfnamefont {Roland}\ \bibnamefont
  {Diehl}} \emph {et~al.},\ }\bibfield  {title} {\enquote {\bibinfo {title}
  {{Radioactive Al-26 and massive stars in the galaxy}},}\ }\href {\doibase
  10.1038/nature04364} {\bibfield  {journal} {\bibinfo  {journal} {Nature}\
  }\textbf {\bibinfo {volume} {439}},\ \bibinfo {pages} {45--47} (\bibinfo
  {year} {2006})},\ \Eprint {http://arxiv.org/abs/astro-ph/0601015}
  {arXiv:astro-ph/0601015} \BibitemShut {NoStop}%
\bibitem [{\citenamefont {{Li}}\ \emph {et~al.}(2011)\citenamefont {{Li}},
  \citenamefont {{Chornock}}, \citenamefont {{Leaman}}, \citenamefont
  {{Filippenko}}, \citenamefont {{Poznanski}}, \citenamefont {{Wang}},
  \citenamefont {{Ganeshalingam}},\ and\ \citenamefont
  {{Mannucci}}}]{2011MNRAS.412.1473L}%
  \BibitemOpen
  \bibfield  {author} {\bibinfo {author} {\bibfnamefont {Weidong}\ \bibnamefont
  {{Li}}}, \bibinfo {author} {\bibfnamefont {Ryan}\ \bibnamefont {{Chornock}}},
  \bibinfo {author} {\bibfnamefont {Jesse}\ \bibnamefont {{Leaman}}}, \bibinfo
  {author} {\bibfnamefont {Alexei~V.}\ \bibnamefont {{Filippenko}}}, \bibinfo
  {author} {\bibfnamefont {Dovi}\ \bibnamefont {{Poznanski}}}, \bibinfo
  {author} {\bibfnamefont {Xiaofeng}\ \bibnamefont {{Wang}}}, \bibinfo {author}
  {\bibfnamefont {Mohan}\ \bibnamefont {{Ganeshalingam}}}, \ and\ \bibinfo
  {author} {\bibfnamefont {Filippo}\ \bibnamefont {{Mannucci}}},\ }\bibfield
  {title} {\enquote {\bibinfo {title} {{Nearby supernova rates from the Lick
  Observatory Supernova Search - III. The rate-size relation, and the rates as
  a function of galaxy Hubble type and colour}},}\ }\href {\doibase
  10.1111/j.1365-2966.2011.18162.x} {\bibfield  {journal} {\bibinfo  {journal}
  {Mon. Not. Roy. Astron. Soc.}\ }\textbf {\bibinfo {volume} {412}},\ \bibinfo
  {pages} {1473--1507} (\bibinfo {year} {2011})},\ \Eprint
  {http://arxiv.org/abs/1006.4613} {arXiv:1006.4613 [astro-ph.SR]} \BibitemShut
  {NoStop}%
\bibitem [{\citenamefont {Rozwadowska}\ \emph {et~al.}(2021)\citenamefont
  {Rozwadowska}, \citenamefont {Vissani},\ and\ \citenamefont
  {Cappellaro}}]{Rozwadowska:2020nab}%
  \BibitemOpen
  \bibfield  {author} {\bibinfo {author} {\bibfnamefont {Karolina}\
  \bibnamefont {Rozwadowska}}, \bibinfo {author} {\bibfnamefont {Francesco}\
  \bibnamefont {Vissani}}, \ and\ \bibinfo {author} {\bibfnamefont {Enrico}\
  \bibnamefont {Cappellaro}},\ }\bibfield  {title} {\enquote {\bibinfo {title}
  {{On the rate of core collapse supernovae in the milky way}},}\ }\href
  {\doibase 10.1016/j.newast.2020.101498} {\bibfield  {journal} {\bibinfo
  {journal} {New Astron.}\ }\textbf {\bibinfo {volume} {83}},\ \bibinfo {pages}
  {101498} (\bibinfo {year} {2021})},\ \Eprint
  {http://arxiv.org/abs/2009.03438} {arXiv:2009.03438 [astro-ph.HE]}
  \BibitemShut {NoStop}%
\bibitem [{\citenamefont {{Hanke}}\ \emph {et~al.}(2013)\citenamefont
  {{Hanke}}, \citenamefont {{M{\"u}ller}}, \citenamefont {{Wongwathanarat}},
  \citenamefont {{Marek}},\ and\ \citenamefont {{Janka}}}]{Hanke:2013}%
  \BibitemOpen
  \bibfield  {author} {\bibinfo {author} {\bibfnamefont {Florian}\ \bibnamefont
  {{Hanke}}}, \bibinfo {author} {\bibfnamefont {Bernhard}\ \bibnamefont
  {{M{\"u}ller}}}, \bibinfo {author} {\bibfnamefont {Annop}\ \bibnamefont
  {{Wongwathanarat}}}, \bibinfo {author} {\bibfnamefont {Andreas}\ \bibnamefont
  {{Marek}}}, \ and\ \bibinfo {author} {\bibfnamefont {Hans-Thomas}\
  \bibnamefont {{Janka}}},\ }\bibfield  {title} {\enquote {\bibinfo {title}
  {{SASI Activity in Three-dimensional Neutrino-hydrodynamics Simulations of
  Supernova Cores}},}\ }\href {\doibase 10.1088/0004-637X/770/1/66} {\bibfield
  {journal} {\bibinfo  {journal} {\apj}\ }\textbf {\bibinfo {volume} {770}},\
  \bibinfo {eid} {66} (\bibinfo {year} {2013})},\ \Eprint
  {http://arxiv.org/abs/1303.6269} {arXiv:1303.6269 [astro-ph.SR]} \BibitemShut
  {NoStop}%
\bibitem [{\citenamefont {Takiwaki}\ \emph {et~al.}(2014)\citenamefont
  {Takiwaki}, \citenamefont {Kotake},\ and\ \citenamefont
  {Suwa}}]{Takiwaki:2013cqa}%
  \BibitemOpen
  \bibfield  {author} {\bibinfo {author} {\bibfnamefont {Tomoya}\ \bibnamefont
  {Takiwaki}}, \bibinfo {author} {\bibfnamefont {Kei}\ \bibnamefont {Kotake}},
  \ and\ \bibinfo {author} {\bibfnamefont {Yudai}\ \bibnamefont {Suwa}},\
  }\bibfield  {title} {\enquote {\bibinfo {title} {{A Comparison of Two- and
  Three-dimensional Neutrino-hydrodynamics simulations of Core-collapse
  Supernovae}},}\ }\href {\doibase 10.1088/0004-637X/786/2/83} {\bibfield
  {journal} {\bibinfo  {journal} {Astrophys. J.}\ }\textbf {\bibinfo {volume}
  {786}},\ \bibinfo {pages} {83} (\bibinfo {year} {2014})},\ \Eprint
  {http://arxiv.org/abs/1308.5755} {arXiv:1308.5755 [astro-ph.SR]} \BibitemShut
  {NoStop}%
\bibitem [{\citenamefont {{Lentz}}\ \emph {et~al.}(2015)\citenamefont
  {{Lentz}}, \citenamefont {{Bruenn}}, \citenamefont {{Hix}}, \citenamefont
  {{Mezzacappa}}, \citenamefont {{Messer}}, \citenamefont {{Endeve}},
  \citenamefont {{Blondin}}, \citenamefont {{Harris}}, \citenamefont
  {{Marronetti}},\ and\ \citenamefont {{Yakunin}}}]{Lentz:2015}%
  \BibitemOpen
  \bibfield  {author} {\bibinfo {author} {\bibfnamefont {Eric~J.}\ \bibnamefont
  {{Lentz}}}, \bibinfo {author} {\bibfnamefont {Stephen~W.}\ \bibnamefont
  {{Bruenn}}}, \bibinfo {author} {\bibfnamefont {W.~Raphael}\ \bibnamefont
  {{Hix}}}, \bibinfo {author} {\bibfnamefont {Anthony}\ \bibnamefont
  {{Mezzacappa}}}, \bibinfo {author} {\bibfnamefont {O.~E.~Bronson}\
  \bibnamefont {{Messer}}}, \bibinfo {author} {\bibfnamefont {Eirik}\
  \bibnamefont {{Endeve}}}, \bibinfo {author} {\bibfnamefont {John~M.}\
  \bibnamefont {{Blondin}}}, \bibinfo {author} {\bibfnamefont {J.~Austin}\
  \bibnamefont {{Harris}}}, \bibinfo {author} {\bibfnamefont {Pedro}\
  \bibnamefont {{Marronetti}}}, \ and\ \bibinfo {author} {\bibfnamefont
  {Konstantin~N.}\ \bibnamefont {{Yakunin}}},\ }\bibfield  {title} {\enquote
  {\bibinfo {title} {{Three-dimensional Core-collapse Supernova Simulated Using
  a 15 M $_{{\ensuremath{\odot}}}$ Progenitor}},}\ }\href {\doibase
  10.1088/2041-8205/807/2/L31} {\bibfield  {journal} {\bibinfo  {journal}
  {Astrophys. J. Lett.}\ }\textbf {\bibinfo {volume} {807}},\ \bibinfo {eid}
  {L31} (\bibinfo {year} {2015})},\ \Eprint {http://arxiv.org/abs/1505.05110}
  {arXiv:1505.05110 [astro-ph.SR]} \BibitemShut {NoStop}%
\bibitem [{\citenamefont {Janka}\ \emph {et~al.}(2016)\citenamefont {Janka},
  \citenamefont {Melson},\ and\ \citenamefont {Summa}}]{Janka:2016fox}%
  \BibitemOpen
  \bibfield  {author} {\bibinfo {author} {\bibfnamefont {H.~Thomas}\
  \bibnamefont {Janka}}, \bibinfo {author} {\bibfnamefont {Tobias}\
  \bibnamefont {Melson}}, \ and\ \bibinfo {author} {\bibfnamefont {Alexander}\
  \bibnamefont {Summa}},\ }\bibfield  {title} {\enquote {\bibinfo {title}
  {{Physics of Core-Collapse Supernovae in Three Dimensions: a Sneak
  Preview}},}\ }\href {\doibase 10.1146/annurev-nucl-102115-044747} {\bibfield
  {journal} {\bibinfo  {journal} {Ann. Rev. Nucl. Part. Sci.}\ }\textbf
  {\bibinfo {volume} {66}},\ \bibinfo {pages} {341--375} (\bibinfo {year}
  {2016})},\ \Eprint {http://arxiv.org/abs/1602.05576} {arXiv:1602.05576
  [astro-ph.SR]} \BibitemShut {NoStop}%
\bibitem [{\citenamefont {O'Connor}\ \emph {et~al.}(2018)\citenamefont
  {O'Connor} \emph {et~al.}}]{OConnor:2018sti}%
  \BibitemOpen
  \bibfield  {author} {\bibinfo {author} {\bibfnamefont {Evan}\ \bibnamefont
  {O'Connor}} \emph {et~al.},\ }\bibfield  {title} {\enquote {\bibinfo {title}
  {{Global Comparison of Core-Collapse Supernova Simulations in Spherical
  Symmetry}},}\ }\href {\doibase 10.1088/1361-6471/aadeae} {\bibfield
  {journal} {\bibinfo  {journal} {J. Phys. G}\ }\textbf {\bibinfo {volume}
  {45}},\ \bibinfo {pages} {104001} (\bibinfo {year} {2018})},\ \Eprint
  {http://arxiv.org/abs/1806.04175} {arXiv:1806.04175 [astro-ph.HE]}
  \BibitemShut {NoStop}%
\bibitem [{\citenamefont {{Bruenn}}\ \emph {et~al.}(2013)\citenamefont
  {{Bruenn}}, \citenamefont {{Mezzacappa}}, \citenamefont {{Hix}},
  \citenamefont {{Lentz}}, \citenamefont {{Messer}}, \citenamefont
  {{Lingerfelt}}, \citenamefont {{Blondin}}, \citenamefont {{Endeve}},
  \citenamefont {{Marronetti}},\ and\ \citenamefont
  {{Yakunin}}}]{2013ApJ...767L...6B}%
  \BibitemOpen
  \bibfield  {author} {\bibinfo {author} {\bibfnamefont {Stephen~W.}\
  \bibnamefont {{Bruenn}}}, \bibinfo {author} {\bibfnamefont {Anthony}\
  \bibnamefont {{Mezzacappa}}}, \bibinfo {author} {\bibfnamefont {W.~Raphael}\
  \bibnamefont {{Hix}}}, \bibinfo {author} {\bibfnamefont {Eric~J.}\
  \bibnamefont {{Lentz}}}, \bibinfo {author} {\bibfnamefont {O.~E.~Bronson}\
  \bibnamefont {{Messer}}}, \bibinfo {author} {\bibfnamefont {Eric~J.}\
  \bibnamefont {{Lingerfelt}}}, \bibinfo {author} {\bibfnamefont {John~M.}\
  \bibnamefont {{Blondin}}}, \bibinfo {author} {\bibfnamefont {Eirik}\
  \bibnamefont {{Endeve}}}, \bibinfo {author} {\bibfnamefont {Pedro}\
  \bibnamefont {{Marronetti}}}, \ and\ \bibinfo {author} {\bibfnamefont
  {Konstantin~N.}\ \bibnamefont {{Yakunin}}},\ }\bibfield  {title} {\enquote
  {\bibinfo {title} {{Axisymmetric Ab Initio Core-collapse Supernova
  Simulations of 12-25M$_\odot$ Stars}},}\ }\href {\doibase
  10.1088/2041-8205/767/1/L6} {\bibfield  {journal} {\bibinfo  {journal}
  {Astrophys. J. Lett.}\ }\textbf {\bibinfo {volume} {767}},\ \bibinfo {eid}
  {L6} (\bibinfo {year} {2013})},\ \Eprint {http://arxiv.org/abs/1212.1747}
  {arXiv:1212.1747 [astro-ph.SR]} \BibitemShut {NoStop}%
\bibitem [{\citenamefont {Bruenn}\ \emph {et~al.}(2016)\citenamefont {Bruenn}
  \emph {et~al.}}]{Bruenn:2014qea}%
  \BibitemOpen
  \bibfield  {author} {\bibinfo {author} {\bibfnamefont {Stephen~W.}\
  \bibnamefont {Bruenn}} \emph {et~al.},\ }\bibfield  {title} {\enquote
  {\bibinfo {title} {{The Development of Explosions in Axisymmetric AB INITIO
  Core-Collapse Supernova Simulations of 12-25 $M_\odot$ Stars}},}\ }\href
  {\doibase 10.3847/0004-637X/818/2/123} {\bibfield  {journal} {\bibinfo
  {journal} {Astrophys. J.}\ }\textbf {\bibinfo {volume} {818}},\ \bibinfo
  {pages} {123} (\bibinfo {year} {2016})},\ \Eprint
  {http://arxiv.org/abs/1409.5779} {arXiv:1409.5779 [astro-ph.SR]} \BibitemShut
  {NoStop}%
\bibitem [{\citenamefont {O'Connor}\ and\ \citenamefont
  {Couch}(2018{\natexlab{a}})}]{OConnor:2015rwy}%
  \BibitemOpen
  \bibfield  {author} {\bibinfo {author} {\bibfnamefont {Evan~P.}\ \bibnamefont
  {O'Connor}}\ and\ \bibinfo {author} {\bibfnamefont {Sean~M.}\ \bibnamefont
  {Couch}},\ }\bibfield  {title} {\enquote {\bibinfo {title} {{Two Dimensional
  Core-Collapse Supernova Explosions Aided by General Relativity with
  Multidimensional Neutrino Transport}},}\ }\href {\doibase
  10.3847/1538-4357/aaa893} {\bibfield  {journal} {\bibinfo  {journal}
  {Astrophys. J.}\ }\textbf {\bibinfo {volume} {854}},\ \bibinfo {pages} {63}
  (\bibinfo {year} {2018}{\natexlab{a}})},\ \Eprint
  {http://arxiv.org/abs/1511.07443} {arXiv:1511.07443 [astro-ph.HE]}
  \BibitemShut {NoStop}%
\bibitem [{\citenamefont {Summa}\ \emph {et~al.}(2016)\citenamefont {Summa},
  \citenamefont {Hanke}, \citenamefont {Janka}, \citenamefont {Melson},
  \citenamefont {Marek},\ and\ \citenamefont {Müller}}]{Summa:2015nyk}%
  \BibitemOpen
  \bibfield  {author} {\bibinfo {author} {\bibfnamefont {Alexander}\
  \bibnamefont {Summa}}, \bibinfo {author} {\bibfnamefont {Florian}\
  \bibnamefont {Hanke}}, \bibinfo {author} {\bibfnamefont {Hans-Thomas}\
  \bibnamefont {Janka}}, \bibinfo {author} {\bibfnamefont {Tobias}\
  \bibnamefont {Melson}}, \bibinfo {author} {\bibfnamefont {Andreas}\
  \bibnamefont {Marek}}, \ and\ \bibinfo {author} {\bibfnamefont {Bernhard}\
  \bibnamefont {Müller}},\ }\bibfield  {title} {\enquote {\bibinfo {title}
  {{Progenitor-dependent Explosion Dynamics in Self-consistent, Axisymmetric
  Simulations of Neutrino-driven Core-collapse Supernovae}},}\ }\href {\doibase
  10.3847/0004-637X/825/1/6} {\bibfield  {journal} {\bibinfo  {journal}
  {Astrophys. J.}\ }\textbf {\bibinfo {volume} {825}},\ \bibinfo {pages} {6}
  (\bibinfo {year} {2016})},\ \Eprint {http://arxiv.org/abs/1511.07871}
  {arXiv:1511.07871 [astro-ph.SR]} \BibitemShut {NoStop}%
\bibitem [{\citenamefont {Kotake}\ \emph {et~al.}(2018)\citenamefont {Kotake},
  \citenamefont {Takiwaki}, \citenamefont {Fischer}, \citenamefont {Nakamura},\
  and\ \citenamefont {Martínez-Pinedo}}]{Kotake:2018ypf}%
  \BibitemOpen
  \bibfield  {author} {\bibinfo {author} {\bibfnamefont {Kei}\ \bibnamefont
  {Kotake}}, \bibinfo {author} {\bibfnamefont {Tomoya}\ \bibnamefont
  {Takiwaki}}, \bibinfo {author} {\bibfnamefont {Tobias}\ \bibnamefont
  {Fischer}}, \bibinfo {author} {\bibfnamefont {Ko}~\bibnamefont {Nakamura}}, \
  and\ \bibinfo {author} {\bibfnamefont {Gabriel}\ \bibnamefont
  {Martínez-Pinedo}},\ }\bibfield  {title} {\enquote {\bibinfo {title}
  {{Impact of Neutrino Opacities on Core-Collapse Supernova Simulations}},}\
  }\href {\doibase 10.3847/1538-4357/aaa716} {\bibfield  {journal} {\bibinfo
  {journal} {Astrophys. J.}\ }\textbf {\bibinfo {volume} {853}},\ \bibinfo
  {pages} {170} (\bibinfo {year} {2018})},\ \Eprint
  {http://arxiv.org/abs/1801.02703} {arXiv:1801.02703 [astro-ph.HE]}
  \BibitemShut {NoStop}%
\bibitem [{\citenamefont {Vartanyan}\ \emph {et~al.}(2018)\citenamefont
  {Vartanyan}, \citenamefont {Burrows}, \citenamefont {Radice}, \citenamefont
  {Skinner},\ and\ \citenamefont {Dolence}}]{Vartanyan:2018xcd}%
  \BibitemOpen
  \bibfield  {author} {\bibinfo {author} {\bibfnamefont {David}\ \bibnamefont
  {Vartanyan}}, \bibinfo {author} {\bibfnamefont {Adam}\ \bibnamefont
  {Burrows}}, \bibinfo {author} {\bibfnamefont {David}\ \bibnamefont {Radice}},
  \bibinfo {author} {\bibfnamefont {M.~Aaron}\ \bibnamefont {Skinner}}, \ and\
  \bibinfo {author} {\bibfnamefont {Joshua}\ \bibnamefont {Dolence}},\
  }\bibfield  {title} {\enquote {\bibinfo {title} {{Revival of the Fittest:
  Exploding Core-Collapse Supernovae from 12 to 25 M$_{\odot}$}},}\ }\href
  {\doibase 10.1093/mnras/sty809} {\bibfield  {journal} {\bibinfo  {journal}
  {Mon. Not. Roy. Astron. Soc.}\ }\textbf {\bibinfo {volume} {477}},\ \bibinfo
  {pages} {3091--3108} (\bibinfo {year} {2018})},\ \Eprint
  {http://arxiv.org/abs/1801.08148} {arXiv:1801.08148 [astro-ph.HE]}
  \BibitemShut {NoStop}%
\bibitem [{\citenamefont {Ott}\ \emph {et~al.}(2018)\citenamefont {Ott},
  \citenamefont {Roberts}, \citenamefont {da~Silva~Schneider}, \citenamefont
  {Fedrow}, \citenamefont {Haas},\ and\ \citenamefont
  {Schnetter}}]{Ott:2017kxl}%
  \BibitemOpen
  \bibfield  {author} {\bibinfo {author} {\bibfnamefont {C.D.}\ \bibnamefont
  {Ott}}, \bibinfo {author} {\bibfnamefont {L.F.}\ \bibnamefont {Roberts}},
  \bibinfo {author} {\bibfnamefont {A.}~\bibnamefont {da~Silva~Schneider}},
  \bibinfo {author} {\bibfnamefont {J.M.}\ \bibnamefont {Fedrow}}, \bibinfo
  {author} {\bibfnamefont {R.}~\bibnamefont {Haas}}, \ and\ \bibinfo {author}
  {\bibfnamefont {E.}~\bibnamefont {Schnetter}},\ }\bibfield  {title} {\enquote
  {\bibinfo {title} {{The Progenitor Dependence of Core-collapse Supernovae
  from Three-dimensional Simulations with Progenitor Models of 12--40
  M$_\odot$}},}\ }\href {\doibase 10.3847/2041-8213/aaa967} {\bibfield
  {journal} {\bibinfo  {journal} {Astrophys. J. Lett.}\ }\textbf {\bibinfo
  {volume} {855}},\ \bibinfo {pages} {L3} (\bibinfo {year} {2018})},\ \Eprint
  {http://arxiv.org/abs/1712.01304} {arXiv:1712.01304 [astro-ph.HE]}
  \BibitemShut {NoStop}%
\bibitem [{\citenamefont {O'Connor}\ and\ \citenamefont
  {Couch}(2018{\natexlab{b}})}]{OConnor:2018tuw}%
  \BibitemOpen
  \bibfield  {author} {\bibinfo {author} {\bibfnamefont {Evan~P.}\ \bibnamefont
  {O'Connor}}\ and\ \bibinfo {author} {\bibfnamefont {Sean~M.}\ \bibnamefont
  {Couch}},\ }\bibfield  {title} {\enquote {\bibinfo {title} {{Exploring
  Fundamentally Three-dimensional Phenomena in High-fidelity Simulations of
  Core-collapse Supernovae}},}\ }\href {\doibase 10.3847/1538-4357/aadcf7}
  {\bibfield  {journal} {\bibinfo  {journal} {Astrophys. J.}\ }\textbf
  {\bibinfo {volume} {865}},\ \bibinfo {pages} {81} (\bibinfo {year}
  {2018}{\natexlab{b}})},\ \Eprint {http://arxiv.org/abs/1807.07579}
  {arXiv:1807.07579 [astro-ph.HE]} \BibitemShut {NoStop}%
\bibitem [{\citenamefont {Glas}\ \emph {et~al.}(2019)\citenamefont {Glas},
  \citenamefont {Just}, \citenamefont {Janka},\ and\ \citenamefont
  {Obergaulinger}}]{Glas:2018oyz}%
  \BibitemOpen
  \bibfield  {author} {\bibinfo {author} {\bibfnamefont {Robert}\ \bibnamefont
  {Glas}}, \bibinfo {author} {\bibfnamefont {Oliver}\ \bibnamefont {Just}},
  \bibinfo {author} {\bibfnamefont {H.~Thomas}\ \bibnamefont {Janka}}, \ and\
  \bibinfo {author} {\bibfnamefont {Martin}\ \bibnamefont {Obergaulinger}},\
  }\bibfield  {title} {\enquote {\bibinfo {title} {{Three-dimensional
  Core-collapse Supernova Simulations with Multidimensional Neutrino Transport
  Compared to the Ray-by-ray-plus Approximation}},}\ }\href {\doibase
  10.3847/1538-4357/ab0423} {\bibfield  {journal} {\bibinfo  {journal}
  {Astrophys. J.}\ }\textbf {\bibinfo {volume} {873}},\ \bibinfo {pages} {45}
  (\bibinfo {year} {2019})},\ \Eprint {http://arxiv.org/abs/1809.10146}
  {arXiv:1809.10146 [astro-ph.HE]} \BibitemShut {NoStop}%
\bibitem [{\citenamefont {Burrows}\ \emph {et~al.}(2020)\citenamefont
  {Burrows}, \citenamefont {Radice}, \citenamefont {Vartanyan}, \citenamefont
  {Nagakura}, \citenamefont {Skinner},\ and\ \citenamefont
  {Dolence}}]{Burrows:2019zce}%
  \BibitemOpen
  \bibfield  {author} {\bibinfo {author} {\bibfnamefont {Adam}\ \bibnamefont
  {Burrows}}, \bibinfo {author} {\bibfnamefont {David}\ \bibnamefont {Radice}},
  \bibinfo {author} {\bibfnamefont {David}\ \bibnamefont {Vartanyan}}, \bibinfo
  {author} {\bibfnamefont {Hiroki}\ \bibnamefont {Nagakura}}, \bibinfo {author}
  {\bibfnamefont {M.~Aaron}\ \bibnamefont {Skinner}}, \ and\ \bibinfo {author}
  {\bibfnamefont {Joshua}\ \bibnamefont {Dolence}},\ }\bibfield  {title}
  {\enquote {\bibinfo {title} {{The Overarching Framework of Core-Collapse
  Supernova Explosions as Revealed by 3D Fornax Simulations}},}\ }\href
  {\doibase 10.1093/mnras/stz3223} {\bibfield  {journal} {\bibinfo  {journal}
  {Mon. Not. Roy. Astron. Soc.}\ }\textbf {\bibinfo {volume} {491}},\ \bibinfo
  {pages} {2715--2735} (\bibinfo {year} {2020})},\ \Eprint
  {http://arxiv.org/abs/1909.04152} {arXiv:1909.04152 [astro-ph.HE]}
  \BibitemShut {NoStop}%
\bibitem [{\citenamefont {{O'Connor}}\ and\ \citenamefont
  {{Ott}}(2013)}]{OConnor:2013}%
  \BibitemOpen
  \bibfield  {author} {\bibinfo {author} {\bibfnamefont {Evan}\ \bibnamefont
  {{O'Connor}}}\ and\ \bibinfo {author} {\bibfnamefont {Christian~D.}\
  \bibnamefont {{Ott}}},\ }\bibfield  {title} {\enquote {\bibinfo {title} {{The
  Progenitor Dependence of the Pre-explosion Neutrino Emission in Core-collapse
  Supernovae}},}\ }\href {\doibase 10.1088/0004-637X/762/2/126} {\bibfield
  {journal} {\bibinfo  {journal} {Astrophys. J.}\ }\textbf {\bibinfo {volume}
  {762}},\ \bibinfo {eid} {126} (\bibinfo {year} {2013})},\ \Eprint
  {http://arxiv.org/abs/1207.1100} {arXiv:1207.1100 [astro-ph.HE]} \BibitemShut
  {NoStop}%
\bibitem [{\citenamefont {Olsen}\ and\ \citenamefont
  {Qian}(2021)}]{Olsen:2021uvt}%
  \BibitemOpen
  \bibfield  {author} {\bibinfo {author} {\bibfnamefont {Jackson}\ \bibnamefont
  {Olsen}}\ and\ \bibinfo {author} {\bibfnamefont {Yong-Zhong}\ \bibnamefont
  {Qian}},\ }\bibfield  {title} {\enquote {\bibinfo {title} {{Comparison of
  simulated neutrino emission models with data on Supernova 1987A}},}\ }\href
  {\doibase 10.1103/PhysRevD.104.123020} {\bibfield  {journal} {\bibinfo
  {journal} {Phys. Rev. D}\ }\textbf {\bibinfo {volume} {104}},\ \bibinfo
  {pages} {123020} (\bibinfo {year} {2021})},\ \Eprint
  {http://arxiv.org/abs/2108.08463} {arXiv:2108.08463 [astro-ph.HE]}
  \BibitemShut {NoStop}%
\bibitem [{\citenamefont {{Woosley}}\ \emph {et~al.}(1988)\citenamefont
  {{Woosley}}, \citenamefont {{Pinto}},\ and\ \citenamefont
  {{Ensman}}}]{Woosley:1988}%
  \BibitemOpen
  \bibfield  {author} {\bibinfo {author} {\bibfnamefont {S.~E.}\ \bibnamefont
  {{Woosley}}}, \bibinfo {author} {\bibfnamefont {Philip~A.}\ \bibnamefont
  {{Pinto}}}, \ and\ \bibinfo {author} {\bibfnamefont {L.}~\bibnamefont
  {{Ensman}}},\ }\bibfield  {title} {\enquote {\bibinfo {title} {{Supernova
  1987A: Six Weeks Later}},}\ }\href {\doibase 10.1086/165908} {\bibfield
  {journal} {\bibinfo  {journal} {Astrophys. J.}\ }\textbf {\bibinfo {volume}
  {324}},\ \bibinfo {pages} {466} (\bibinfo {year} {1988})}\BibitemShut
  {NoStop}%
\bibitem [{\citenamefont {Mezzacappa}(2005)}]{Mezzacappa:2005ju}%
  \BibitemOpen
  \bibfield  {author} {\bibinfo {author} {\bibfnamefont {Anthony}\ \bibnamefont
  {Mezzacappa}},\ }\bibfield  {title} {\enquote {\bibinfo {title}
  {{Ascertaining the Core Collapse Supernova Mechanism: the State of the Art
  and the Road Ahead}},}\ }\href {\doibase
  10.1146/annurev.nucl.55.090704.151608} {\bibfield  {journal} {\bibinfo
  {journal} {Ann. Rev. Nucl. Part. Sci.}\ }\textbf {\bibinfo {volume} {55}},\
  \bibinfo {pages} {467--515} (\bibinfo {year} {2005})}\BibitemShut {NoStop}%
\bibitem [{\citenamefont {{Smartt}}(2009)}]{Smartt:2009}%
  \BibitemOpen
  \bibfield  {author} {\bibinfo {author} {\bibfnamefont {Stephen~J.}\
  \bibnamefont {{Smartt}}},\ }\bibfield  {title} {\enquote {\bibinfo {title}
  {{Progenitors of Core-Collapse Supernovae}},}\ }\href {\doibase
  10.1146/annurev-astro-082708-101737} {\bibfield  {journal} {\bibinfo
  {journal} {Ann. Rev. Astron. Astrophys.}\ }\textbf {\bibinfo {volume} {47}},\
  \bibinfo {pages} {63--106} (\bibinfo {year} {2009})},\ \Eprint
  {http://arxiv.org/abs/0908.0700} {arXiv:0908.0700 [astro-ph.SR]} \BibitemShut
  {NoStop}%
\bibitem [{\citenamefont {Janka}(2012)}]{Janka:2012wk}%
  \BibitemOpen
  \bibfield  {author} {\bibinfo {author} {\bibfnamefont {Hans-Thomas}\
  \bibnamefont {Janka}},\ }\bibfield  {title} {\enquote {\bibinfo {title}
  {{Explosion Mechanisms of Core-Collapse Supernovae}},}\ }\href {\doibase
  10.1146/annurev-nucl-102711-094901} {\bibfield  {journal} {\bibinfo
  {journal} {Ann. Rev. Nucl. Part. Sci.}\ }\textbf {\bibinfo {volume} {62}},\
  \bibinfo {pages} {407--451} (\bibinfo {year} {2012})},\ \Eprint
  {http://arxiv.org/abs/1206.2503} {arXiv:1206.2503 [astro-ph.SR]} \BibitemShut
  {NoStop}%
\bibitem [{\citenamefont {Burrows}\ and\ \citenamefont
  {Vartanyan}(2021)}]{Burrows:2020qrp}%
  \BibitemOpen
  \bibfield  {author} {\bibinfo {author} {\bibfnamefont {Adam}\ \bibnamefont
  {Burrows}}\ and\ \bibinfo {author} {\bibfnamefont {David}\ \bibnamefont
  {Vartanyan}},\ }\bibfield  {title} {\enquote {\bibinfo {title}
  {{Core-Collapse Supernova Explosion Theory}},}\ }\href {\doibase
  10.1038/s41586-020-03059-w} {\bibfield  {journal} {\bibinfo  {journal}
  {Nature}\ }\textbf {\bibinfo {volume} {589}},\ \bibinfo {pages} {29--39}
  (\bibinfo {year} {2021})},\ \Eprint {http://arxiv.org/abs/2009.14157}
  {arXiv:2009.14157 [astro-ph.SR]} \BibitemShut {NoStop}%
\bibitem [{\citenamefont {{Colgate}}\ and\ \citenamefont
  {{White}}(1966)}]{Colgate:1966}%
  \BibitemOpen
  \bibfield  {author} {\bibinfo {author} {\bibfnamefont {Stirling~A.}\
  \bibnamefont {{Colgate}}}\ and\ \bibinfo {author} {\bibfnamefont
  {Richard~H.}\ \bibnamefont {{White}}},\ }\bibfield  {title} {\enquote
  {\bibinfo {title} {{The Hydrodynamic Behavior of Supernovae Explosions}},}\
  }\href {\doibase 10.1086/148549} {\bibfield  {journal} {\bibinfo  {journal}
  {\apj}\ }\textbf {\bibinfo {volume} {143}},\ \bibinfo {pages} {626} (\bibinfo
  {year} {1966})}\BibitemShut {NoStop}%
\bibitem [{\citenamefont {{Bethe}}\ and\ \citenamefont
  {{Wilson}}(1985)}]{Bethe:1985}%
  \BibitemOpen
  \bibfield  {author} {\bibinfo {author} {\bibfnamefont {H.~A.}\ \bibnamefont
  {{Bethe}}}\ and\ \bibinfo {author} {\bibfnamefont {J.~R.}\ \bibnamefont
  {{Wilson}}},\ }\bibfield  {title} {\enquote {\bibinfo {title} {{Revival of a
  stalled supernova shock by neutrino heating}},}\ }\href {\doibase
  10.1086/163343} {\bibfield  {journal} {\bibinfo  {journal} {\apj}\ }\textbf
  {\bibinfo {volume} {295}},\ \bibinfo {pages} {14--23} (\bibinfo {year}
  {1985})}\BibitemShut {NoStop}%
\bibitem [{\citenamefont {Pons}\ \emph {et~al.}(1999)\citenamefont {Pons},
  \citenamefont {Reddy}, \citenamefont {Prakash}, \citenamefont {Lattimer},\
  and\ \citenamefont {Miralles}}]{Pons:1998mm}%
  \BibitemOpen
  \bibfield  {author} {\bibinfo {author} {\bibfnamefont {J.~A.}\ \bibnamefont
  {Pons}}, \bibinfo {author} {\bibfnamefont {S.}~\bibnamefont {Reddy}},
  \bibinfo {author} {\bibfnamefont {M.}~\bibnamefont {Prakash}}, \bibinfo
  {author} {\bibfnamefont {J.~M.}\ \bibnamefont {Lattimer}}, \ and\ \bibinfo
  {author} {\bibfnamefont {J.~A.}\ \bibnamefont {Miralles}},\ }\bibfield
  {title} {\enquote {\bibinfo {title} {{Evolution of protoneutron stars}},}\
  }\href {\doibase 10.1086/306889} {\bibfield  {journal} {\bibinfo  {journal}
  {Astrophys. J.}\ }\textbf {\bibinfo {volume} {513}},\ \bibinfo {pages} {780}
  (\bibinfo {year} {1999})},\ \Eprint {http://arxiv.org/abs/astro-ph/9807040}
  {arXiv:astro-ph/9807040} \BibitemShut {NoStop}%
\bibitem [{\citenamefont {Nakazato}\ \emph {et~al.}(2013)\citenamefont
  {Nakazato}, \citenamefont {Sumiyoshi}, \citenamefont {Suzuki}, \citenamefont
  {Totani}, \citenamefont {Umeda},\ and\ \citenamefont
  {Yamada}}]{Nakazato:2012qf}%
  \BibitemOpen
  \bibfield  {author} {\bibinfo {author} {\bibfnamefont {Ken'ichiro}\
  \bibnamefont {Nakazato}}, \bibinfo {author} {\bibfnamefont {Kohsuke}\
  \bibnamefont {Sumiyoshi}}, \bibinfo {author} {\bibfnamefont {Hideyuki}\
  \bibnamefont {Suzuki}}, \bibinfo {author} {\bibfnamefont {Tomonori}\
  \bibnamefont {Totani}}, \bibinfo {author} {\bibfnamefont {Hideyuki}\
  \bibnamefont {Umeda}}, \ and\ \bibinfo {author} {\bibfnamefont {Shoichi}\
  \bibnamefont {Yamada}},\ }\bibfield  {title} {\enquote {\bibinfo {title}
  {{Supernova Neutrino Light Curves and Spectra for Various Progenitor Stars:
  From Core Collapse to Proto-neutron Star Cooling}},}\ }\href {\doibase
  10.1088/0067-0049/205/1/2} {\bibfield  {journal} {\bibinfo  {journal}
  {Astrophys. J. Suppl.}\ }\textbf {\bibinfo {volume} {205}},\ \bibinfo {pages}
  {2} (\bibinfo {year} {2013})},\ \Eprint {http://arxiv.org/abs/1210.6841}
  {arXiv:1210.6841 [astro-ph.HE]} \BibitemShut {NoStop}%
\bibitem [{\citenamefont {Nakazato}\ and\ \citenamefont
  {Suzuki}(2019)}]{Nakazato:2019ojk}%
  \BibitemOpen
  \bibfield  {author} {\bibinfo {author} {\bibfnamefont {Ken'ichiro}\
  \bibnamefont {Nakazato}}\ and\ \bibinfo {author} {\bibfnamefont {Hideyuki}\
  \bibnamefont {Suzuki}},\ }\bibfield  {title} {\enquote {\bibinfo {title}
  {{Cooling timescale for protoneutron stars and properties of nuclear matter:
  Effective mass and symmetry energy at high densities}},}\ }\href {\doibase
  10.3847/1538-4357/ab1d4b} {\bibfield  {journal} {\bibinfo  {journal}
  {Astrophys. J.}\ }\textbf {\bibinfo {volume} {878}},\ \bibinfo {pages} {25}
  (\bibinfo {year} {2019})},\ \Eprint {http://arxiv.org/abs/1905.00014}
  {arXiv:1905.00014 [astro-ph.HE]} \BibitemShut {NoStop}%
\bibitem [{\citenamefont {Li}\ \emph {et~al.}(2021)\citenamefont {Li},
  \citenamefont {Roberts},\ and\ \citenamefont {Beacom}}]{Li:2020ujl}%
  \BibitemOpen
  \bibfield  {author} {\bibinfo {author} {\bibfnamefont {Shirley~Weishi}\
  \bibnamefont {Li}}, \bibinfo {author} {\bibfnamefont {Luke~F.}\ \bibnamefont
  {Roberts}}, \ and\ \bibinfo {author} {\bibfnamefont {John~F.}\ \bibnamefont
  {Beacom}},\ }\bibfield  {title} {\enquote {\bibinfo {title} {{Exciting
  Prospects for Detecting Late-Time Neutrinos from Core-Collapse
  Supernovae}},}\ }\href {\doibase 10.1103/PhysRevD.103.023016} {\bibfield
  {journal} {\bibinfo  {journal} {Phys. Rev. D}\ }\textbf {\bibinfo {volume}
  {103}},\ \bibinfo {pages} {023016} (\bibinfo {year} {2021})},\ \Eprint
  {http://arxiv.org/abs/2008.04340} {arXiv:2008.04340 [astro-ph.HE]}
  \BibitemShut {NoStop}%
\bibitem [{\citenamefont {Alekseev}\ \emph {et~al.}(1988)\citenamefont
  {Alekseev}, \citenamefont {Alekseeva}, \citenamefont {Krivosheina},\ and\
  \citenamefont {Volchenko}}]{Alekseev:1988gp}%
  \BibitemOpen
  \bibfield  {author} {\bibinfo {author} {\bibfnamefont {E.~N.}\ \bibnamefont
  {Alekseev}}, \bibinfo {author} {\bibfnamefont {L.~N.}\ \bibnamefont
  {Alekseeva}}, \bibinfo {author} {\bibfnamefont {I.~V.}\ \bibnamefont
  {Krivosheina}}, \ and\ \bibinfo {author} {\bibfnamefont {V.~I.}\ \bibnamefont
  {Volchenko}},\ }\bibfield  {title} {\enquote {\bibinfo {title} {{Detection of
  the Neutrino Signal From {SN1987A} in the {LMC} Using the Inr Baksan
  Underground Scintillation Telescope}},}\ }\href {\doibase
  10.1016/0370-2693(88)91651-6} {\bibfield  {journal} {\bibinfo  {journal}
  {Phys. Lett. B}\ }\textbf {\bibinfo {volume} {205}},\ \bibinfo {pages}
  {209--214} (\bibinfo {year} {1988})}\BibitemShut {NoStop}%
\bibitem [{\citenamefont {Loredo}\ and\ \citenamefont
  {Lamb}(2002)}]{Loredo:2001rx}%
  \BibitemOpen
  \bibfield  {author} {\bibinfo {author} {\bibfnamefont {Thomas~J.}\
  \bibnamefont {Loredo}}\ and\ \bibinfo {author} {\bibfnamefont {Don~Q.}\
  \bibnamefont {Lamb}},\ }\bibfield  {title} {\enquote {\bibinfo {title}
  {{Bayesian analysis of neutrinos observed from supernova SN-1987A}},}\ }\href
  {\doibase 10.1103/PhysRevD.65.063002} {\bibfield  {journal} {\bibinfo
  {journal} {Phys. Rev. D}\ }\textbf {\bibinfo {volume} {65}},\ \bibinfo
  {pages} {063002} (\bibinfo {year} {2002})},\ \Eprint
  {http://arxiv.org/abs/astro-ph/0107260} {arXiv:astro-ph/0107260} \BibitemShut
  {NoStop}%
\bibitem [{\citenamefont {Burrows}\ and\ \citenamefont
  {Lattimer}(1987)}]{Burrows:1987zz}%
  \BibitemOpen
  \bibfield  {author} {\bibinfo {author} {\bibfnamefont {Adam}\ \bibnamefont
  {Burrows}}\ and\ \bibinfo {author} {\bibfnamefont {James~M.}\ \bibnamefont
  {Lattimer}},\ }\bibfield  {title} {\enquote {\bibinfo {title} {{Neutrinos
  from SN 1987A}},}\ }\href {\doibase 10.1086/184938} {\bibfield  {journal}
  {\bibinfo  {journal} {Astrophys. J. Lett.}\ }\textbf {\bibinfo {volume}
  {318}},\ \bibinfo {pages} {L63--L68} (\bibinfo {year} {1987})}\BibitemShut
  {NoStop}%
\bibitem [{\citenamefont {{Bruenn}}(1987)}]{Bruenn:1987}%
  \BibitemOpen
  \bibfield  {author} {\bibinfo {author} {\bibfnamefont {Stephen~W.}\
  \bibnamefont {{Bruenn}}},\ }\bibfield  {title} {\enquote {\bibinfo {title}
  {{Neutrinos from SN1987A and current models of stellar-core collapse}},}\
  }\href {\doibase 10.1103/PhysRevLett.59.938} {\bibfield  {journal} {\bibinfo
  {journal} {\prl}\ }\textbf {\bibinfo {volume} {59}},\ \bibinfo {pages}
  {938--941} (\bibinfo {year} {1987})}\BibitemShut {NoStop}%
\bibitem [{\citenamefont {{Sato}}\ and\ \citenamefont
  {{Suzuki}}(1987)}]{Suzuki:1987}%
  \BibitemOpen
  \bibfield  {author} {\bibinfo {author} {\bibfnamefont {Katsuhiko}\
  \bibnamefont {{Sato}}}\ and\ \bibinfo {author} {\bibfnamefont {Hideyuki}\
  \bibnamefont {{Suzuki}}},\ }\bibfield  {title} {\enquote {\bibinfo {title}
  {{Analysis of neutrino burst from the supernova 1987A in the Large Magellanic
  Cloud}},}\ }\href {\doibase 10.1103/PhysRevLett.58.2722} {\bibfield
  {journal} {\bibinfo  {journal} {\prl}\ }\textbf {\bibinfo {volume} {58}},\
  \bibinfo {pages} {2722--2725} (\bibinfo {year} {1987})}\BibitemShut {NoStop}%
\bibitem [{\citenamefont {Pumo}\ \emph {et~al.}(2023)\citenamefont {Pumo},
  \citenamefont {Cosentino}, \citenamefont {Pastorello}, \citenamefont
  {Benetti}, \citenamefont {Cherubini}, \citenamefont {Manic\`o},\ and\
  \citenamefont {Zampieri}}]{Pumo:2023qoy}%
  \BibitemOpen
  \bibfield  {author} {\bibinfo {author} {\bibfnamefont {M.~L.}\ \bibnamefont
  {Pumo}}, \bibinfo {author} {\bibfnamefont {S.~P.}\ \bibnamefont {Cosentino}},
  \bibinfo {author} {\bibfnamefont {A.}~\bibnamefont {Pastorello}}, \bibinfo
  {author} {\bibfnamefont {S.}~\bibnamefont {Benetti}}, \bibinfo {author}
  {\bibfnamefont {S.}~\bibnamefont {Cherubini}}, \bibinfo {author}
  {\bibfnamefont {G.}~\bibnamefont {Manic\`o}}, \ and\ \bibinfo {author}
  {\bibfnamefont {L.}~\bibnamefont {Zampieri}},\ }\bibfield  {title} {\enquote
  {\bibinfo {title} {{Long-rising Type II supernovae resembling supernova 1987A
  \textendash{} I. A comparative study through scaling relations}},}\ }\href
  {\doibase 10.1093/mnras/stad861} {\bibfield  {journal} {\bibinfo  {journal}
  {Mon. Not. Roy. Astron. Soc.}\ }\textbf {\bibinfo {volume} {521}},\ \bibinfo
  {pages} {4801--4818} (\bibinfo {year} {2023})},\ \Eprint
  {http://arxiv.org/abs/2303.10478} {arXiv:2303.10478 [astro-ph.HE]}
  \BibitemShut {NoStop}%
\bibitem [{\citenamefont {{Hillebrandt}}\ \emph {et~al.}(1987)\citenamefont
  {{Hillebrandt}}, \citenamefont {{Hoeflich}}, \citenamefont {{Weiss}},\ and\
  \citenamefont {{Truran}}}]{Hillenbrandt:1987}%
  \BibitemOpen
  \bibfield  {author} {\bibinfo {author} {\bibfnamefont {W.}~\bibnamefont
  {{Hillebrandt}}}, \bibinfo {author} {\bibfnamefont {P.}~\bibnamefont
  {{Hoeflich}}}, \bibinfo {author} {\bibfnamefont {A.}~\bibnamefont {{Weiss}}},
  \ and\ \bibinfo {author} {\bibfnamefont {J.~W.}\ \bibnamefont {{Truran}}},\
  }\bibfield  {title} {\enquote {\bibinfo {title} {{Explosion of a blue
  supergiant: a model for supernova SN1987A}},}\ }\href {\doibase
  10.1038/327597a0} {\bibfield  {journal} {\bibinfo  {journal} {\nat}\ }\textbf
  {\bibinfo {volume} {327}},\ \bibinfo {pages} {597--600} (\bibinfo {year}
  {1987})}\BibitemShut {NoStop}%
\bibitem [{\citenamefont {{Woosley}}\ \emph {et~al.}(1987)\citenamefont
  {{Woosley}}, \citenamefont {{Pinto}}, \citenamefont {{Martin}},\ and\
  \citenamefont {{Weaver}}}]{Woosley:1987}%
  \BibitemOpen
  \bibfield  {author} {\bibinfo {author} {\bibfnamefont {S.~E.}\ \bibnamefont
  {{Woosley}}}, \bibinfo {author} {\bibfnamefont {P.~A.}\ \bibnamefont
  {{Pinto}}}, \bibinfo {author} {\bibfnamefont {P.~G.}\ \bibnamefont
  {{Martin}}}, \ and\ \bibinfo {author} {\bibfnamefont {Thomas~A.}\
  \bibnamefont {{Weaver}}},\ }\bibfield  {title} {\enquote {\bibinfo {title}
  {{Supernova 1987A in the Large Magellanic Cloud: The Explosion of a
  approximately 20 M$_{sun}$ Star Which Has Experienced Mass Loss?}}}\ }\href
  {\doibase 10.1086/165402} {\bibfield  {journal} {\bibinfo  {journal} {\apj}\
  }\textbf {\bibinfo {volume} {318}},\ \bibinfo {pages} {664} (\bibinfo {year}
  {1987})}\BibitemShut {NoStop}%
\bibitem [{\citenamefont {{Saio}}\ \emph {et~al.}(1988)\citenamefont {{Saio}},
  \citenamefont {{Kato}},\ and\ \citenamefont {{Nomoto}}}]{Saio:1988}%
  \BibitemOpen
  \bibfield  {author} {\bibinfo {author} {\bibfnamefont {Hideyuki}\
  \bibnamefont {{Saio}}}, \bibinfo {author} {\bibfnamefont {Mariko}\
  \bibnamefont {{Kato}}}, \ and\ \bibinfo {author} {\bibfnamefont {Ken'ichi}\
  \bibnamefont {{Nomoto}}},\ }\bibfield  {title} {\enquote {\bibinfo {title}
  {{Why Did the Progenitor of SN 1987A Undergo the Blue-Red-Blue Evolution?}}}\
  }\href {\doibase 10.1086/166565} {\bibfield  {journal} {\bibinfo  {journal}
  {\apj}\ }\textbf {\bibinfo {volume} {331}},\ \bibinfo {pages} {388} (\bibinfo
  {year} {1988})}\BibitemShut {NoStop}%
\bibitem [{\citenamefont {{Podsiadlowski}}(1992)}]{Posiadlowski:1992}%
  \BibitemOpen
  \bibfield  {author} {\bibinfo {author} {\bibfnamefont {Philipp}\ \bibnamefont
  {{Podsiadlowski}}},\ }\bibfield  {title} {\enquote {\bibinfo {title} {{The
  Progenitor of SN 1987A}},}\ }\href {\doibase 10.1086/133043} {\bibfield
  {journal} {\bibinfo  {journal} {Publications of the Astronomical Society of
  the Pacific}\ }\textbf {\bibinfo {volume} {104}},\ \bibinfo {pages} {717}
  (\bibinfo {year} {1992})}\BibitemShut {NoStop}%
\bibitem [{\citenamefont {Menon}\ and\ \citenamefont
  {Heger}(2017)}]{Menon:2017hva}%
  \BibitemOpen
  \bibfield  {author} {\bibinfo {author} {\bibfnamefont {Athira}\ \bibnamefont
  {Menon}}\ and\ \bibinfo {author} {\bibfnamefont {Alexander}\ \bibnamefont
  {Heger}},\ }\bibfield  {title} {\enquote {\bibinfo {title} {{The quest for
  blue supergiants: binary merger models for the evolution of the progenitor of
  SN 1987A}},}\ }\href {\doibase 10.1093/mnras/stx818} {\bibfield  {journal}
  {\bibinfo  {journal} {Mon. Not. Roy. Astron. Soc.}\ }\textbf {\bibinfo
  {volume} {469}},\ \bibinfo {pages} {4649--4664} (\bibinfo {year} {2017})},\
  \Eprint {http://arxiv.org/abs/1703.04918} {arXiv:1703.04918 [astro-ph.SR]}
  \BibitemShut {NoStop}%
\bibitem [{\citenamefont {Urushibata}\ \emph {et~al.}(2018)\citenamefont
  {Urushibata}, \citenamefont {Takahashi}, \citenamefont {Umeda},\ and\
  \citenamefont {Yoshida}}]{Urushibata:2017hnl}%
  \BibitemOpen
  \bibfield  {author} {\bibinfo {author} {\bibfnamefont {T.}~\bibnamefont
  {Urushibata}}, \bibinfo {author} {\bibfnamefont {K.}~\bibnamefont
  {Takahashi}}, \bibinfo {author} {\bibfnamefont {H.}~\bibnamefont {Umeda}}, \
  and\ \bibinfo {author} {\bibfnamefont {T.}~\bibnamefont {Yoshida}},\
  }\bibfield  {title} {\enquote {\bibinfo {title} {{A progenitor model of SN
  1987A based on the slow-merger scenario}},}\ }\href {\doibase
  10.1093/mnrasl/slx166} {\bibfield  {journal} {\bibinfo  {journal} {Mon. Not.
  Roy. Astron. Soc.}\ }\textbf {\bibinfo {volume} {473}},\ \bibinfo {pages}
  {L101--L105} (\bibinfo {year} {2018})},\ \Eprint
  {http://arxiv.org/abs/1705.04084} {arXiv:1705.04084 [astro-ph.SR]}
  \BibitemShut {NoStop}%
\bibitem [{\citenamefont {Nakamura}\ \emph {et~al.}(2022)\citenamefont
  {Nakamura}, \citenamefont {Takiwaki},\ and\ \citenamefont
  {Kotake}}]{Nakamura:2022zlc}%
  \BibitemOpen
  \bibfield  {author} {\bibinfo {author} {\bibfnamefont {Ko}~\bibnamefont
  {Nakamura}}, \bibinfo {author} {\bibfnamefont {Tomoya}\ \bibnamefont
  {Takiwaki}}, \ and\ \bibinfo {author} {\bibfnamefont {Kei}\ \bibnamefont
  {Kotake}},\ }\bibfield  {title} {\enquote {\bibinfo {title}
  {{Three-dimensional simulation of a core-collapse supernova for a binary star
  progenitor of SN~1987A}},}\ }\href {\doibase 10.1093/mnras/stac1586}
  {\bibfield  {journal} {\bibinfo  {journal} {Mon. Not. Roy. Astron. Soc.}\
  }\textbf {\bibinfo {volume} {514}},\ \bibinfo {pages} {3941--3952} (\bibinfo
  {year} {2022})},\ \Eprint {http://arxiv.org/abs/2202.06295} {arXiv:2202.06295
  [astro-ph.HE]} \BibitemShut {NoStop}%
\bibitem [{\citenamefont {Woosley}\ and\ \citenamefont
  {Heger}(2007)}]{Woosley:2007as}%
  \BibitemOpen
  \bibfield  {author} {\bibinfo {author} {\bibfnamefont {S.~E.}\ \bibnamefont
  {Woosley}}\ and\ \bibinfo {author} {\bibfnamefont {Alexander}\ \bibnamefont
  {Heger}},\ }\bibfield  {title} {\enquote {\bibinfo {title} {{Nucleosynthesis
  and Remnants in Massive Stars of Solar Metallicity}},}\ }\href {\doibase
  10.1016/j.physrep.2007.02.009} {\bibfield  {journal} {\bibinfo  {journal}
  {Phys. Rept.}\ }\textbf {\bibinfo {volume} {442}},\ \bibinfo {pages}
  {269--283} (\bibinfo {year} {2007})},\ \Eprint
  {http://arxiv.org/abs/astro-ph/0702176} {arXiv:astro-ph/0702176} \BibitemShut
  {NoStop}%
\bibitem [{\citenamefont {Keil}\ \emph {et~al.}(2003)\citenamefont {Keil},
  \citenamefont {Raffelt},\ and\ \citenamefont {Janka}}]{Keil:2002in}%
  \BibitemOpen
  \bibfield  {author} {\bibinfo {author} {\bibfnamefont {Mathias~Th.}\
  \bibnamefont {Keil}}, \bibinfo {author} {\bibfnamefont {Georg~G.}\
  \bibnamefont {Raffelt}}, \ and\ \bibinfo {author} {\bibfnamefont
  {Hans-Thomas}\ \bibnamefont {Janka}},\ }\bibfield  {title} {\enquote
  {\bibinfo {title} {{Monte Carlo study of supernova neutrino spectra
  formation}},}\ }\href {\doibase 10.1086/375130} {\bibfield  {journal}
  {\bibinfo  {journal} {Astrophys. J.}\ }\textbf {\bibinfo {volume} {590}},\
  \bibinfo {pages} {971--991} (\bibinfo {year} {2003})},\ \Eprint
  {http://arxiv.org/abs/astro-ph/0208035} {arXiv:astro-ph/0208035} \BibitemShut
  {NoStop}%
\bibitem [{\citenamefont {Jegerlehner}\ \emph {et~al.}(1996)\citenamefont
  {Jegerlehner}, \citenamefont {Neubig},\ and\ \citenamefont
  {Raffelt}}]{Jegerlehner:1996kx}%
  \BibitemOpen
  \bibfield  {author} {\bibinfo {author} {\bibfnamefont {Beat}\ \bibnamefont
  {Jegerlehner}}, \bibinfo {author} {\bibfnamefont {Frank}\ \bibnamefont
  {Neubig}}, \ and\ \bibinfo {author} {\bibfnamefont {Georg}\ \bibnamefont
  {Raffelt}},\ }\bibfield  {title} {\enquote {\bibinfo {title} {{Neutrino
  oscillations and the supernova SN1987A signal}},}\ }\href {\doibase
  10.1103/PhysRevD.54.1194} {\bibfield  {journal} {\bibinfo  {journal} {Phys.
  Rev. D}\ }\textbf {\bibinfo {volume} {54}},\ \bibinfo {pages} {1194--1203}
  (\bibinfo {year} {1996})},\ \Eprint {http://arxiv.org/abs/astro-ph/9601111}
  {arXiv:astro-ph/9601111} \BibitemShut {NoStop}%
\bibitem [{\citenamefont {Lunardini}\ and\ \citenamefont
  {Smirnov}(2004)}]{Lunardini:2004bj}%
  \BibitemOpen
  \bibfield  {author} {\bibinfo {author} {\bibfnamefont {Cecilia}\ \bibnamefont
  {Lunardini}}\ and\ \bibinfo {author} {\bibfnamefont {Alexei~Yu.}\
  \bibnamefont {Smirnov}},\ }\bibfield  {title} {\enquote {\bibinfo {title}
  {{Neutrinos from SN1987A: Flavor conversion and interpretation of
  results}},}\ }\href {\doibase 10.1016/j.astropartphys.2004.05.005} {\bibfield
   {journal} {\bibinfo  {journal} {Astropart. Phys.}\ }\textbf {\bibinfo
  {volume} {21}},\ \bibinfo {pages} {703--720} (\bibinfo {year} {2004})},\
  \Eprint {http://arxiv.org/abs/hep-ph/0402128} {arXiv:hep-ph/0402128}
  \BibitemShut {NoStop}%
\bibitem [{\citenamefont {Costantini}\ \emph {et~al.}(2007)\citenamefont
  {Costantini}, \citenamefont {Ianni}, \citenamefont {Pagliaroli},\ and\
  \citenamefont {Vissani}}]{Costantini:2006xd}%
  \BibitemOpen
  \bibfield  {author} {\bibinfo {author} {\bibfnamefont {Maria~Laura}\
  \bibnamefont {Costantini}}, \bibinfo {author} {\bibfnamefont {Aldo}\
  \bibnamefont {Ianni}}, \bibinfo {author} {\bibfnamefont {Giulia}\
  \bibnamefont {Pagliaroli}}, \ and\ \bibinfo {author} {\bibfnamefont
  {Francesco}\ \bibnamefont {Vissani}},\ }\bibfield  {title} {\enquote
  {\bibinfo {title} {{Is there a problem with low energy SN1987A neutrinos?}}}\
  }\href {\doibase 10.1088/1475-7516/2007/05/014} {\bibfield  {journal}
  {\bibinfo  {journal} {JCAP}\ }\textbf {\bibinfo {volume} {05}},\ \bibinfo
  {pages} {014} (\bibinfo {year} {2007})},\ \Eprint
  {http://arxiv.org/abs/astro-ph/0608399} {arXiv:astro-ph/0608399} \BibitemShut
  {NoStop}%
\bibitem [{\citenamefont {Pagliaroli}\ \emph {et~al.}(2009)\citenamefont
  {Pagliaroli}, \citenamefont {Vissani}, \citenamefont {Costantini},\ and\
  \citenamefont {Ianni}}]{Pagliaroli:2008ur}%
  \BibitemOpen
  \bibfield  {author} {\bibinfo {author} {\bibfnamefont {G.}~\bibnamefont
  {Pagliaroli}}, \bibinfo {author} {\bibfnamefont {F.}~\bibnamefont {Vissani}},
  \bibinfo {author} {\bibfnamefont {M.~L.}\ \bibnamefont {Costantini}}, \ and\
  \bibinfo {author} {\bibfnamefont {A.}~\bibnamefont {Ianni}},\ }\bibfield
  {title} {\enquote {\bibinfo {title} {{Improved analysis of SN1987A
  antineutrino events}},}\ }\href {\doibase
  10.1016/j.astropartphys.2008.12.010} {\bibfield  {journal} {\bibinfo
  {journal} {Astropart. Phys.}\ }\textbf {\bibinfo {volume} {31}},\ \bibinfo
  {pages} {163--176} (\bibinfo {year} {2009})},\ \Eprint
  {http://arxiv.org/abs/0810.0466} {arXiv:0810.0466 [astro-ph]} \BibitemShut
  {NoStop}%
\bibitem [{\citenamefont {Vissani}(2015)}]{Vissani:2014doa}%
  \BibitemOpen
  \bibfield  {author} {\bibinfo {author} {\bibfnamefont {Francesco}\
  \bibnamefont {Vissani}},\ }\bibfield  {title} {\enquote {\bibinfo {title}
  {{Comparative analysis of SN1987A antineutrino fluence}},}\ }\href {\doibase
  10.1088/0954-3899/42/1/013001} {\bibfield  {journal} {\bibinfo  {journal} {J.
  Phys. G}\ }\textbf {\bibinfo {volume} {42}},\ \bibinfo {pages} {013001}
  (\bibinfo {year} {2015})},\ \Eprint {http://arxiv.org/abs/1409.4710}
  {arXiv:1409.4710 [astro-ph.HE]} \BibitemShut {NoStop}%
\bibitem [{\citenamefont {Vogel}\ and\ \citenamefont
  {Beacom}(1999)}]{Vogel:1999zy}%
  \BibitemOpen
  \bibfield  {author} {\bibinfo {author} {\bibfnamefont {P.}~\bibnamefont
  {Vogel}}\ and\ \bibinfo {author} {\bibfnamefont {John~F.}\ \bibnamefont
  {Beacom}},\ }\bibfield  {title} {\enquote {\bibinfo {title} {{Angular
  distribution of neutron inverse beta decay, $\bar\nu_e + p \rightarrow e^+ +
  n$}},}\ }\href {\doibase 10.1103/PhysRevD.60.053003} {\bibfield  {journal}
  {\bibinfo  {journal} {Phys. Rev. D}\ }\textbf {\bibinfo {volume} {60}},\
  \bibinfo {pages} {053003} (\bibinfo {year} {1999})},\ \Eprint
  {http://arxiv.org/abs/hep-ph/9903554} {arXiv:hep-ph/9903554} \BibitemShut
  {NoStop}%
\bibitem [{\citenamefont {Strumia}\ and\ \citenamefont
  {Vissani}(2003)}]{Strumia:2003zx}%
  \BibitemOpen
  \bibfield  {author} {\bibinfo {author} {\bibfnamefont {Alessandro}\
  \bibnamefont {Strumia}}\ and\ \bibinfo {author} {\bibfnamefont {Francesco}\
  \bibnamefont {Vissani}},\ }\bibfield  {title} {\enquote {\bibinfo {title}
  {{Precise quasielastic neutrino/nucleon cross-section}},}\ }\href {\doibase
  10.1016/S0370-2693(03)00616-6} {\bibfield  {journal} {\bibinfo  {journal}
  {Phys. Lett. B}\ }\textbf {\bibinfo {volume} {564}},\ \bibinfo {pages}
  {42--54} (\bibinfo {year} {2003})},\ \Eprint
  {http://arxiv.org/abs/astro-ph/0302055} {arXiv:astro-ph/0302055} \BibitemShut
  {NoStop}%
\bibitem [{\citenamefont {{Panagia}}(1999)}]{1999IAUS..190..549P}%
  \BibitemOpen
  \bibfield  {author} {\bibinfo {author} {\bibfnamefont {N.}~\bibnamefont
  {{Panagia}}},\ }\bibfield  {title} {\enquote {\bibinfo {title} {{Distance to
  SN 1987 A and the LMC}},}\ }in\ \href@noop {} {\emph {\bibinfo {booktitle}
  {New Views of the Magellanic Clouds}}},\ Vol.\ \bibinfo {volume} {190},\
  \bibinfo {editor} {edited by\ \bibinfo {editor} {\bibfnamefont {Y.~H.}\
  \bibnamefont {{Chu}}}, \bibinfo {editor} {\bibfnamefont {N.}~\bibnamefont
  {{Suntzeff}}}, \bibinfo {editor} {\bibfnamefont {J.}~\bibnamefont
  {{Hesser}}}, \ and\ \bibinfo {editor} {\bibfnamefont {D.}~\bibnamefont
  {{Bohlender}}}}\ (\bibinfo {year} {1999})\ p.\ \bibinfo {pages}
  {549}\BibitemShut {NoStop}%
\bibitem [{\citenamefont {Yuksel}\ and\ \citenamefont
  {Beacom}(2007)}]{Yuksel:2007mn}%
  \BibitemOpen
  \bibfield  {author} {\bibinfo {author} {\bibfnamefont {Hasan}\ \bibnamefont
  {Yuksel}}\ and\ \bibinfo {author} {\bibfnamefont {John~F.}\ \bibnamefont
  {Beacom}},\ }\bibfield  {title} {\enquote {\bibinfo {title} {{Neutrino
  Spectrum from SN 1987A and from Cosmic Supernovae}},}\ }\href {\doibase
  10.1103/PhysRevD.76.083007} {\bibfield  {journal} {\bibinfo  {journal} {Phys.
  Rev. D}\ }\textbf {\bibinfo {volume} {76}},\ \bibinfo {pages} {083007}
  (\bibinfo {year} {2007})},\ \Eprint {http://arxiv.org/abs/astro-ph/0702613}
  {arXiv:astro-ph/0702613} \BibitemShut {NoStop}%
\bibitem [{\citenamefont {Nagakura}\ and\ \citenamefont
  {Hotokezaka}(2021)}]{Nagakura:2020gls}%
  \BibitemOpen
  \bibfield  {author} {\bibinfo {author} {\bibfnamefont {Hiroki}\ \bibnamefont
  {Nagakura}}\ and\ \bibinfo {author} {\bibfnamefont {Kenta}\ \bibnamefont
  {Hotokezaka}},\ }\bibfield  {title} {\enquote {\bibinfo {title} {{Non-thermal
  neutrinos created by shock acceleration in successful and failed
  core-collapse supernova}},}\ }\href {\doibase 10.1093/mnras/stab040}
  {\bibfield  {journal} {\bibinfo  {journal} {Mon. Not. Roy. Astron. Soc.}\
  }\textbf {\bibinfo {volume} {502}},\ \bibinfo {pages} {89--107} (\bibinfo
  {year} {2021})},\ \Eprint {http://arxiv.org/abs/2010.15136} {arXiv:2010.15136
  [astro-ph.HE]} \BibitemShut {NoStop}%
\bibitem [{\citenamefont {Duan}\ \emph {et~al.}(2010)\citenamefont {Duan},
  \citenamefont {Fuller},\ and\ \citenamefont {Qian}}]{Duan:2010bg}%
  \BibitemOpen
  \bibfield  {author} {\bibinfo {author} {\bibfnamefont {Huaiyu}\ \bibnamefont
  {Duan}}, \bibinfo {author} {\bibfnamefont {George~M.}\ \bibnamefont
  {Fuller}}, \ and\ \bibinfo {author} {\bibfnamefont {Yong-Zhong}\ \bibnamefont
  {Qian}},\ }\bibfield  {title} {\enquote {\bibinfo {title} {{Collective
  Neutrino Oscillations}},}\ }\href {\doibase
  10.1146/annurev.nucl.012809.104524} {\bibfield  {journal} {\bibinfo
  {journal} {Ann. Rev. Nucl. Part. Sci.}\ }\textbf {\bibinfo {volume} {60}},\
  \bibinfo {pages} {569--594} (\bibinfo {year} {2010})},\ \Eprint
  {http://arxiv.org/abs/1001.2799} {arXiv:1001.2799 [hep-ph]} \BibitemShut
  {NoStop}%
\bibitem [{\citenamefont {Horiuchi}\ and\ \citenamefont
  {Kneller}(2018)}]{Horiuchi:2018ofe}%
  \BibitemOpen
  \bibfield  {author} {\bibinfo {author} {\bibfnamefont {Shunsaku}\
  \bibnamefont {Horiuchi}}\ and\ \bibinfo {author} {\bibfnamefont {James~P}\
  \bibnamefont {Kneller}},\ }\bibfield  {title} {\enquote {\bibinfo {title}
  {{What can be learned from a future supernova neutrino detection?}}}\ }\href
  {\doibase 10.1088/1361-6471/aaa90a} {\bibfield  {journal} {\bibinfo
  {journal} {J. Phys. G}\ }\textbf {\bibinfo {volume} {45}},\ \bibinfo {pages}
  {043002} (\bibinfo {year} {2018})},\ \Eprint
  {http://arxiv.org/abs/1709.01515} {arXiv:1709.01515 [astro-ph.HE]}
  \BibitemShut {NoStop}%
\bibitem [{\citenamefont {Tamborra}\ and\ \citenamefont
  {Shalgar}(2021)}]{Tamborra:2020cul}%
  \BibitemOpen
  \bibfield  {author} {\bibinfo {author} {\bibfnamefont {Irene}\ \bibnamefont
  {Tamborra}}\ and\ \bibinfo {author} {\bibfnamefont {Shashank}\ \bibnamefont
  {Shalgar}},\ }\bibfield  {title} {\enquote {\bibinfo {title} {{New
  Developments in Flavor Evolution of a Dense Neutrino Gas}},}\ }\href
  {\doibase 10.1146/annurev-nucl-102920-050505} {\bibfield  {journal} {\bibinfo
   {journal} {Ann. Rev. Nucl. Part. Sci.}\ }\textbf {\bibinfo {volume} {71}},\
  \bibinfo {pages} {165--188} (\bibinfo {year} {2021})},\ \Eprint
  {http://arxiv.org/abs/2011.01948} {arXiv:2011.01948 [astro-ph.HE]}
  \BibitemShut {NoStop}%
\bibitem [{\citenamefont {Capozzi}\ and\ \citenamefont
  {Saviano}(2022)}]{Capozzi:2022slf}%
  \BibitemOpen
  \bibfield  {author} {\bibinfo {author} {\bibfnamefont {Francesco}\
  \bibnamefont {Capozzi}}\ and\ \bibinfo {author} {\bibfnamefont {Ninetta}\
  \bibnamefont {Saviano}},\ }\bibfield  {title} {\enquote {\bibinfo {title}
  {{Neutrino Flavor Conversions in High-Density Astrophysical and Cosmological
  Environments}},}\ }\href {\doibase 10.3390/universe8020094} {\bibfield
  {journal} {\bibinfo  {journal} {Universe}\ }\textbf {\bibinfo {volume} {8}},\
  \bibinfo {pages} {94} (\bibinfo {year} {2022})},\ \Eprint
  {http://arxiv.org/abs/2202.02494} {arXiv:2202.02494 [hep-ph]} \BibitemShut
  {NoStop}%
\bibitem [{\citenamefont {Richers}\ and\ \citenamefont
  {Sen}(2022)}]{Richers:2022zug}%
  \BibitemOpen
  \bibfield  {author} {\bibinfo {author} {\bibfnamefont {Sherwood}\
  \bibnamefont {Richers}}\ and\ \bibinfo {author} {\bibfnamefont {Manibrata}\
  \bibnamefont {Sen}},\ }\bibfield  {title} {\enquote {\bibinfo {title} {{Fast
  Flavor Transformations}},}\ }\href@noop {} {\  (\bibinfo {year} {2022})},\
  \Eprint {http://arxiv.org/abs/2207.03561} {arXiv:2207.03561 [astro-ph.HE]}
  \BibitemShut {NoStop}%
\bibitem [{\citenamefont {Dighe}\ and\ \citenamefont
  {Smirnov}(2000)}]{Dighe:1999bi}%
  \BibitemOpen
  \bibfield  {author} {\bibinfo {author} {\bibfnamefont {Amol~S.}\ \bibnamefont
  {Dighe}}\ and\ \bibinfo {author} {\bibfnamefont {Alexei~Yu.}\ \bibnamefont
  {Smirnov}},\ }\bibfield  {title} {\enquote {\bibinfo {title} {{Identifying
  the neutrino mass spectrum from the neutrino burst from a supernova}},}\
  }\href {\doibase 10.1103/PhysRevD.62.033007} {\bibfield  {journal} {\bibinfo
  {journal} {Phys. Rev. D}\ }\textbf {\bibinfo {volume} {62}},\ \bibinfo
  {pages} {033007} (\bibinfo {year} {2000})},\ \Eprint
  {http://arxiv.org/abs/hep-ph/9907423} {arXiv:hep-ph/9907423} \BibitemShut
  {NoStop}%
\bibitem [{\citenamefont {Ehring}\ \emph
  {et~al.}(2023{\natexlab{a}})\citenamefont {Ehring}, \citenamefont {Abbar},
  \citenamefont {Janka}, \citenamefont {Raffelt},\ and\ \citenamefont
  {Tamborra}}]{Ehring:2023lcd}%
  \BibitemOpen
  \bibfield  {author} {\bibinfo {author} {\bibfnamefont {Jakob}\ \bibnamefont
  {Ehring}}, \bibinfo {author} {\bibfnamefont {Sajad}\ \bibnamefont {Abbar}},
  \bibinfo {author} {\bibfnamefont {Hans-Thomas}\ \bibnamefont {Janka}},
  \bibinfo {author} {\bibfnamefont {Georg}\ \bibnamefont {Raffelt}}, \ and\
  \bibinfo {author} {\bibfnamefont {Irene}\ \bibnamefont {Tamborra}},\
  }\bibfield  {title} {\enquote {\bibinfo {title} {{Fast neutrino flavor
  conversion in core-collapse supernovae: A parametric study in 1D models}},}\
  }\href {\doibase 10.1103/PhysRevD.107.103034} {\bibfield  {journal} {\bibinfo
   {journal} {Phys. Rev. D}\ }\textbf {\bibinfo {volume} {107}},\ \bibinfo
  {pages} {103034} (\bibinfo {year} {2023}{\natexlab{a}})},\ \Eprint
  {http://arxiv.org/abs/2301.11938} {arXiv:2301.11938 [astro-ph.HE]}
  \BibitemShut {NoStop}%
\bibitem [{\citenamefont {Nagakura}(2023)}]{Nagakura:2023mhr}%
  \BibitemOpen
  \bibfield  {author} {\bibinfo {author} {\bibfnamefont {Hiroki}\ \bibnamefont
  {Nagakura}},\ }\bibfield  {title} {\enquote {\bibinfo {title} {{Roles of Fast
  Neutrino-Flavor Conversion on the Neutrino-Heating Mechanism of Core-Collapse
  Supernova}},}\ }\href {\doibase 10.1103/PhysRevLett.130.211401} {\bibfield
  {journal} {\bibinfo  {journal} {Phys. Rev. Lett.}\ }\textbf {\bibinfo
  {volume} {130}},\ \bibinfo {pages} {211401} (\bibinfo {year} {2023})},\
  \Eprint {http://arxiv.org/abs/2301.10785} {arXiv:2301.10785 [astro-ph.HE]}
  \BibitemShut {NoStop}%
\bibitem [{\citenamefont {Ehring}\ \emph
  {et~al.}(2023{\natexlab{b}})\citenamefont {Ehring}, \citenamefont {Abbar},
  \citenamefont {Janka}, \citenamefont {Raffelt},\ and\ \citenamefont
  {Tamborra}}]{Ehring:2023abs}%
  \BibitemOpen
  \bibfield  {author} {\bibinfo {author} {\bibfnamefont {Jakob}\ \bibnamefont
  {Ehring}}, \bibinfo {author} {\bibfnamefont {Sajad}\ \bibnamefont {Abbar}},
  \bibinfo {author} {\bibfnamefont {Hans-Thomas}\ \bibnamefont {Janka}},
  \bibinfo {author} {\bibfnamefont {Georg}\ \bibnamefont {Raffelt}}, \ and\
  \bibinfo {author} {\bibfnamefont {Irene}\ \bibnamefont {Tamborra}},\
  }\bibfield  {title} {\enquote {\bibinfo {title} {{Fast Neutrino Flavor
  Conversions Can Help and Hinder Neutrino-Driven Explosions}},}\ }\href
  {\doibase 10.1103/PhysRevLett.131.061401} {\bibfield  {journal} {\bibinfo
  {journal} {Phys. Rev. Lett.}\ }\textbf {\bibinfo {volume} {131}},\ \bibinfo
  {pages} {061401} (\bibinfo {year} {2023}{\natexlab{b}})},\ \Eprint
  {http://arxiv.org/abs/2305.11207} {arXiv:2305.11207 [astro-ph.HE]}
  \BibitemShut {NoStop}%
\bibitem [{\citenamefont {{Woosley}}\ \emph {et~al.}(1986)\citenamefont
  {{Woosley}}, \citenamefont {{Wilson}},\ and\ \citenamefont
  {{Mayle}}}]{Woosley:1986}%
  \BibitemOpen
  \bibfield  {author} {\bibinfo {author} {\bibfnamefont {S.~E.}\ \bibnamefont
  {{Woosley}}}, \bibinfo {author} {\bibfnamefont {J.~R.}\ \bibnamefont
  {{Wilson}}}, \ and\ \bibinfo {author} {\bibfnamefont {R.}~\bibnamefont
  {{Mayle}}},\ }\bibfield  {title} {\enquote {\bibinfo {title} {{Gravitational
  Collapse and the Cosmic Antineutrino Background}},}\ }\href {\doibase
  10.1086/163968} {\bibfield  {journal} {\bibinfo  {journal} {\apj}\ }\textbf
  {\bibinfo {volume} {302}},\ \bibinfo {pages} {19} (\bibinfo {year}
  {1986})}\BibitemShut {NoStop}%
\bibitem [{\citenamefont {Vartanyan}\ \emph
  {et~al.}(2019{\natexlab{a}})\citenamefont {Vartanyan}, \citenamefont
  {Burrows}, \citenamefont {Radice}, \citenamefont {Skinner},\ and\
  \citenamefont {Dolence}}]{Vartanyan:2018iah}%
  \BibitemOpen
  \bibfield  {author} {\bibinfo {author} {\bibfnamefont {David}\ \bibnamefont
  {Vartanyan}}, \bibinfo {author} {\bibfnamefont {Adam}\ \bibnamefont
  {Burrows}}, \bibinfo {author} {\bibfnamefont {David}\ \bibnamefont {Radice}},
  \bibinfo {author} {\bibfnamefont {Aaron~M.}\ \bibnamefont {Skinner}}, \ and\
  \bibinfo {author} {\bibfnamefont {Joshua}\ \bibnamefont {Dolence}},\
  }\bibfield  {title} {\enquote {\bibinfo {title} {{A Successful 3D
  Core-Collapse Supernova Explosion Model}},}\ }\href {\doibase
  10.1093/mnras/sty2585} {\bibfield  {journal} {\bibinfo  {journal} {Mon. Not.
  Roy. Astron. Soc.}\ }\textbf {\bibinfo {volume} {482}},\ \bibinfo {pages}
  {351--369} (\bibinfo {year} {2019}{\natexlab{a}})},\ \Eprint
  {http://arxiv.org/abs/1809.05106} {arXiv:1809.05106 [astro-ph.HE]}
  \BibitemShut {NoStop}%
\bibitem [{\citenamefont {Burrows}\ \emph {et~al.}(2019)\citenamefont
  {Burrows}, \citenamefont {Radice},\ and\ \citenamefont
  {Vartanyan}}]{Burrows:2019rtd}%
  \BibitemOpen
  \bibfield  {author} {\bibinfo {author} {\bibfnamefont {Adam}\ \bibnamefont
  {Burrows}}, \bibinfo {author} {\bibfnamefont {David}\ \bibnamefont {Radice}},
  \ and\ \bibinfo {author} {\bibfnamefont {David}\ \bibnamefont {Vartanyan}},\
  }\bibfield  {title} {\enquote {\bibinfo {title} {{Three-dimensional supernova
  explosion simulations of 9-, 10-, 11-, 12-, and 13-M\ensuremath{\odot}
  stars}},}\ }\href {\doibase 10.1093/mnras/stz543} {\bibfield  {journal}
  {\bibinfo  {journal} {Mon. Not. Roy. Astron. Soc.}\ }\textbf {\bibinfo
  {volume} {485}},\ \bibinfo {pages} {3153--3168} (\bibinfo {year} {2019})},\
  \Eprint {http://arxiv.org/abs/1902.00547} {arXiv:1902.00547 [astro-ph.SR]}
  \BibitemShut {NoStop}%
\bibitem [{\citenamefont {Vartanyan}\ \emph
  {et~al.}(2019{\natexlab{b}})\citenamefont {Vartanyan}, \citenamefont
  {Burrows},\ and\ \citenamefont {Radice}}]{Vartanyan:2019ssu}%
  \BibitemOpen
  \bibfield  {author} {\bibinfo {author} {\bibfnamefont {David}\ \bibnamefont
  {Vartanyan}}, \bibinfo {author} {\bibfnamefont {Adam}\ \bibnamefont
  {Burrows}}, \ and\ \bibinfo {author} {\bibfnamefont {David}\ \bibnamefont
  {Radice}},\ }\bibfield  {title} {\enquote {\bibinfo {title} {{Temporal and
  Angular Variations of 3D Core-Collapse Supernova Emissions and their Physical
  Correlations}},}\ }\href {\doibase 10.1093/mnras/stz2307} {\bibfield
  {journal} {\bibinfo  {journal} {Mon. Not. Roy. Astron. Soc.}\ }\textbf
  {\bibinfo {volume} {489}},\ \bibinfo {pages} {2227--2246} (\bibinfo {year}
  {2019}{\natexlab{b}})},\ \Eprint {http://arxiv.org/abs/1906.08787}
  {arXiv:1906.08787 [astro-ph.HE]} \BibitemShut {NoStop}%
\bibitem [{\citenamefont {Nagakura}\ \emph {et~al.}(2019)\citenamefont
  {Nagakura}, \citenamefont {Burrows}, \citenamefont {Radice},\ and\
  \citenamefont {Vartanyan}}]{Nagakura:2019gmh}%
  \BibitemOpen
  \bibfield  {author} {\bibinfo {author} {\bibfnamefont {Hiroki}\ \bibnamefont
  {Nagakura}}, \bibinfo {author} {\bibfnamefont {Adam}\ \bibnamefont
  {Burrows}}, \bibinfo {author} {\bibfnamefont {David}\ \bibnamefont {Radice}},
  \ and\ \bibinfo {author} {\bibfnamefont {David}\ \bibnamefont {Vartanyan}},\
  }\bibfield  {title} {\enquote {\bibinfo {title} {{Towards an Understanding of
  the Resolution Dependence of Core-Collapse Supernova Simulations}},}\ }\href
  {\doibase 10.1093/mnras/stz2730} {\bibfield  {journal} {\bibinfo  {journal}
  {Mon. Not. Roy. Astron. Soc.}\ }\textbf {\bibinfo {volume} {490}},\ \bibinfo
  {pages} {4622--4637} (\bibinfo {year} {2019})},\ \Eprint
  {http://arxiv.org/abs/1905.03786} {arXiv:1905.03786 [astro-ph.HE]}
  \BibitemShut {NoStop}%
\bibitem [{\citenamefont {Warren}\ \emph {et~al.}(2020)\citenamefont {Warren},
  \citenamefont {Couch}, \citenamefont {O'Connor},\ and\ \citenamefont
  {Morozova}}]{Warren:2019lgb}%
  \BibitemOpen
  \bibfield  {author} {\bibinfo {author} {\bibfnamefont {MacKenzie~L.}\
  \bibnamefont {Warren}}, \bibinfo {author} {\bibfnamefont {Sean~M.}\
  \bibnamefont {Couch}}, \bibinfo {author} {\bibfnamefont {Evan~P.}\
  \bibnamefont {O'Connor}}, \ and\ \bibinfo {author} {\bibfnamefont
  {Viktoriya}\ \bibnamefont {Morozova}},\ }\bibfield  {title} {\enquote
  {\bibinfo {title} {{Constraining Properties of the Next Nearby Core-collapse
  Supernova with Multimessenger Signals}},}\ }\href {\doibase
  10.3847/1538-4357/ab97b7} {\bibfield  {journal} {\bibinfo  {journal}
  {Astrophys. J.}\ }\textbf {\bibinfo {volume} {898}},\ \bibinfo {pages} {139}
  (\bibinfo {year} {2020})},\ \Eprint {http://arxiv.org/abs/1912.03328}
  {arXiv:1912.03328 [astro-ph.HE]} \BibitemShut {NoStop}%
\bibitem [{\citenamefont {Baxter}\ \emph {et~al.}(2022)\citenamefont {Baxter}
  \emph {et~al.}}]{SNEWS:2021ewj}%
  \BibitemOpen
  \bibfield  {author} {\bibinfo {author} {\bibfnamefont {Amanda~L.}\
  \bibnamefont {Baxter}} \emph {et~al.} (\bibinfo {collaboration} {SNEWS}),\
  }\bibfield  {title} {\enquote {\bibinfo {title} {{SNEWPY: A Data Pipeline
  from Supernova Simulations to Neutrino Signals}},}\ }\href {\doibase
  10.3847/1538-4357/ac350f} {\bibfield  {journal} {\bibinfo  {journal}
  {Astrophys. J.}\ }\textbf {\bibinfo {volume} {925}},\ \bibinfo {pages} {107}
  (\bibinfo {year} {2022})},\ \Eprint {http://arxiv.org/abs/2109.08188}
  {arXiv:2109.08188 [astro-ph.IM]} \BibitemShut {NoStop}%
\bibitem [{\citenamefont {{Tamborra}}\ \emph {et~al.}(2014)\citenamefont
  {{Tamborra}}, \citenamefont {{Raffelt}}, \citenamefont {{Hanke}},
  \citenamefont {{Janka}},\ and\ \citenamefont {{M{\"u}ller}}}]{Tamborra:2014}%
  \BibitemOpen
  \bibfield  {author} {\bibinfo {author} {\bibfnamefont {Irene}\ \bibnamefont
  {{Tamborra}}}, \bibinfo {author} {\bibfnamefont {Georg}\ \bibnamefont
  {{Raffelt}}}, \bibinfo {author} {\bibfnamefont {Florian}\ \bibnamefont
  {{Hanke}}}, \bibinfo {author} {\bibfnamefont {Hans-Thomas}\ \bibnamefont
  {{Janka}}}, \ and\ \bibinfo {author} {\bibfnamefont {Bernhard}\ \bibnamefont
  {{M{\"u}ller}}},\ }\bibfield  {title} {\enquote {\bibinfo {title} {{Neutrino
  emission characteristics and detection opportunities based on
  three-dimensional supernova simulations}},}\ }\href {\doibase
  10.1103/PhysRevD.90.045032} {\bibfield  {journal} {\bibinfo  {journal}
  {\prd}\ }\textbf {\bibinfo {volume} {90}},\ \bibinfo {eid} {045032} (\bibinfo
  {year} {2014})},\ \Eprint {http://arxiv.org/abs/1406.0006} {arXiv:1406.0006
  [astro-ph.SR]} \BibitemShut {NoStop}%
\bibitem [{\citenamefont {{Takiwaki}}\ and\ \citenamefont
  {{Kotake}}(2018)}]{Takiwaki:2018}%
  \BibitemOpen
  \bibfield  {author} {\bibinfo {author} {\bibfnamefont {Tomoya}\ \bibnamefont
  {{Takiwaki}}}\ and\ \bibinfo {author} {\bibfnamefont {Kei}\ \bibnamefont
  {{Kotake}}},\ }\bibfield  {title} {\enquote {\bibinfo {title} {{Anisotropic
  emission of neutrino and gravitational-wave signals from rapidly rotating
  core-collapse supernovae}},}\ }\href {\doibase 10.1093/mnrasl/sly008}
  {\bibfield  {journal} {\bibinfo  {journal} {Mon. Not. Roy. Astron. Soc.}\
  }\textbf {\bibinfo {volume} {475}},\ \bibinfo {pages} {L91--L95} (\bibinfo
  {year} {2018})},\ \Eprint {http://arxiv.org/abs/1711.01905} {arXiv:1711.01905
  [astro-ph.HE]} \BibitemShut {NoStop}%
\bibitem [{\citenamefont {Lin}\ \emph {et~al.}(2020)\citenamefont {Lin},
  \citenamefont {Lunardini}, \citenamefont {Zanolin}, \citenamefont {Kotake},\
  and\ \citenamefont {Richardson}}]{Lin:2019wwm}%
  \BibitemOpen
  \bibfield  {author} {\bibinfo {author} {\bibfnamefont {Zidu}\ \bibnamefont
  {Lin}}, \bibinfo {author} {\bibfnamefont {Cecilia}\ \bibnamefont
  {Lunardini}}, \bibinfo {author} {\bibfnamefont {Michele}\ \bibnamefont
  {Zanolin}}, \bibinfo {author} {\bibfnamefont {Kei}\ \bibnamefont {Kotake}}, \
  and\ \bibinfo {author} {\bibfnamefont {Colter}\ \bibnamefont {Richardson}},\
  }\bibfield  {title} {\enquote {\bibinfo {title} {{Detectability of standing
  accretion shock instabilities activity in supernova neutrino signals}},}\
  }\href {\doibase 10.1103/PhysRevD.101.123028} {\bibfield  {journal} {\bibinfo
   {journal} {Phys. Rev. D}\ }\textbf {\bibinfo {volume} {101}},\ \bibinfo
  {pages} {123028} (\bibinfo {year} {2020})},\ \Eprint
  {http://arxiv.org/abs/1911.10656} {arXiv:1911.10656 [astro-ph.HE]}
  \BibitemShut {NoStop}%
\bibitem [{\citenamefont {Walk}\ \emph {et~al.}(2019)\citenamefont {Walk},
  \citenamefont {Tamborra}, \citenamefont {Janka},\ and\ \citenamefont
  {Summa}}]{Walk:2019ier}%
  \BibitemOpen
  \bibfield  {author} {\bibinfo {author} {\bibfnamefont {Laurie}\ \bibnamefont
  {Walk}}, \bibinfo {author} {\bibfnamefont {Irene}\ \bibnamefont {Tamborra}},
  \bibinfo {author} {\bibfnamefont {Hans-Thomas}\ \bibnamefont {Janka}}, \ and\
  \bibinfo {author} {\bibfnamefont {Alexander}\ \bibnamefont {Summa}},\
  }\bibfield  {title} {\enquote {\bibinfo {title} {{Effects of the standing
  accretion-shock instability and the lepton-emission self-sustained asymmetry
  in the neutrino emission of rotating supernovae}},}\ }\href {\doibase
  10.1103/PhysRevD.100.063018} {\bibfield  {journal} {\bibinfo  {journal}
  {Phys. Rev. D}\ }\textbf {\bibinfo {volume} {100}},\ \bibinfo {pages}
  {063018} (\bibinfo {year} {2019})},\ \Eprint
  {http://arxiv.org/abs/1901.06235} {arXiv:1901.06235 [astro-ph.HE]}
  \BibitemShut {NoStop}%
\bibitem [{\citenamefont {Bollig}\ \emph {et~al.}(2021)\citenamefont {Bollig},
  \citenamefont {Yadav}, \citenamefont {Kresse}, \citenamefont {Janka},
  \citenamefont {M\"uller},\ and\ \citenamefont {Heger}}]{Bollig:2020phc}%
  \BibitemOpen
  \bibfield  {author} {\bibinfo {author} {\bibfnamefont {Robert}\ \bibnamefont
  {Bollig}}, \bibinfo {author} {\bibfnamefont {Naveen}\ \bibnamefont {Yadav}},
  \bibinfo {author} {\bibfnamefont {Daniel}\ \bibnamefont {Kresse}}, \bibinfo
  {author} {\bibfnamefont {H.~Th.}\ \bibnamefont {Janka}}, \bibinfo {author}
  {\bibfnamefont {Bernhard}\ \bibnamefont {M\"uller}}, \ and\ \bibinfo {author}
  {\bibfnamefont {Alexander}\ \bibnamefont {Heger}},\ }\bibfield  {title}
  {\enquote {\bibinfo {title} {{Self-consistent 3D Supernova Models From
  \ensuremath{-}7 Minutes to +7 s: A 1-bethe Explosion of a \ensuremath{\sim}19
  $M_\odot$ Progenitor}},}\ }\href {\doibase 10.3847/1538-4357/abf82e}
  {\bibfield  {journal} {\bibinfo  {journal} {Astrophys. J.}\ }\textbf
  {\bibinfo {volume} {915}},\ \bibinfo {pages} {28} (\bibinfo {year} {2021})},\
  \Eprint {http://arxiv.org/abs/2010.10506} {arXiv:2010.10506 [astro-ph.HE]}
  \BibitemShut {NoStop}%
\bibitem [{\citenamefont {Fiorillo}\ \emph {et~al.}(2023)\citenamefont
  {Fiorillo}, \citenamefont {Heinlein}, \citenamefont {Janka}, \citenamefont
  {Raffelt}, \citenamefont {Vitagliano},\ and\ \citenamefont
  {Bollig}}]{Fiorillo:2023frv}%
  \BibitemOpen
  \bibfield  {author} {\bibinfo {author} {\bibfnamefont {Damiano F.~G.}\
  \bibnamefont {Fiorillo}}, \bibinfo {author} {\bibfnamefont {Malte}\
  \bibnamefont {Heinlein}}, \bibinfo {author} {\bibfnamefont {Hans-Thomas}\
  \bibnamefont {Janka}}, \bibinfo {author} {\bibfnamefont {Georg}\ \bibnamefont
  {Raffelt}}, \bibinfo {author} {\bibfnamefont {Edoardo}\ \bibnamefont
  {Vitagliano}}, \ and\ \bibinfo {author} {\bibfnamefont {Robert}\ \bibnamefont
  {Bollig}},\ }\bibfield  {title} {\enquote {\bibinfo {title} {{Supernova
  simulations confront SN 1987A neutrinos}},}\ }\href {\doibase
  10.1103/PhysRevD.108.083040} {\bibfield  {journal} {\bibinfo  {journal}
  {Phys. Rev. D}\ }\textbf {\bibinfo {volume} {108}},\ \bibinfo {pages}
  {083040} (\bibinfo {year} {2023})},\ \Eprint
  {http://arxiv.org/abs/2308.01403} {arXiv:2308.01403 [astro-ph.HE]}
  \BibitemShut {NoStop}%
\bibitem [{\citenamefont {{Farmer}}\ \emph {et~al.}(2016)\citenamefont
  {{Farmer}}, \citenamefont {{Fields}}, \citenamefont {{Petermann}},
  \citenamefont {{Dessart}}, \citenamefont {{Cantiello}}, \citenamefont
  {{Paxton}},\ and\ \citenamefont {{Timmes}}}]{2016ApJS..227...22F}%
  \BibitemOpen
  \bibfield  {author} {\bibinfo {author} {\bibfnamefont {R.}~\bibnamefont
  {{Farmer}}}, \bibinfo {author} {\bibfnamefont {C.~E.}\ \bibnamefont
  {{Fields}}}, \bibinfo {author} {\bibfnamefont {I.}~\bibnamefont
  {{Petermann}}}, \bibinfo {author} {\bibfnamefont {Luc}\ \bibnamefont
  {{Dessart}}}, \bibinfo {author} {\bibfnamefont {M.}~\bibnamefont
  {{Cantiello}}}, \bibinfo {author} {\bibfnamefont {B.}~\bibnamefont
  {{Paxton}}}, \ and\ \bibinfo {author} {\bibfnamefont {F.~X.}\ \bibnamefont
  {{Timmes}}},\ }\bibfield  {title} {\enquote {\bibinfo {title} {{On Variations
  Of Pre-supernova Model Properties}},}\ }\href {\doibase
  10.3847/1538-4365/227/2/22} {\bibfield  {journal} {\bibinfo  {journal}
  {Astrophys. J. Suppl.}\ }\textbf {\bibinfo {volume} {227}},\ \bibinfo {eid}
  {22} (\bibinfo {year} {2016})},\ \Eprint {http://arxiv.org/abs/1611.01207}
  {arXiv:1611.01207 [astro-ph.SR]} \BibitemShut {NoStop}%
\bibitem [{\citenamefont {Sukhbold}\ \emph {et~al.}(2016)\citenamefont
  {Sukhbold}, \citenamefont {Ertl}, \citenamefont {Woosley}, \citenamefont
  {Brown},\ and\ \citenamefont {Janka}}]{Sukhbold:2015wba}%
  \BibitemOpen
  \bibfield  {author} {\bibinfo {author} {\bibfnamefont {Tuguldur}\
  \bibnamefont {Sukhbold}}, \bibinfo {author} {\bibfnamefont {T.}~\bibnamefont
  {Ertl}}, \bibinfo {author} {\bibfnamefont {S.~E.}\ \bibnamefont {Woosley}},
  \bibinfo {author} {\bibfnamefont {Justin~M.}\ \bibnamefont {Brown}}, \ and\
  \bibinfo {author} {\bibfnamefont {H.~T.}\ \bibnamefont {Janka}},\ }\bibfield
  {title} {\enquote {\bibinfo {title} {{Core-Collapse Supernovae from 9 to 120
  Solar Masses Based on Neutrino-powered Explosions}},}\ }\href {\doibase
  10.3847/0004-637X/821/1/38} {\bibfield  {journal} {\bibinfo  {journal}
  {Astrophys. J.}\ }\textbf {\bibinfo {volume} {821}},\ \bibinfo {pages} {38}
  (\bibinfo {year} {2016})},\ \Eprint {http://arxiv.org/abs/1510.04643}
  {arXiv:1510.04643 [astro-ph.HE]} \BibitemShut {NoStop}%
\bibitem [{\citenamefont {Burrows}()}]{Burrows}%
  \BibitemOpen
  \bibfield  {author} {\bibinfo {author} {\bibfnamefont {Adam}\ \bibnamefont
  {Burrows}},\ }\href@noop {} {}\bibinfo {note} {Private
  communication}\BibitemShut {NoStop}%
\bibitem [{\citenamefont {Roberts}\ and\ \citenamefont
  {Reddy}(2016)}]{Roberts:2016rsf}%
  \BibitemOpen
  \bibfield  {author} {\bibinfo {author} {\bibfnamefont {Luke~F.}\ \bibnamefont
  {Roberts}}\ and\ \bibinfo {author} {\bibfnamefont {Sanjay}\ \bibnamefont
  {Reddy}},\ }\bibfield  {title} {\enquote {\bibinfo {title} {{Neutrino
  Signatures From Young Neutron Stars}},}\ }\href {\doibase
  10.1007/978-3-319-21846-5\_5} {\  (\bibinfo {year} {2016}),\
  10.1007/978-3-319-21846-5\_5},\ \Eprint {http://arxiv.org/abs/1612.03860}
  {arXiv:1612.03860 [astro-ph.HE]} \BibitemShut {NoStop}%
\bibitem [{\citenamefont {James}(2006)}]{James:2006zz}%
  \BibitemOpen
  \bibfield  {author} {\bibinfo {author} {\bibfnamefont {Frederick}\
  \bibnamefont {James}},\ }\href@noop {} {\emph {\bibinfo {title} {{Statistical
  methods in experimental physics}}}}\ (\bibinfo  {publisher} {World
  Scientific},\ \bibinfo {year} {2006})\BibitemShut {NoStop}%
\end{thebibliography}
%


\end{document}